\documentclass[aps,prc,twocolumn,amsmath,amssymb,superscriptaddress,showpacs,showkeys]{revtex4}

\usepackage{graphicx}
\usepackage{dcolumn}
\usepackage{bm}

\usepackage{amsmath}
\usepackage{amsfonts}
\usepackage{amssymb}
\usepackage{gensymb}	
\usepackage{enumitem}
\usepackage{enumerate}

\begin{document}

\newcommand{\beq}{\begin{equation}}
\newcommand{\eps}{\epsilon}
\newcommand{\eeq}{\end{equation}}
\newcommand{\bea}{\begin{eqnarray}}
\newcommand{\eea}{\end{eqnarray}}

\title{Production of Energetic 
Light Fragments
in CEM, LAQGSM, and MCNP6}

\author{Stepan G. Mashnik}
\email[]{mashnik@lanl.gov}
\affiliation{Los Alamos National Laboratory, Los Alamos, NM 87545, USA}
\author{Leslie M. Kerby}
\email[]{kerblesl@isu.edu}
\affiliation{Los Alamos National Laboratory, Los Alamos, NM 87545, USA}
\affiliation{Idaho State University, Pocatello, ID 83201, USA}
\affiliation{Idaho National Laboratory, Idaho Falls, ID 83402, USA}
\author{Konstantin~K.~Gudima}
\affiliation{Institute of Applied Physics, Academy of Science of Moldova, 
Chi\c{s}in\u{a}u, Moldova}

\author{Arnold J. Sierk}
\email[]{t2ajs@lanl.gov}
\affiliation{Los Alamos National Laboratory, Los Alamos, NM 87545, USA}
\author{Jeffrey S. Bull}
\affiliation{Los Alamos National Laboratory, Los Alamos, NM 87545, USA}
\author{Michael R. James}
\affiliation{Los Alamos National Laboratory, Los Alamos, NM 87545, USA}

\date{\today}

\begin{abstract}
We extend the
cascade-exciton model (CEM), and the Los Alamos
version of the quark-gluon string model (LAQGSM), event generators
of the Monte-Carlo N-particle transport code version 6 (MCNP6),
to describe production of energetic light fragments (LF) heavier than
$^4$He from various nuclear reactions induced by particles and
nuclei at energies up to about 1 TeV/nucleon.
In these models, energetic LF can be produced via 
Fermi break-up, preequilibrium emission, and coalescence of cascade particles.
Initially, we study several variations of the Fermi break-up
model and choose the best option for these models. Then, we extend
the modified exciton model (MEM) used by these codes to 
account for a possibility of multiple emission of up
to 66 types of particles and LF (up to $^{28}$Mg) at the preequilibrium
stage of reactions. Then, we expand the coalescence model
to allow coalescence of LF from nucleons emitted at the intranuclear
cascade stage of reactions and from lighter clusters, up to fragments with mass
numbers $A \le 7$, in the case of CEM, and $A \le 12$, in the case
of LAQGSM. Next, we modify MCNP6 to allow calculating and outputting
spectra of LF and heavier products with arbitrary mass and charge numbers.
The improved version of CEM is implemented into MCNP6.
Finally, we test the improved versions of CEM,
LAQGSM, and MCNP6 on a variety of measured nuclear reactions.  The modified
codes give an improved description of energetic LF
from particle- and nucleus-induced reactions; showing a
good agreement with a variety of available experimental data. They have an
improved predictive power compared to the previous versions and can be used
as reliable tools in simulating applications involving such types of reactions.
\end{abstract}



\maketitle

\section{Introduction}
\label{chap:introduction}
The Los Alamos National Laboratory (LANL)
Monte-Carlo N-particle transport code MCNP6 \cite{MCNP6}
uses by default the latest version of the cascade-exciton model
(CEM), CEM03.03 \cite{CEM03.03, CEM, Trieste08}, as its event generator
to simulate reactions induced by nucleons, pions, and 
photons of energies up to 4.5 GeV
and the Los Alamos version of the quark-gluon string model 
(LAQGSM), LAQGSM03.03 \cite{Trieste08, LAQGSM, LAQGSM03.03}, to simulate 
such reactions at higher energies, as well as reactions induced by other
elementary particles and by nuclei with energies up to $\sim 1$ TeV/nucleon.

MCNP6 is used around the world by several thousands of users
in applications ranging from radiation protection and dosimetry,
nuclear-reactor design, nuclear criticality safety, detector design and analysis,
de-contamination and de-commissioning, accelerator applications, medical physics,
space research, and beyond. This is why it is important that MCNP6 predicts
as well as possible arbitrary nuclear reactions,
including production of energetic light fragments (LF).

At lower energies, MCNP6 uses tables of evaluated
nuclear data (referred to as ``data libraries''), 
while for higher energies ($>150$ MeV), MCNP6 uses 
CEM03.03 and LAQGSM03.03 as mentioned above,
as well as by default for some reactions, or when chosen 
by users, the Bertini intranuclear cascade (INC)~\cite{BertiniINC}, 
ISABEL \cite{ISABEL}, or the INC
developed  at Liege (INCL) by Cugnon and colleagues
from CEA/Saclay, France, version INCL4.2
\cite{INCL4.2}, merged with
the evaporation/fission and Fermi break-up models available in MCNP6
(see details in \cite{MCNP6}). 

Emission of energetic heavy clusters from nuclear reactions
play a critical role in several applications, including electronics performance
in space, human radiation dosages in space or other extreme radiation
environments, proton and heavy-ion therapy in cancer treatment, accelerator
and shielding applications, and more. 

Understanding the production of LF
is very interesting also from a scientific point of view, as there is still
uncertainty about the dependences of the different reaction mechanisms 
on the energy of the inducing particle, the
mass number of the target, and the type and emission energy of the fragments.
To the best of our knowledge, none of the currently available simulation tools
are able to accurately predict emission of LF from
arbitrary reactions. This research may help to understand better the
mechanisms of nuclear reactions at intermediate and high energies. 

This work focuses significantly on the emission of high-energy LF at the
preequilibrium stage of nuclear reactions, as considered in these models.
However, high-energy LF can be produced by other reaction mechanisms. For
example, Cugnon et al. have extended their Li\`{e}ge intranuclear cascade (INCL)
code to consider emission of LF heavier than $^4$He during the cascade
stage of reactions via coalescence of several nucleons at the nuclear periphery
\cite{Cugnon}. But INCL has not yet been generalized across all types
of nuclear reactions; it does not work yet for heavy-ion induced reactions
and is currently limited to incident energies only below several GeV/nucleon
The most advanced versions of INCL so far published work only
for projectiles with $A \le 18$ and at incident energies below
15--20 GeV/nucleon.

Emission of $^7$Be at the preequilibrium stage (described by a hybrid exciton
model and coalescence pick-up model) was studied by A. Yu. Konobeyev and Yu. A.
Korovin two decades ago \cite{Konobeyev}. Preequilibrium emission
of helium and lithium ions was discussed in Ref. \cite{Uozumi}. 
Preequilibrium emission of light fragments was also studied within the CEM in 
2002 \cite{SantaFe2002}, but that project was never completed. 

Besides preequilibrium emission, energetic fragments can be produced also 
via Fermi break-up \cite{Fermi} and
multifragmentation processes, as described, e.g., by the statistical
multifragmentation model (SMM)~\cite{SMM}.

Finally, energetic LF can also be produced at the earliest
stages of nuclear reactions
as described by various versions of quantum molecular dynamics (QMD)
(see, e.g., Ref.\ \cite{QMD} and references therein).
QMD is a very promising approach to describe nuclear reactions
and may become in the future one of the most important ``workhorses'' 
in transport codes.  However, as of today,
it does not have as good a predictive power as do simpler and much faster
INC-type models. We are not aware of any publication where
LF spectra are predicted well by a version of QMD. In addition, as was
determined by three international comparisons of models and codes for
spallation reaction applications performed since 1992 under the auspices of 
the Nuclear Energy Agency of the Organization for Economic Co-operation
and Development (NEA/OECD) and 
International Atomic Energy Agency
(IAEA), generally, all tested versions of QMD showed
a worse predictive power in comparison with INC-type models.
In addition, current QMD codes are about 100--300 times slower than INC-type
event generators, which makes them less practical for complex 
simulation applications, even those using the currently available supercomputers.
Therefore, QMD is not yet so widely used in realistic nuclear simulation\
applications (see details and references, e.g., in \cite{Spall-Handbook, David2015}).

Lastly, the authors of most of the recent
measurements of LF spectra analyze their experimental data using different
simplified approaches assuming emission of LF from moving sources
(see, e.g., Refs.\ \cite{Machner, Budzanowski, BudzanowskiNi}). Such
simplified moving-source prescriptions are fitted to describe as well as possible
only their own measured LF spectra, and have not been developed further to become
universal models with predictive power for spectra of LF from arbitrary
reactions. Such approaches cannot describe at all many other
characteristics of nuclear reactions, like the yields and energies of spallation
products, fission-fragment production, etc., therefore can not
be used as event generators in transport codes.

For detailed information on spallation reactions, models, and researches, see the book
{\it Handbook of Spallation Research}, by Filges and Goldenbaum~\cite{Spall-Handbook}. 
A useful recent summary paper by David, on spallation models,
is available in Ref.\ \cite{David2015}.

\begin{figure}[htp]
\centering
\includegraphics[trim = 1mm 2mm 7mm 0.5mm, width=80mm]{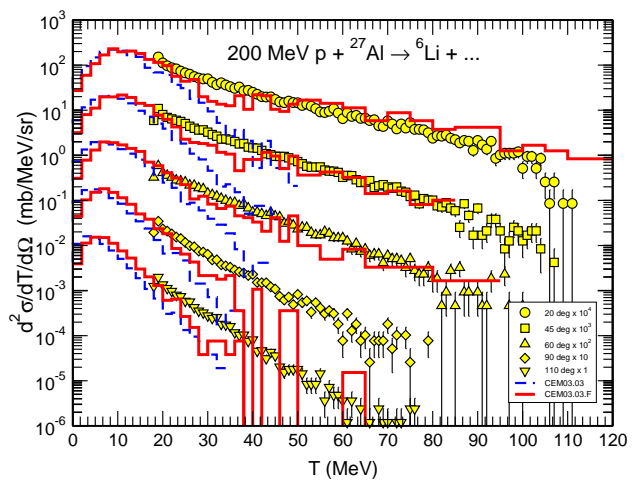}   
\caption{(Color online)
Comparison of experimental $^6$Li spectra at 20, 45, 60, 90, and 110 degrees 
by Machner et al.\ \cite{Machner}
(symbols) with calculations by the unmodified CEM03.03 (dashed 
histograms) and results by the newly revised CEM03.03.F 
(solid histograms), as discussed in the text.}
\label{fig:p200AlCompOld}
\end{figure}

The CEM and LAQGSM event generators in MCNP6 describe quite
well the spectra of emitted particles and of
fragments with sizes up to $^{4}$He across a broad range of
target masses and incident energies (up to $\sim 5$~GeV for CEM and up to
$\sim 1$~TeV/nucleon for LAQGSM),
as well as the yields of most spallation and fission products
(see, e.g., Refs.\ \cite{Trieste08,EPJ-Plus2011,Tokyo2014} and references
therein).
However, as shown by dashed histograms
in Fig.\ \ref{fig:p200AlCompOld}, these models
do not predict well the high-energy tails of LF spectra heavier than $^4$He. 

This is true for other projectiles, incident energies,
and target mass numbers for all fragments heavier than $^4$He.  At lower
energies of ejectiles ($\lesssim 25$~MeV), CEM describes well the data, but for
intermediate energies ($\gtrsim 25$~MeV) the CEM predictions fall off sharply.
This is because the only currently included mechanism for producing $^6$Li
fragments is evaporation, which considers emission of LF (up to $^{28}$Mg).
At higher energies ($\gtrsim 25$~MeV), the fragments should
largely be produced by these models at the preequilibrium stage which would
require an improved modified exciton model (MEM), as well as
a contribution from the coalescence of nucleons produced in the INC with
$A > 4$.  Neither the MEM nor the coalescence model
used by the 03.03 versions of CEM and LAQGSM
considers these heavier fragments.

The aim of this work is to extend the precompound model in these event
generators to include such processes, leading to an increase of predictive power
for LF production in MCNP6. This entails upgrading the MEM
currently used at the preequilibrium stage in CEM and LAQGSM. It also includes
verifying and extending the coalescence and the Fermi break-up models used in the
precompound stages of spallation reactions within CEM and LAQGSM.

\section{CEM and LAQGSM overview} 

Details, examples of results, and useful references to different versions of CEM
and LAQGSM may be found in a recent lecture \cite{Trieste08}.

The cascade-exciton model of nuclear reactions was proposed more than 30
years ago at the Laboratory of Theoretical Physics, JINR, Dubna, USSR by Gudima,
Mashnik, and Toneev \cite{CEM}. It is based on the standard (non
time-dependent) Dubna intranuclear cascade model \cite{Barashenkov1972,
Barashenkov1973} and the modified exciton model \cite{MEM, MODEX}.
The code LAQGSM03.03 is the latest modification~\cite{LAQGSM03.03} of LAQGSM
\cite{LAQGSM}, which in its turn is an improvement of the quark-gluon string
model (QGSM) \cite{Amelin}. It describes reactions induced by both particles and
nuclei at incident energies up to about 1 TeV/nucleon.

The basic versions of both the CEM and LAQGSM event generators are the so-called
``03.03'' versions, namely CEM03.03 \cite{CEM03.03, CEM, Trieste08} and
LAQGSM03.03 \cite{Trieste08, LAQGSM, LAQGSM03.03}. The CEM code calculates
nuclear reactions induced only by nucleons, pions, and photons. It assumes that the
reactions occur in three stages. The
first stage is the INC, in which primary particles can be re-scattered and
produce secondary particles several times prior to absorption by, or escape from,
the nucleus. When the cascade stage of a reaction is completed, CEM uses the
coalescence model to create high-energy $d$, $t$, $^3$He, and $^4$He by
final-state interactions among emitted cascade nucleons outside the target.
The emission of the cascade particles determines the particle-hole configuration,
$Z$, $A$, and the excitation energy that comprise the starting conditions for the
second, preequilibrium stage of the reaction. The subsequent relaxation of the
nuclear excitation is treated in terms of an improved version of the MEM
of preequilibrium decay, followed by the equilibrium
evaporation/fission stage described using a modification of the 
generalized evaporation model (GEM) code GEM2 by Furihata \cite{GEM2}.

Generally, all three components may contribute to experimentally measured
particle spectra and other distributions. But if the residual nuclei after the
INC have atomic numbers with $A \leq A_{Fermi} = 12$, CEM uses the Fermi break-up
model to calculate their further disintegration instead of using the
preequilibrium and evaporation models. Fermi break-up, which estimates the
probabilities of various final states by calculating the approximate phase
space available for each configuration, is much faster to calculate and gives
results very similar to those from using the continuation of the more
detailed models for lighter nuclei. LAQGSM also describes nuclear reactions,
as a three-stage process: an INC, followed by preequilibrium emission of particles
during the equilibration of the excited residual nuclei formed after the INC,
followed by evaporation of particles from and/or fission of the compound nuclei.
LAQGSM was developed with a primary focus on describing reactions induced by
nuclei, as well as induced by most elementary particles, at high energies, up to
about 1 TeV/nucleon. The INC of LAQGSM is completely different from that in CEM.
LAQGSM also considers Fermi break-up of nuclei with $A \leq 12$ produced after
the cascade, and the coalescence model to create high-energy $d$, $t$, $^3$He,
and $^4$He from nucleons emitted during the INC.

From this brief overview of these models, it is clear that energetic LF can
only be produced in this approach through one of the following three processes:
Fermi break-up, preequilibrium emission, and coalescence.
Below, we explore each of these mechanisms.

Many people participated in the development of CEM and LAQGSM
over their more than 40-year history.  Contributors to the ``03.03'' versions are
S. G. Mashnik, K. K. Gudima, A. J. Sierk, 
\framebox{
R. E. Prael, 
}
M. I. Baznat, and N. V. Mokhov. L. M. Kerby joined these efforts
recently, primarily to extend the precompound models of CEM
and LAQGSM by accounting for possible emission of LF
heavier than $^4$He, specifically up to $^{28}$Mg.

For more details on the physics of CEM and LAQGSM, see Ref.~\cite{Trieste08} 
and references therein.

\section{Fermi Break-up}
\label{FBU}
Generally, after the fast INC stage of a nuclear reaction, a much slower
evaporation/fission stage follows, with or without taking into account
an intermediate preequilibrium stage between the INC and 
the equilibrated evaporation/fission.
Such a picture is well grounded in cases of heavy nuclei,
as both evaporation and fission models are based on 
statistical assumptions, requiring a large number of nucleons.
Naturally, in the case of light nuclei with only a few nucleons, 
statistical models are less well justified.  
In addition, such light nuclei like carbon and oxygen exhibit
considerable alpha-particle clustering, not accounted for in
evaporation/fission models. This is why in the case of light
excited nuclei, their deexcitation is often calculated using
the so called ``Fermi break-up'' model, suggested initially by Fermi
\cite{Fermi}.

It is impossible to measure all nuclear data needed
for applications involving light target nuclei; 
therefore, Monte-Carlo transport codes are usually used
to simulate fragmentation reactions. It is important that
available transport codes predict such reactions as well as possible. 
For this reason, efforts have been made recently to investigate the
validity and performance of, and to improve where possible, nuclear reaction
models used by such transport codes
as GEANT4 (e.g., \cite{GEANT4}), SHIELD-HIT (e.g., \cite{SHIELD}), PHITS 
(e.g., \cite{PHITS}), as well as MCNP6 (e.g., ~\cite{NIMA2014, NIMB2015}). 

Deexcitation of light nuclei with $A \leq A_{\rm Fermi}$ remaining after the INC
is described in CEM and LAQGSM only with the Fermi break-up model, where 
$A_{\rm Fermi}$ is a ``cut-off value'' fixed in the models. The value of
$A_{\rm Fermi}$ is a model parameter, not a physical characteristic of nuclear
reactions. Actually, the initial version of the Fermi break-up model incorporated
into CEM and LAQGSM (see details in Ref.\ \cite{Trieste08}) used $A \leq A_
{\rm Fermi} =
16$, just as $A_{\rm Fermi} = 16$ is used currently in GEANT4 (see \cite{GEANT4})
and in SHIELD-HIT (see \cite{SHIELD}). But that initial version
of the Fermi break-up model had some problems and caused code crashes in some 
cases (see details in Ref. \cite{Trieste08}).
To avoid unphysical results and code crashes, we chose the expedient of using
$A_{\rm Fermi} = 12$ in both CEM and LAQGSM. Later, the problems in the Fermi
break-up model were fixed in the codes, 
but the value of $A_{\rm Fermi}$ was not changed at
that time, nor was how its value affects the final results of these codes
studied. We address this in our current work, calculating spectra of emitted
particles and LF, and yields of all possible products from various reactions
using different values for $A_{\rm Fermi}$.

One of the most difficult tasks for any theoretical model is to predict cross
sections of arbitrary products as functions of the incident energy of the
projectile initiating the reaction, i.e., ``excitation functions.'' Therefore, we
start the study by comparing the available experimental data on
excitation functions of products from several proton-induced reactions on light
nuclei at intermediate energies with predictions by MCNP6 using its default event
generator for such reactions, CEM03.03, as well as with results calculated by
CEM03.03 used as a stand-alone code.

\begin{figure}
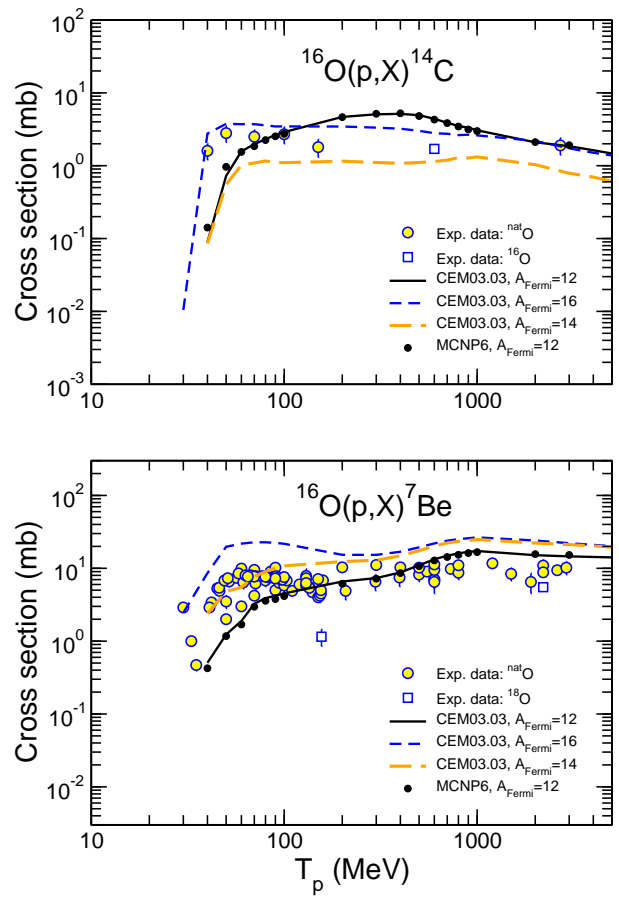

\centering
\includegraphics[width=80mm]{C14.eps}                     
\includegraphics[width=80mm]{Be7.eps}
\caption{ (Color online)
Excitation functions for the
 production of $^{14}$C and $^{7}$Be, calculated with
 CEM03.03 using the ``standard'' version of the Fermi break-up model
 ($A_{\rm Fermi} = 12$) and with cut-off values $A_{\rm Fermi}$ of 14 and 16
(lines), as well as with MCNP6 using CEM03.03 ($A_{\rm Fermi} = 12$; solid\
 points) compared with experimental data (open symbols), as indicated.
  Experimental data are from the T16 Lib compilation \cite{T16}.}
\label{fig:Fermi8}
\vspace*{-4mm}
\end{figure}

We show as examples two excitation functions, 
for proton-induced reactions on $^{16}$O.
Many more results can be found in Ref.~\cite{NIMA2014}. 
Fig.~\ref{fig:Fermi8} presents results for the reaction $p$ + $^{16}$O. Most of
the experimental data for these reactions were measured on $^{nat}$O targets,
with only a few data points obtained for pure $^{16}$O; all the calculations use
$^{16}$O. For these reactions, we perform three sets of calculations, using
$A_{\rm Fermi} = 12$, $14$, and $16$ in CEM03.03. The general
agreement/disagreement of the results with available measured data for oxygen is
very similar to what was displayed in Ref.~\cite{NIMA2014} for $p$ + $^{14}$N,
$^{27}$Al, or $^{nat}$Si. 

Our results demonstrate very good agreement between the excitation
functions simulated by MCNP6 using CEM03.03 and calculations by the stand-alone
CEM03.03, and a reasonable agreement with most of the available experimental
data.  This serves as a validation of MCNP6 and demonstrates there are no
problems with the incorporation of CEM03.03 into MCNP6 or with the simulations
of these reactions by either code.

The observed discrepancies between some calculated
excitation functions and measured data at energies below 20 MeV are not of
concern for our current emphasis:.
As a default, MCNP6 uses data libraries at such low energies and
never uses CEM03.03 or other event generators, when data libraries are available,
as is the case for the reactions studied here. By contrast, CEM uses its INC to
simulate the first stage of nuclear reactions, and the INC is not expected to
work properly at such low energies (see details in \cite{CEM03.03, Trieste08}).

Results calculated with the values of $A_{\rm Fermi} = 12$, 14, and 16 all agree
reasonably well with available data, taking into account that all calculations,
at all energies and for all reactions are done with the default versions of these
codes, without varying any parameters. However, in some cases, there
are significant differences between excitation functions calculated with
$A_{\rm Fermi} = 12$ and $16$.

For many cases, a better description of the heavier fragments occurs for
$A_{\rm Fermi} = 16$ or $14$, and usually the LF are better
described using $A_{Fermi} = 12$. However, the
model with any of these values agrees reasonably well with the measured data,
especially for LF with $Z \leq 4$ (e. g.,~\cite{NIMA2014}). For LF with
$Z > 4$, it is difficult to determine 
which value agrees better with the data:
$A_{\rm Fermi} = 12$ or $A_{\rm Fermi} = 16$. Light fragments with $Z = 3$ and 4
are described a little better with $A_{\rm Fermi} = 12$. As discussed below,
preequilibrium emission described with an extended
version of the MEM (not accounted for in the calculations shown in 
Fig.~\ref{fig:Fermi8}), can be important and may change the final CEM results for
such reactions; therefore, we do not make yet a final decision about which
value of the Fermi break-up cut-off works better; keeping the previous value of 12.

After analyzing all excitation functions for the light targets where Fermi
break-up dominates,
for which we found reliable experimental data, we then study spectra of 
particles and LF from proton-induced reactions on light nuclei, where
the Fermi break-up mechanism should manifest itself most clearly.  We show only
two examples of double differential spectra.  Many more examples are
presented in Ref.~\cite{NIMA2014}, some of which
address different reaction mechanisms for fragment production, with some
involving more than one mechanism in the production of the same LF in a given
reaction.

\begin{figure}
\centering
\includegraphics[width=60mm,angle=-90]{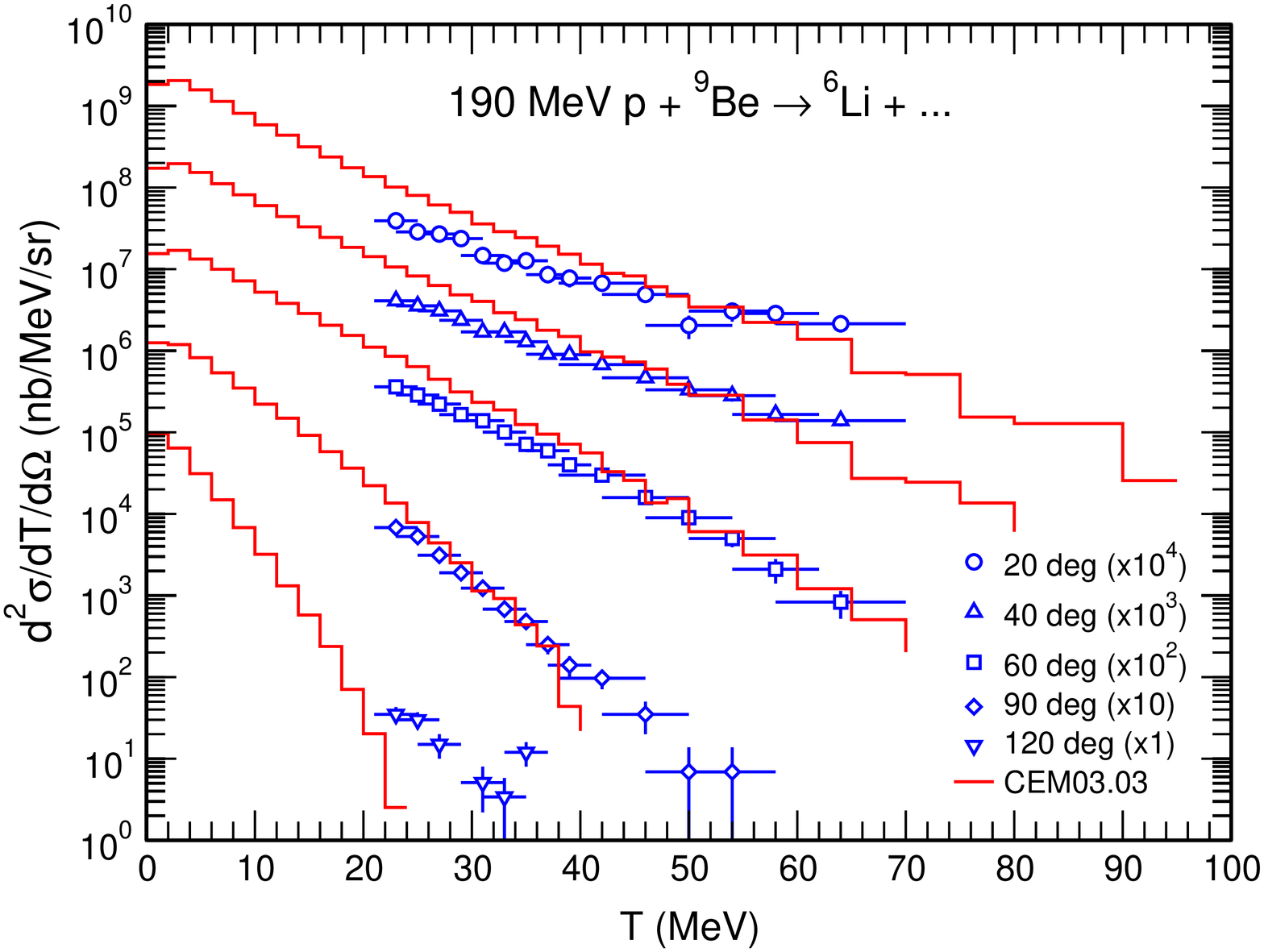}       
\includegraphics[width=60mm,angle=-90]{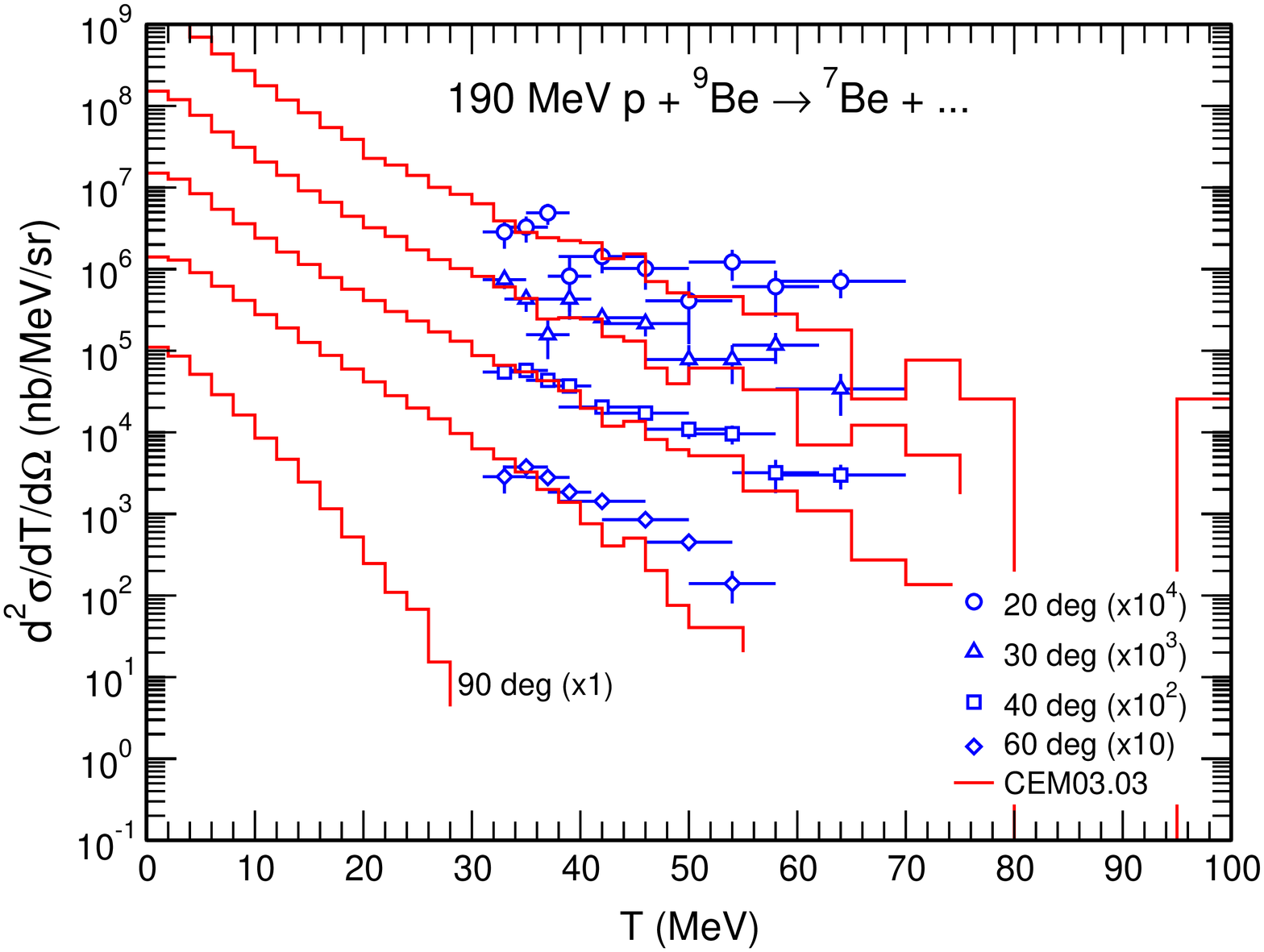}
\caption{ (Color online)
Examples of measured $^6$Li and $^7$Be double-differential spectra from $p$ +
 $^9$Be at 190 MeV \cite{Green} (open symbols), compared to CEM results 
 (histograms).}
\label{fig:Fermi22}
\end{figure}

Fig.~\ref{fig:Fermi22} shows examples of measured $^6$Li and $^7$Be spectra
from $p$ + $^9$Be at 190 MeV~\cite{Green},
compared to CEM results. Because $^9$Be has a mass number
$A < A_{\rm Fermi} = 12$, all the LF from these reactions are calculated
either as fragments from the Fermi break-up of the excited nuclei remaining after
the initial INC stage, or as residual nuclei after emission of several 
particles from the $^9$Be target nucleus during the INC. No preequilibrium
or evaporation mechanisms are considered.  There is a reasonably good agreement
of the CEM predictions with the measured spectra from
all reactions we tested, at different incident energies, from
different light target nuclei, and for all products 
where we found experimental data:
protons, complex particles, and LF heavier than $^4$He
(see examples of more results and details in Ref. \cite{NIMA2014}).

\begin{figure}[]
\centering
\hspace*{0.5mm}
\includegraphics[width=60mm,angle=-90]{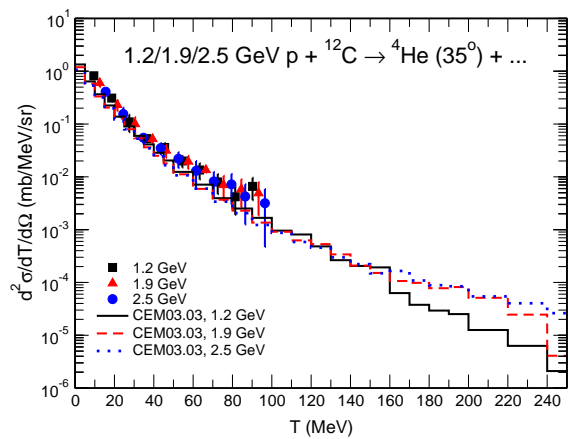}             
\caption{
(Color online) $^4$He spectra at 35\degree for 1.2/1.9/2.5 GeV $p$ + $^{12}$C 
measured by
 M. Fidelus of the PISA collaboration \cite{Fidelus} (solid symbols) compared to
 calculations by CEM03.03 (histograms).}
\label{LimFrag1}
\end{figure}

As a particular case, we test how well the Fermi break-up model used
in these codes describes so-called ``limiting fragmentation'' reactions.
The limiting fragmentation hypothesis, first proposed by Benecke 
et al.\ \cite{Benecke}, suggests that fragmentation cross sections reach asymptotic
values at sufficiently high incident-projectile energies. In other words,
above a given bombarding energy, both the differential and total production cross
sections remain constant. 
Fig.~\ref{LimFrag1} illustrates the validity of
the limiting fragmentation hypothesis for the $^4$He spectra at 35 degrees
from 1.2/1.9/2.5 GeV $p$ + $^{12}$C reactions measured by M. Fidelus
of the PISA collaboration \cite{Fidelus}.

In Refs.\ \cite{Leslie-thesis, RMPmanuscript}, we show similar
results calculated by MCNP6 using CEM03.03, as well as a comparison of
MCNP6 results with the yields 
(total production cross sections) of all measured fragments, from
protons to $^{12}$N, from the same reactions. 

We conclude that the limiting fragmentation hypothesis is supported by
measurements for $p$ + $^{12}$C interactions, and predicted by these models,
in which the Fermi break-up mechanism plays a major role.

An independent test of the Fermi break-up model used in CEM03.03 and LAQGSM03.03
was performed recently by Konobeyev and Fischer \cite{KonFisch2014} for the Fall
2014 Nuclear Data Week. These authors calculated with MCNP6 using its Bertini
\cite{BertiniINC}, ISABEL~\cite{ISABEL}, INCL4.2+ABLA
\cite{INCL4.2, ABLA}, and CEM03.03 event generators~\cite{CEM03.03}, as well as
with the TALYS code \cite{TALYS}, all the experimental spectra of $^3$He and
$^4$He measured in Ref.\ \cite{Green} from the reaction 190 MeV $p$ + $^9$Be; all
spectra of $p$, $d$, $t$, $^3$He, and $^4$He from the reaction of 300 MeV $p$ +
$^9$Be\ \cite{Green}, as well as all neutron spectra from interactions of 113 MeV
protons with $^9$Be\ \cite{Meier1989} and from 256 MeV $p$ + 
$^9$Be\ \cite{Meier1992}. As is often done in the literature, to get quantitative
estimates of the degree of fidelity to data of the spectra calculated by
different models, the authors performed a detailed
statistical analysis using nine different ``deviation factors,'' namely, $H$,
$R^{CE}$, $R^{EC}$, $<F>$, $S$, $L$, $P_{2.0}$, $P_{10.0}$, and $N_x$. The
definition of each can be found in Ref.\ \cite{KonFisch2014}. 
The authors found that results by CEM03.03 for these particular
reactions agree better with the experimental data than all the other models
tested. As $^9$Be has a mass number of 9, all these reactions are calculated
using the INC followed by the Fermi break-up model. The better
results from CEM03.03 in comparison to the other models prove that the Fermi
break-up model used by CEM03.03 and LAQGSM03.03 in MCNP6 is reliable and
can be used with confidence as a good predictive tool for various nuclear
applications and academic studies. 

\section{Extending the Preequilibrium Model} 
\label{preeq-exp}

The preequilibrium interaction stage of nuclear reactions is considered by the
current CEM and LAQGSM in the framework of the latest version of the
MEM \cite{MEM, MODEX}, as described in Refs.\ \cite{CEM03.03,Trieste08}.
At the preequilibrium stage of a reaction, CEM03.03 and LAQGSM03.03 take into
account all possible nuclear transitions changing the number of excitons $n$
with $\Delta n = +2$, -2, and 0, as well as all possible multiple subsequent
emissions of $n$, $p$, $d$, $t$, $^3$He, and $^4$He. The corresponding system 
of master
equations describing the behavior of a nucleus at the preequilibrium stage is
solved by the Monte-Carlo technique \cite{CEM}. In this section, 
we extend the MEM to include the possibility of emitting heavy
clusters, with $A > 4$, up to $^{28}$Mg. 

The probability of finding the nuclear system at time $t$ in the $E\alpha$
state, $P(E,\alpha,t)$, is describe in MEM by the differential equation:
\begin{equation}
\begin{split}
\frac{\delta P(E,\alpha,t)}{\delta t} = & \sum_{\alpha' \neq \alpha}[\lambda(E\alpha,E\alpha')P(E,\alpha',t) \\
            & - \lambda(E\alpha',E\alpha)P(E,\alpha,t)] .
\label{Master}
\end{split}
\end{equation}
Here $\lambda(E\alpha,E\alpha')$ is the energy-conserving probability rate,
defined in the first-order of the 
time-dependent perturbation theory as
\begin{equation}
\lambda (E\alpha,E\alpha') = \frac{2\pi}{\hbar} |<E\alpha|V|E\alpha'>|^2 \omega_{\alpha}(E) ,
\label{LambdaGeneral}
\end{equation}
where $\hbar$ is Planck's constant divided by $2\pi$.
The matrix element $<E\alpha|V|E\alpha'>$ is believed to be a smooth function of
energy, and $\omega_\alpha(E)$ is the density of the final states of the system.
We note that Eq.~(\ref{Master}) is derived assuming that the memory time
$\tau_{mem}$ of the system is small compared to the characteristic time for
intranuclear transitions $\hbar / \lambda(E\alpha,E\alpha')$ but, on the other
hand, Eq.~(\ref{Master}) itself is applicable for times $t \gg \hbar
/ \lambda(E\alpha,E\alpha')$. Due to the condition $\tau_{mem} \gg \hbar /
\lambda(E\alpha,E\alpha')$, being described by Eq.~(\ref{Master}), the random
process is a Markovian one.  

The MEM \cite{MEM, MODEX} utilized by CEM
and LAQGSM uses effectively the relationship of the master equation
(\ref{Master}) with Markovian random processes. Indeed, an attainment of the
statistical equilibration described by Eq.~(\ref{Master}) is an example of a
discontinuous Markovian process: the temporal variable changes continuously
and at a random moment the state of the system changes by a discontinuous jump;
the behavior of the system at the next moment being completely defined by its
present state. As long as the transition probabilities 
$\lambda(E\alpha,E\alpha')$ are time-independent, the waiting time for the system
in the $E\alpha$ state has an exponential distribution (Poisson flow) with the
average lifetime $\hbar / \Lambda(\alpha,E) = \hbar / 
\sum_{\alpha'}{\lambda(E\alpha,E\alpha')}$. This prompts a simple method of solving
the related system of Eq.~(\ref{Master}): simulation of the random process by the
Monte-Carlo technique. In this treatment, it is possible to generalize the
exciton model to all nuclear transitions with $\Delta n = 0, \pm 2$, and the
multiple emission of particles and to depletion of nuclear states due to particle
emission. In this case the system (\ref{Master}) becomes~\cite{CEM}:
\begin{equation}
\begin{split}
\frac{\delta P(E,\alpha,t)}{\delta t} & = -\Lambda(n,E)P(E,n,t) \\
	& + \lambda_+(n-2,E)P(E,n-2,t) \\
	& + \lambda_0(n,E)P(E,n,t) \\ 
	& + \lambda_-(n+2,E)P(E,n+2,t) \\
	& + \sum_j \int dT \int dE' \lambda_j (n,E,T) \\
	& \times P(E',n+n_j,t)\delta(E' -E-B_j-T).
\end{split}
\label{Probability}
\end{equation}
With the master equation~(\ref{Probability}), we can find the particle
emission rates $\lambda_j$ and the exciton transition rates $\lambda_+$,
$\lambda_0$, and $\lambda_-$.

According to the detailed balance principle, the emission width $\Gamma _{j}$,
can be estimated as~\cite{CEM}: 
\begin{equation}
\Gamma_{j}(p,h,E) = \int_{V_j^c}^{E-B_j} \lambda_j (p,h,E,T)dT ,
\label{GammaLambda}
\end{equation}
where the partial transmission probabilities, $\lambda_j$, are equal to
\begin{equation}
\begin{split}
\lambda_j (p,h,E,T) = & \frac{2s_j + 1}{\pi^2\hbar^3} \mu_j \frac{\omega 
(p-1,h,E-B_j-T)}{\omega (p,h,E)} \\
& \times \Re (p,h) T \sigma_j^{inv} (T) \mbox{ ,}
\end{split}
\label{LambdaTransmission}
\end{equation}
where $p$, $h$, $E$, and $\omega$
are the number of particle excitons, the number of hole excitons, the excitation
energy of the excited nucleus, and the
level density of its $n$-exciton state,
while
$s_j$, $B_j$, $V_j^c$, $\mu_j$, $T$, and $\sigma_j^{inv}$ 
 are the spin, the binding energy,
the Coulomb barrier, the reduced mass,
the kinetic energy, and the  inverse cross section
 of the emitted particle $j$, respectively.
The factor $\Re_j(p,h)$ ensures the condition for
the exciton chosen to be the particle of type $j$ and can easily be 
calculated by the Monte-Carlo technique.

Eq.~(\ref{LambdaTransmission}) describes the emission of neutrons and
protons only (an extension of
Eq.~(\ref{LambdaTransmission}) for the case of complex particles can be
found in Ref.~\cite{CEM}).
For complex particles, the level density formula $\omega$ becomes more
complicated and an extra factor $\gamma_j$ must be introduced (e. g., \cite{CEM}):
\begin{equation}
\gamma_j \approx p_j^3 (\frac{p_j}{A})^{p_j - 1} .
\label{GammaBeta}
\end{equation}

Eq.~(\ref{GammaBeta}) for $\gamma_j$ is actually only a rough
estimation that is refined
in CEM03.03
by parameterizing it over a mesh of residual nuclear
energy and mass number (e.g., \cite{CEM03.03}). 

Assuming an equidistant level scheme with the single-particle density $g$,
the level density of the $n$-exciton state is~\cite{Ericson}
\begin{equation}
\omega(p,h,E) = \frac{g (gE)^{p+h-1}}{p! h! (p+h-1)!} \mbox{ .}
\label{OmegaGeneral}
\end{equation}
This expression should be substituted into Eq.~(\ref{LambdaTransmission}) to obtain
the transmission rates $\lambda_j$. 

According to Eq.~(\ref{LambdaGeneral}), for a preequilibrium nucleus with
excitation energy $E$ and number of excitons $n=p+h$, the partial transition
probabilities changing the exciton number by $\Delta n$ are
\begin{equation}
\lambda_{\Delta n} (p,h,E) =
\frac{2\pi}{\hbar}|M_{\Delta n}|^2 \omega_{\Delta n} (p,h,E) \mbox{ .}
\label{LambdaTransitionGeneral}
\end{equation}
For these transition rates, one needs the number of states, $\omega$, taking
into account the selection rules for intranuclear exciton-exciton scattering.
The appropriate formulae have been derived by Williams~\cite{Williams} and later
corrected for the exclusion principle and indistinguishability of identical
excitons in Refs.~\cite{Williams2,Ribansky}:
\begin{equation}
\begin{split}
\omega_+ (p,h,E) = & \frac{1}{2} g \frac{[gE-{\cal A}(p+1,h+1)]^2} {n+1} \\
	& \times \biggl[ \frac{gE - {\cal A}(p+1,h+1)}{gE - {\cal A}(p,h)} \biggr]
^{n-1} \mbox{ ,} \\
\omega_0 (p,h,E) = & \frac{1}{2} g \frac{[gE-{\cal A}(p,h)]}{n} \\
	& \times [p(p-1)+4ph+h(h-1)] \mbox{ ,} \\
\omega_- (p,h,E) = & \frac{1}{2} gph(n-2) \mbox{ ,}
\label{OmegaTransition}
\end{split}
\end{equation}
where ${\cal A}(p,h) = (p^2 +h^2 +p-h)/4 - h/2$. By neglecting the difference of
matrix elements with different $\Delta n$, $M_+ = M_- = M_0 = M$, we estimate the
value of $M$ for a given nuclear state by associating the $\lambda_+ (p,h,E)$
transitions with the probability for quasi-free scattering of a nucleon above
the Fermi level on a nucleon of the target nucleus. Therefore, we have
\begin{equation}
\begin{split}
\frac{ < \sigma (v_{rel}) v_{rel} >}{V_{int}} = & \frac{\pi}{\hbar} |M|^2
 \frac{g [ gE-{\cal A}(p+1,h+1)]}{n+1} \\
	& \times
         \biggl[ \frac{gE - {\cal A}(p+1,h+1)}{gE - {\cal A}(p,h)} \biggr] ^{n-1} \mbox{ ,}
\label{SigmaAverage}
\end{split}
\end{equation}
where $V_{int}$ is the interaction volume estimated as $V_{int} = {4 \over 3}
\pi (2 r_c + \lambda / 2 \pi)^3$, with the de Broglie wave length $\lambda /
2 \pi$ corresponding to the relative velocity $v_{rel} = \sqrt{2 T_{rel} /m_N}$.
$m_N$ is the mass of interacting excitons (nucleons) and $T_{rel}$ is their 
relative kinetic energy.
A value of the order of the nucleon radius is used for $r_c$ in the CEM:
$r_c = 0.6$ fm.

The averaging on the left-hand side of Eq.~(\ref{SigmaAverage}) is carried out
over all excited states, taking into account the exclusion principle. Combining
~(\ref{LambdaTransitionGeneral}), (\ref{OmegaTransition}), and
(\ref{SigmaAverage}) we finally get for the transition rates:
\begin{equation}
\begin{split}
\lambda_+ (p,h,E) = & \frac{ < \sigma (v_{rel}) v_{rel} >}{V_{int}} \mbox{ ,} \\
\lambda_0 (p,h,E) = & \frac{ < \sigma (v_{rel}) v_{rel} >}{V_{int}} \biggl[
 \frac{gE - {\cal A}(p,h)}{gE - {\cal A}(p+1,h+1)} \biggr] ^{n+1} \\
	& \times \frac{n+1}{n} \frac{p(p-1)+4ph+h(h-1)}{gE-{\cal A}(p,h)} \mbox{ ,} \\
\lambda_- (p,h,E) = & \frac{ < \sigma (v_{rel}) v_{rel} >}{V_{int}} \biggl[
 \frac{gE - {\cal A}(p,h)}{gE - {\cal A}(p+1,h+1)} \biggr] ^{n+1} \\
	& \times \frac{ph(n+1)(n-2)}{[gE-{\cal A}(p,h)]^2} \mbox{ .}	
\label{LambdaTransition}
\end{split}
\end{equation}

The CEM predicts angular distributions for preequilibrium particles that are
forward-peaked in the laboratory system. For instance, CEM03.03 assumes that a
nuclear state with a given excitation energy $E$ should be specified not only
by the exciton number $n$ but also by the momentum direction $\Omega$. Following
Ref.~\cite{Mantzouranis}, the master equation (Eq.~(\ref{Probability})) can be 
generalized for this case provided that the angular dependence for the transition
rates $\lambda_+$, $\lambda_0$, and $\lambda_-$ (Eq.~(\ref{LambdaTransition}))
may be factorized. In accordance with Eq.~(\ref{SigmaAverage}), in the CEM it is
assumed that
\begin{equation}
<\sigma> \to <\sigma> F(\Omega) \mbox{ ,}
\label{SigmaFactor}
\end{equation}
where
\begin{equation}
F(\Omega) = {d \sigma^{free}/ d \Omega \over
\int d \Omega '  d \sigma^{free} / d \Omega '} \mbox{ .} 
\label{Factor}
\end{equation}
The scattering cross section $ d \sigma^{free}/ d \Omega$ is assumed to be
isotropic in the reference frame of the interacting excitons, thus resulting in
an asymmetry in both the nucleus center-of-mass and laboratory frames. The
angular distributions of preequilibrium complex particles are assumed to be
similar to those for the nucleons in each nuclear state \cite{CEM}.

This calculational scheme is easily realized by the Monte-Carlo technique. It
provides a good description of double-differential spectra of preequilibrium
nucleons and a not-so-good but still reasonable description of complex-particle
spectra from different types of nuclear reactions at incident energies from tens
of MeV to several GeV. 

For incident energies below about 200 MeV, Kalbach
has developed a phenomenological systematics for preequilibrium-particle angular
distributions by fitting available measured spectra of nucleons and complex
particles \cite{Kalbach88}. As the Kalbach systematics are based
on measured spectra, they describe very well the double-differential spectra of
preequilibrium particles and generally provide a better agreement of calculated
preequilibrium complex-particle spectra with data than does the CEM approach
based on Eqs.~(\ref{SigmaFactor}, \ref{Factor}). Therefore, CEM03.03 incorporates
the Kalbach systematics \cite{Kalbach88} to describe angular distributions of
both preequilibrium nucleons and complex particles at incident energies up to
210 MeV. At higher energies, CEM03.03 uses the CEM approach based on
Eqs.~(\ref{SigmaFactor}, \ref{Factor}).

As the MEM uses a Monte-Carlo technique to
solve the master equations describing the behavior of the nucleus at the
preequilibrium stage (see details in~\cite{CEM}), it is relatively easy to
extend the number of types of possible LF that can be emitted during this stage.
For this, we have only to extend the loop in the CEM03.03 code calculating
$\Gamma_j$ for $j$ from 1 to 6 (i.e., for the emission of
$n$, $p$, $d$, $t$, $^3$He, and $^4$He) 
to a larger value, in this case, up to $j = 66$, to account for
the possibility of preequilibrium emission of up to 66 types of particles 
and LF. Of course, in this extended loop, we have to calculate the emission
width $\Gamma_j$ for all $j$ values.
This entails calculating Coulomb barriers, binding energies,
reduced masses, inverse cross sections, and condensation probabilities for all
66 types of particles and LF. As this extended CEM03.03 is intended to allow
production of energetic light fragments, we subsequently refer to it as
CEM03.03F, where ``F'' stands for energetic {\bf f}ragments. We also refer later
to a similarly extended version of LAQGSM03.03 as LAQGSM03.03F.
The list of all particles and LF that can be emitted during the 
preequilibrium stage of a nuclear reaction calculated with 
CEM03.03F is provided below in Tab.~\ref{Particles}.

\begin{table}[here]
\caption{The list of particles and  light fragments that can be emitted
during the preequilibrium stage of reactions in the extended MEM.}
\centering
\begin{tabular}{rlllllll}
\hline\hline 
 $Z_j$\hspace{2mm} & \multicolumn{7}{l} {Ejectiles} \\
\hline
0\hspace{2mm}  & $n$     &         &         &         &         &         &         \\
1\hspace{2mm}  & $p$     &\hspace{1mm}   $d$ &\hspace{1mm}   $t$ &         &         &         &         \\
2\hspace{2mm}  &$^{3 }$He&\hspace{1mm}$^{4 }$He&\hspace{1mm}$^{6 }$He&\hspace{1mm}$^{8 }$He&         &         &         \\
3\hspace{2mm}  &$^{6 }$Li&\hspace{1mm}$^{7 }$Li&\hspace{1mm}$^{8 }$Li&\hspace{1mm}$^{9 }$Li&         &         &         \\
4\hspace{2mm}  &$^{7 }$Be&\hspace{1mm}$^{9 }$Be&\hspace{1mm}$^{10}$Be&\hspace{1mm}$^{11}$Be&\hspace{1mm}$^{12}$Be&         &         \\
5\hspace{2mm}  &$^{8 }$B &\hspace{1mm}$^{10}$B &\hspace{1mm}$^{11}$B &\hspace{1mm}$^{12}$B &$\hspace{1mm}^{13}$B &         &         \\
6\hspace{2mm}  &$^{10}$C &\hspace{1mm}$^{11}$C &\hspace{1mm}$^{12}$C &\hspace{1mm}$^{13}$C &\hspace{1mm}$^{14}$C &\hspace{1mm}$^{15}$C &\hspace{1mm}$^{16}$C \\
7\hspace{2mm}  &$^{12}$N &\hspace{1mm}$^{13}$N &\hspace{1mm}$^{14}$N &\hspace{1mm}$^{15}$N &\hspace{1mm}$^{16}$N &\hspace{1mm}$^{17}$N &         \\
8\hspace{2mm}  &$^{14}$O &\hspace{1mm}$^{15}$O &\hspace{1mm}$^{16}$O &\hspace{1mm}$^{17}$O &\hspace{1mm}$^{18}$O &\hspace{1mm}$^{19}$O &\hspace{1mm}$^{20}$O \\
9\hspace{2mm}  &$^{17}$F &\hspace{1mm}$^{18}$F &\hspace{1mm}$^{19}$F &\hspace{1mm}$^{20}$F &\hspace{1mm}$^{21}$F &         &         \\
10\hspace{2mm} &$^{18}$Ne&\hspace{1mm}$^{19}$Ne&\hspace{1mm}$^{20}$Ne&\hspace{1mm}$^{21}$Ne&\hspace{1mm}$^{22}$Ne&\hspace{1mm}$^{23}$Ne&\hspace{1mm}$^{24}$Ne\\
11\hspace{2mm} &$^{21}$Na&\hspace{1mm}$^{22}$Na&\hspace{1mm}$^{23}$Na&\hspace{1mm}$^{24}$Na&\hspace{1mm}$^{25}$Na&         &         \\
12\hspace{2mm} &$^{22}$Mg&\hspace{1mm}$^{23}$Mg&\hspace{1mm}$^{24}$Mg&
\hspace{1mm}$^{25}$Mg&\hspace{1mm}$^{26}$Mg&\hspace{1mm}$^{27}$Mg&
\hspace{1mm}$^{28}$Mg\\
\hline\hline 
\end{tabular}
\label{Particles}
\end{table}

As can be seen from Eq.\ \ref{LambdaTransmission},
the inverse cross sections used by these models at the 
preequilibrium stage (and at the evaporation/fission stage) 
have a significant impact on the calculated particle width, and affect
greatly the final results and the accuracy of the MCNP6,
MCNPX\ \cite{MCNPX} and MARS15\ \cite{MARS15} 
transport codes, which use these models as their event generators. 
This is why it is necessary to use as good as possible approximations
for the inverse cross sections in the extended models.

The unmodified codes use the inverse cross sections $\sigma_{inv}$, 
from Dostrovsky's formulas \cite{Dostrovsky} for all emitted nucleons and the
complex particles ($d$, $t$, $^3$He, and $^4$He):
\begin{equation}
\sigma_{inv} (\eps) = \sigma_{g} \alpha  \left(
1 + {\beta \over \eps} \right) \mbox{ ,}
\label{eq:Dost}
\end{equation}
which is often written as
$$
\sigma_{inv} (\eps) =
\begin{cases}
\sigma_g c_n (1 + b/ \eps)& \mbox{for neutrons} \cr
\sigma_g c_j (1- V_j/ \eps)& \mbox{for charged particles ,}\cr
\end{cases}
$$
where $\sigma_g = \pi R^2_d$ [fm$^2$] is the geometrical cross section.
``$d$'' denotes the ``daughter'' nucleus with mass and charge
numbers $A_d$ and $Z_d$ produced from the ``parent'' nucleus ``$i$''
with mass and charge
numbers $A_i$ and $Z_i$ after the emission of the particle ``$j$''
with
mass and charge
numbers $A_j$ and $Z_j$
and kinetic energy $\eps$; $R_d = r_0 A_d^{1/3}$, and $r_0 = 1.5$ fm.
$\alpha$ and $\beta$ are defined as:
$$ \alpha = 0.76 + 2.2 A_d \mbox{ MeV,} $$
$$ \beta = {2.12 A_d^{-2/3} - 0.05 \over 0.76 + 2.2 A_d^{-1/3} } 
\mbox{ MeV,} $$ 
and $c_j$ is estimated by interpolation of the tabulated
values published in Ref.\ \cite{Dostrovsky}.

The Coulomb barrier (in MeV) is estimated as:
\beq
V_j = k_j Z_j Z_d e^2 / R_c \mbox{ ,}
\label{eq:Evap38}
\eeq
where $R_c = r_0 (A_d^{1/3} + A_j^{1/3})$, $r_0 = 1.5$ fm, and
the penetrability coefficients $k_j$ are calculated via
interpolation of the tabulated values published in Ref. \cite{Dostrovsky}.

At the evaporation/fission stage of reactions described by CEM0.03 and
LAQGSM0.03, which use an extension of the generalized evaporation model code GEM2
by Furihata~\cite{GEM2}, the inverse cross sections are calculated with the
same functional form, but using different constants from those in the
original approximations \cite{Dostrovsky}. 
We label those different inverse cross sections as ``GEM2''.

The Dostrovsky model is very old. It was not intended for use above about 50
MeV/nucleon, and is not very suitable for emission of fragments heavier than
$^4$He. Better total-reaction-cross-section models 
that can be used as an estimate for inverse cross sections
are available today, most notably the NASA model \cite{NASA},
the approximations by Barashenkov and Polanski \cite{BP},
and those by Kalbach \cite{Kalbach}.
A quite complete list of references on modern  
total-reaction-cross-section models, as well as on recent publications
where these models are compared with each other and with available
experimental data can be found in Ref.~\cite{NIMB2015}.

We have performed recently an extensive comparison of of the NASA \cite{NASA},
Tsang et al.\ \cite{Tsang}, Dostrovsky et al.\ \cite{Dostrovsky}, Barashenkov
and Polanski \cite{BP}, GEM2 \cite{GEM2}, and Kalbach \cite{Kalbach} systematics
for total reaction ({\it inverse}) cross sections
(see also the older works \cite{SantaFe2002, CEM03.02, Prael} 
with similar comparisons). We conclude
that the NASA approach is superior, in general, to the other available models
(see Ref.~\cite{SantaFe2002, CEM03.02, Prael, NIMB2015} for
the details of these findings). This is why we implement the NASA
inverse cross sections into the MEM to be used at the preequilibrium stage
of reactions.

The NASA approximation as described by Eq.~(\ref{eq:NASA})
attempts to simulate several
quantum-mechanical effects, such as the optical potential for neutrons
(with the parameter $X_m$) and collective effects like Pauli blocking (through
the quantity $\delta_T$).  (For more details, see Ref.~\cite{NASA}.)

\begin{equation}
\sigma_{NASA} = \pi r^2_0 (A_P^{1/3} + A_T^{1/3} + \delta_T )^2 
(1 - R_c \frac{B_T}{T_{cm}})X_m \mbox{ ,}
\label{eq:NASA}
\end{equation}
where 
$r_0$, $A_P$, $A_T$, $\delta_T$, $R_c$, $B_T$, $T_{cm}$, and  $X_m$ 
are a constant used to calculate the radii of nuclei,
the mass number of the projectile nucleus,
the mass number of the target nucleus,
an energy-dependent parameter,
a system-dependent Coulomb multiplier,
the energy-dependent Coulomb barrier,
the colliding system center-of-momentum energy, and
an optical model multiplier used for neutron-induced reactions, respectively.

In the case of neutron-induced reactions, we can not use the unmodified NASA
systematics to approximate the inverse cross sections for neutrons, as 
that model, while being much better at predicting the total reaction cross
section throughout most of the energy region of the data, falls to zero at
low energies. Since neutrons have no Coulomb barrier and
are emitted at even very low energies, a finite neutron cross section at
very low energies is needed. For these
low-energy neutrons, we use the Kalbach systematics\ \cite{Kalbach}, which
prove to be a very good approximation for the inverse cross sections of low-energy
neutrons, as discussed in Refs.~\cite{SantaFe2002, NIMB2015, FY2014}.
In CEM03.03F, we use the Kalbach systematics \cite{Kalbach} to replace the NASA
inverse cross sections \cite{NASA} for low-energy neutrons, similar
to what was suggested and done in Ref.\ \cite{SantaFe2002} for the code CEM2k.
In other words, at neutron energies around the maximum cross
section and below, the calculation
uses Kalbach systematics, and switches to the NASA model for the higher
neutron-energy range. The Kalbach systematics are scaled 
in CEM03.03F to match the NASA
model results at the transition point (depending on the nucleus)
so as not to have a discontinuity.  Transition points
and scaling factors are obtained for all possible residual nuclei, by mass
number; they are fixed in the code and are used in all subsequent
calculations. (Ref.\ \cite{FY2014} provides tables of these.)

Examples of inverse cross sections for the emission of neutrons together
with discussions and relevant references can be found in 
Refs.\ \cite{SantaFe2002,NIMB2015,FY2014}. We limit ourselves to
one example with inverse cross sections for the emission of protons,
and one example of inverse cross section for $^{12}$C.

\begin{figure}[h!]
\centering
\includegraphics[width=70mm]{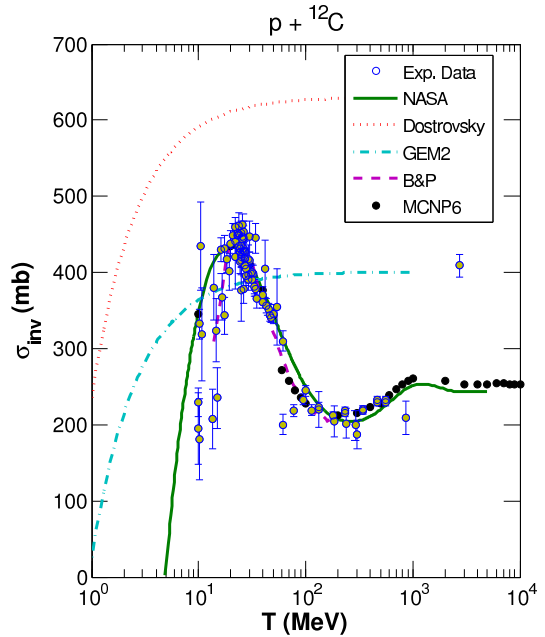}                    
\caption{(Color online) Reaction cross section for
  $p$ + $^{12}$C, as calculated by the NASA, Dostrovsky, GEM2, and BP
  models (solid and broken lines). The black dots are cross sections calculated
  by MCNP6, and the circles are experimental data \cite{Carlson}.}
\label{fig:p+C}
\end{figure}

Fig.\ \ref{fig:p+C} illustrates calculated total reaction cross sections for
p + $^{12}$C using the NASA, Dostrovsky, GEM2, and
Barashenkov and Polanski (BP) models, compared to calculations by MCNP6 and
experimental data. The NASA model appears to be superior to the Dostrovsky-like
models. 
 
\begin{figure}[h!]
\centering
\includegraphics[width=70mm]{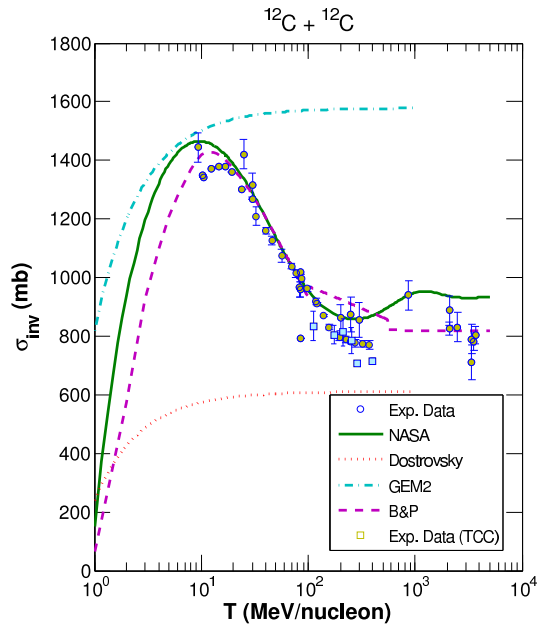}                  
\caption{(Color online) Reaction cross section for $^{12}$C + $^{12}$C
 as calculated by the NASA, Dostrovsky, GEM2, and BP models (solid and
 broken lines). The circles are experimental data
 \cite{CC} and the squares are total
 charge-changing cross section (TCC) measurements \cite{TCC}.}
\label{fig:12C+12C}
\end{figure}

Fig.\ \ref{fig:12C+12C} displays the total reaction cross section for
$^{12}$C + $^{12}$C, as calculated by the NASA, Dostrovsky, GEM2, and BP model,
compared to experimental data and to measured total charge-changing (TCC)
cross sections. TCC cross sections should be $5$--$10\%$ less than total reaction
cross sections, as TCC cross sections do not include neutron removal.
The NASA cross-section model fits the experimentally measured data, in general,
better than the other models tested. See Ref.\ \cite{FY2014} for results of other
heavy-ion-induced reactions.

Many similar results for the emission of nucleons, complex particles, and LF
heavier than $^4$He can be found in Refs.~\cite{SantaFe2002, NIMB2015, CEM03.02,
FY2014}.

The partial transmission probability $\lambda_j$, the probability that an
ejectile of the type $j$ will be emitted with kinetic energy $T$,
is given in Eq.~(\ref{LambdaTransmission}). This form is valid only for
neutrons and protons. An extended form appropriate for the case of complex
particles and LF is (see Ref.\ \cite{CEM}):
\begin{equation}
\begin{split}
\lambda_{j}(p,h,E,T) = & \gamma_j \frac{2s_j + 1}{\pi^2\hbar^3} \mu_j \Re (p,h)
 \frac{\omega (p_j,0,T+B_j)}{g_j} \\
& \times \frac{\omega (p-p_j,h,E-B_j-T)}{\omega (p,h,E)} T \sigma_j^{inv} (T) \mbox{ ,}
\end{split}
\label{eq:lambda_j}
\end{equation}
where
\begin{equation}
g_j = \frac{V (2\mu_j)^{3/2}}{4\pi^2\hbar^3} (2s_j + 1) (T+B_j)^{1/2} \mbox{ .}
\label{eq:g_j}
\end{equation}
Details on Eq.~(\ref{eq:g_j}) can be found in Ref.\ \cite{WuChang}. $\gamma_j$ is the
probability that the proper number of particle excitons will coalesce to form a
type $j$ fragment (also called $\gamma_\beta$ in a number of earlier
publications; see, e.g., Refs.\ \cite{Betak76, Blideanu, WuChang}). 
 
In the standard CEM03.03, the
Dostrovsky form of the inverse cross section is simple enough so that for
neutrons and protons the integral
from Eq.~(\ref{GammaLambda}) 
 can be done analytically. However, for complex
particles, the level density $\omega$ becomes too complicated (see details in
Refs.\ \cite{CEM, CEM03.03, Trieste08}); therefore, the integral is
evaluated numerically. In this case, a 6-point Gauss-Legendre quadrature is used when
the exciton number is 15 or less, and a 6-point Gauss-Laguerre quadrature is used
when the number of excitons is greater than 15. 

We adopt for CEM03.03F the NASA form of the cross sections, which removes the 
possibility of analytic integration, so the integral is always calculated
numerically. We use an 8-point Gauss-Legendre quadrature when the number of excitons is
15 or less, and an 8-point Gauss-Laguerre quadrature when the number of excitons
is greater than 15. (See Ref.\ \cite{NIMB2015} for details.)

These integration methods are sufficient for these models
because individual $\Gamma_j$ precision
is not extremely important for choosing what type of particle/LF $j$ will be
emitted. In contrast to analytical preequilibrium models, the Monte-Carlo method
employed by CEM uses the ratios of $\Gamma_j$ to the sum of $\Gamma_j$ over
all $j$. That is, if we estimate all $\Gamma_j$ with the same percentage error,
the final choice of the type $j$ of particle/LF to be emitted as simulated by
CEM would be the same as if we would calculate all $\Gamma_j$ exactly. We think
that this is the main reason why CEM provides quite reasonable results using the
old Dostrovsky approximation for inverse cross sections, in spite of the
significant difference of the Dostrovsky inverse cross sections from those
now used.  The ratios of the individual widths to the total width were
approximated better than each individual width, because the errors in each
channel have the same sign. This is illustrated in Fig.\ \ref{fig:p+C}.
(See more examples and discussion in Ref.\ \cite{NIMB2015}.)

We observe that the condensation probability $\gamma_j$ could be calculated
more physically from first principles, if one were studying only this problem.
But such a calculation is not feasible in these event generators
given practical Monte-Carlo computational time
limitations. $\gamma_j$ is, therefore, estimated as the overlap integral of the
wave function of independent nucleons with that of the complex particle (see
details in \cite{CEM}), as shown in Eq.\ \ref{GammaBeta}.

As noted above, Eq.\ \ref{GammaBeta} is a rather crude estimate. 
As is frequently done (see e.g.,
Refs.\ \cite{WuChang, Blideanu}), the values of $\gamma_j$ are taken from fitting
the theoretical preequilibrium spectra to the experimental ones. 
In CEM03.03F, to
improve the description of preequilibrium complex-particle emission, we estimate
$\gamma_j$ by multiplying the estimate provided by Eq.~(\ref{GammaBeta}) by
empirical coefficients $F_j (A,Z,T_0)$, whose values are fitted to available
nucleon-induced experimental complex-particle spectra. Therefore, the new
equation for $\gamma_j$ using this empirical coefficient is 

\begin{equation}
\gamma_j = F_j p_j^3 \left(\frac{p_j}{A}\right)^{p_j-1} .
\label{Eq:gammaj}
\end{equation}

The values of $F_j$ for $d$, $t$, $^3$He, and $^4$He used by the original 
CEM03.03 need to be re-fitted after the current upgrades
to the inverse-cross-sections and the coalescence model (discussed below).
Then, values of $F_j$ need
to be obtained for heavy clusters up to $^{28}$Mg, once the model is extended
to emit these heavy clusters.  This was done for all possible target nuclei.
We have developed a universal approximation, or a ``numerical model,''
for $F_j (A,Z,T_0)$ to be used in CEM03.03F. All details of this part of
our work can be found in Refs.\ \cite{Leslie-thesis,RMPmanuscript,FY2014}.

Once a fragment of type $j$ has been randomly chosen for emission, the kinetic
energy of this fragment needs to be determined. This is done by sampling the
kinetic energy from the $\lambda_j$ distribution, Eq.~(\ref{eq:lambda_j}),
using the NASA cross section as the $\sigma_j^{inv}(T)$. 

The energy-dependence of $\lambda_j$ for the new
inverse cross sections is more complicated than that
that arising from the simple Dostrovsky form used in the original
CEM03.03. This affects the method we choose to randomly 
sample $T_j$ $(\equiv T)$ from the correct spectrum.

To sample $T_j$ uniformly from the $\lambda_j$ distribution using the Monte-Carlo 
method, we must first find the maximum of $\lambda_j$. In CEM03.03, this
is done analytically using the derivative of $\lambda_j$ with respect to $T_j$,
due to the simple nature of the energy-dependence in the Dostrovsky
systematics. The NASA cross section energy dependence is much more
complicated; therefore, we find the maximum of $\lambda_j$ numerically
using the Golden-Section method. This also provides us the flexibility to modify
the cross-section model in the future without needing to modify the kinetic
energy algorithm.

After finding the maximum value of $\lambda_j$, the kinetic energy of the emitted
fragment $j$ is uniformly sampled from the $\lambda_j$ distribution using the
rejection technique from a Gamma distribution (shape parameter $\alpha = 2$) as
the comparison function. 
(See Ref.~\cite{NumericalRecipes} for a description of the Gamma distribution.)

Fig.\ 57 of Ref.\ \cite{FY2014} illustrates results for the
probability of emitting $^6$Li with a given kinetic energy $T_{Li}$
simulated by CEM03.03F and the original CEM03.03.
Probabilities from the $\lambda_j$ distributions with the NASA inverse cross
sections differ slightly from those with the Dostrovsky inverse cross sections,
just as expected.  Technical details with many illustrative figures 
on this part of our work can by found in Refs.\ \cite{Leslie-thesis, FY2014}.

\section{Coalescence Model} 
\label{Chapter:Coalescence}

As previously described, one of the three possible mechanisms CEM and
LAQGSM use to produce energetic LF is the coalescence of nucleons emitted
during the INC as well as of the already
coalesced lighter fragments into heavier clusters.

When the cascade stage of a reaction is completed, CEM03.03
and LAQGSM03.03 use the
coalescence model described in Refs.\ \cite{DCM, Gudima:83a}
to ``create" high-energy $d$, $t$, $^3$He, and $^4$He by
final-state interactions among emitted cascade nucleons, already outside 
of the target nucleus. In contrast to most other
coalescence models for heavy-ion-induced reactions,
where complex-particle spectra are estimated simply by
convolving the measured or calculated inclusive spectra of nucleons
with corresponding fitted coefficients, 
CEM03.03 and LAQGSM03.03 use in their simulations of
particle coalescence real information about all emitted cascade nucleons
and do not use integrated spectra. 
These models assume that
all the cascade nucleons having differences in their momenta 
smaller than $p_c$ and the correct isotopic content form an appropriate
composite particle. This means that the formation probability for,
e.g. a deuteron is
\bea
W_d(\vec p,b) &=& \int \int d \vec p_p  d \vec p_n
\rho^C(\vec p_p,b) \rho^C(\vec p_n,b) \nonumber \\
&\times& \delta(\vec p_p + \vec p_n - \vec p)
\Theta(p_c - |\vec p_p - \vec p_n|) ,
\label{eq:Coal72}
\eea
where the particle density in momentum space is related to the
one-particle distribution function $f$ by
\begin{equation}
\rho^C(\vec p,b) = \int d \vec r f^C (\vec r, \vec p,b) .
\label{eq:Coal73}
\end{equation}
Here, $b$ is the impact parameter for the projectile interacting
with the target nucleus and the
superscript index $C$ shows that only cascade nucleons are taken into
account for the coalescence process. The coalescence radii $p_c$
were fitted for each composite particle in Ref.\ \cite{DCM}
to describe available data for the reaction Ne+U at 1.04 GeV/nucleon,
but the fitted values turned out to be quite universal and were
subsequently found to satisfactorily describe high-energy  
complex-particle production for a variety of reactions induced 
both by particles and nuclei at incident energies up to 
about 400 GeV/nucleon,
when describing nuclear reactions with different versions of
LAQGSM \cite{LAQGSM, LAQGSM03.03, CEM03.02}
or with its predecessor, the quark-gluon string
model (QGSM) \cite{Amelin}.
These parameters (in units of [MeV/c]) are:
\begin{equation}
p_c(d) = 90 \mbox{; }
p_c(t) =  p_c(^3{\mbox He}) = 108 \mbox{; }
p_c(^4{\mbox He}) = 115 \mbox{.}
\label{eq:Coal74}
\end{equation}

As the INC of CEM is different from those of LAQGSM or QGSM, 
it is natural to expect different best values for $p_c$ as well.
Recent studies have shown
(see e.g., Refs.\ \cite{CEM03.03, Trieste08} and references therein)
that the values of parameters $p_c$ defined by Eq.\ (\ref{eq:Coal74})
are also good for CEM03.03 for projectile particles 
with kinetic energies $T_0$ lower than 300 MeV and equal to or above 
1 GeV. For incident energies in the interval
300 MeV $< T_0 \le 1$ GeV, a better overall agreement with the
available experimental data is obtained by using values of $p_c$
equal to 150, 175, and 175 MeV/c for $d$, $t$($^3$He), and $^4$He,
respectively. These values of $p_c$ are fixed as defaults in CEM03.03.
If several cascade nucleons are chosen to coalesce into composite
particles, they are removed from the distributions of nucleons and
do not contribute further to such nucleon characteristics as spectra,
multiplicities, etc.

In comparison with the initial version \cite{DCM, Gudima:83a},
in CEM03.03 and LAQGSM03.03, several coalescence routines have been 
changed and have been tested against a large variety of 
measured data on nucleon- and nucleus-induced reactions at different 
incident energies. Many examples with results by CEM03.03 and LAQGSM03.03
for reactions where the contribution from 
the coalescence mechanism is important and can be easily
seen may be found in e.g., Refs.\ \cite{Trieste08, EPJ-Plus2011}. 

Note that following the coalescence idea used by these models,
the latest versions of the INCL code (e.g., \cite{Cugnon})
also consider (by different means) the coalescence of nucleons 
in the very outskirts of the nuclear surface 
into light fragments during the INC stages of reactions. 
In a way, the coalescence of INCL is similar to the one
considered by CEM and LAQGSM as proposed in Ref.\ \cite{DCM,
Gudima:83a}, with the main difference being that INCL considers coalescence
of INC nucleons on the border of a nucleus, just barely inside the target
nucleus, while CEM and LAQGSM coalesce INC nucleons and lighter
clusters already outside the nucleus. 

The standard ``03.03'' versions of both CEM and LAQGSM consider coalescence
of only $d$, $t$, $^3$He, and $^4$He. Here we extend the coalescence model
in CEM to account for the coalescence of heavier clusters, with mass numbers 
up to $A = 7$, and up to $A = 12$ in LAQGSM. The extended coalescence
model of CEM is described below, while the one of LAQGSM, in the next 
section.

The coalescence model of CEM first checks all nucleons to form 2-nucleon pairs,
as their momenta permit. It then checks if an alpha particle can be formed
from two 2-nucleon pairs (either from two n-p pairs or from an n-n and a p-p
pair). After this it checks to see if any of the two-nucleon pairs left can
combine with another nucleon to form either tritium or $^3$He. And lastly, it
checks to see if any of these three-nucleon groups (tritium or $^3$He) can
coalesce with another nucleon to form $^4$He. 

The extended coalescence model takes these two-nucleon pairs, three-nucleon 
(tritium or $^3$He only) groups, and $^4$He to see if they can coalesce to form
heavier clusters. $^4$He can coalesce with a 3-nucleon group to form either
$^7$Be or $^7$Li. Two 3-nucleon groups can coalesce to form either $^6$Li or
$^6$He. And $^4$He can coalesce with a 2-nucleon pair to form either $^6$Li or
$^6$He. All coalesced nucleons are removed from the distributions of nucleons
and lighter fragments so that the coalescence model conserves both atomic and
mass numbers.  For additional details of the extended coalescence model in CEM,
see Ref.\ \cite{Coal2015}.

$p_c$ determines how dissimilar the momenta of nucleons can be and still
coalesce.  Naturally, after the extension of the coalescence model
in CEM to account LF heavier than $^4$He, we had to redefine $p_c$, to 
include a value for heavy clusters, or LF: $p_c(LF)$.  In CEM03.03F,
the new $p_c$'s for incident energies, $T$, less than 300 MeV or greater than
1000 MeV are:
\begin{eqnarray}
p_c(d) & = & 90 \mbox{ MeV/c ;} \nonumber \\
p_c(t) & = & p_c(^3{\mbox He}) = 108 \mbox{ MeV/c ;} \\
p_c(^4{\mbox He}) & = & 130 \mbox{ MeV/c ;} \nonumber \\
p_c(LF) & = & 175 \mbox{ MeV/c .} \nonumber 
\label{eq:exp_coalescence1}
\end{eqnarray}
For $300$ MeV $< T < 1000$ MeV they are:
\begin{eqnarray}
p_c(d) & = & 150 \mbox{ MeV/c ;} \nonumber \\
p_c(t) & = & p_c(^3{\mbox He}) = 175 \mbox{ MeV/c ;} \\
p_c(^4{\mbox He}) & = & 205 \mbox{ MeV/c ;} \nonumber \\
p_c(LF) & = & 250 \mbox{ MeV/c .} \nonumber 
\label{eq:exp_coalescence2}
\end{eqnarray}
$p_c(^4{\mbox He})$ was increased compared to the original
$p_c$ values defined by Eq.\ (\ref{eq:Coal74}) because 
too many alpha particles were lost (coalesced into heavy clusters);
therefore, this was compensated for by coalescing more $^4$He.  

\begin{figure}[htb!]
\centering
\includegraphics[width=3.5in]{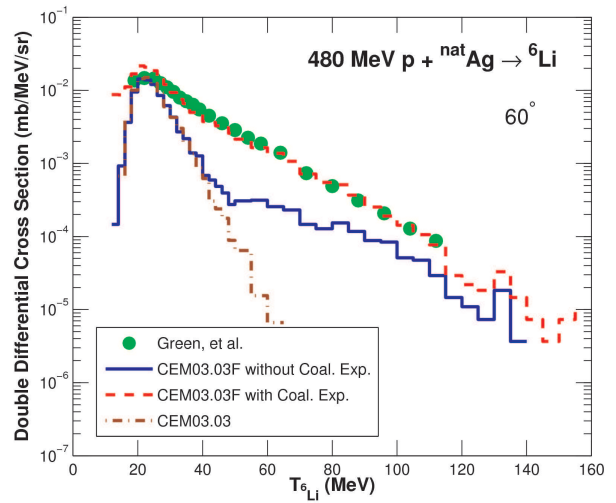}               
\caption{
(Color online) Comparison of experimental measurements of the reaction 480 MeV
  $p$ + $^{nat}$Ag $\rightarrow$ $^6$Li at 60\degree ~by Green et al.\ \cite{Green480} 
(circles), with simulations from the original CEM03.03 (dashed-dotted
 line), CEM03.03F without the coalescence extension (solid line) and
 CEM03.03F with the coalescence extension (dashed line).}
\label{fig:p480AgLi6}
\end{figure}

As an example,
Fig.\ \ref{fig:p480AgLi6} displays experimental measurements of the reaction 480
MeV $p$ + $^{nat}$Ag $\rightarrow$ $^6$Li by Green et al.\ \cite{Green480},
compared with simulations from CEM03.03F without the coalescence
extension (i.e., with the extended preequilibrium model and
improved inverse cross sections but using the old coalescence model), 
CEM03.03F with the coalescence extension, and the original CEM03.03. 
Even without the coalescence extension, CEM03.03F (which contains the extended
preequilibrium model and improved inverse cross sections) yields
much better results than the original CEM03.03 without these improvements. Adding the
coalescence extension produces even better results.

This example also highlights how coalescence can produce heavy clusters not
only at high energies, but also at low and moderate energies, thus improving
agreement with experimental data in all these energy regions.

Similar results for many other reactions where the coalescence mechanism is
important and easily seen can be found in 
Refs.\ \cite{Leslie-thesis, RMPmanuscript, Coal2015}.

\section{LAQGSM Extension} 

LAQGSM \cite{Trieste08, LAQGSM, LAQGSM03.03}
is a very powerful predictive tool for heavy-ion-induced reactions and/or
nuclear reactions induced by particles at high energies ($>$ several GeV/nucleon). 
MCNP6 uses it as its default event generator to simulate 
all heavy-ion induced reactions as well as 
reactions induced by particles at energies above 4.5 GeV
(above 1.2 GeV, in the case of photonuclear reactions).

The INC of LAQGSM03.03 is described with a recently improved version 
\cite{LAQGSM03.03, Varenna06}
of the time-dependent intranuclear cascade model developed 
initially at JINR in Dubna, often referred to in the literature as 
the Dubna intranuclear cascade model, DCM (see \cite{DCM}
and references 
therein). The DCM models interactions of fast cascade particles 
(``participants") with nucleon spectators of both the target and 
projectile nuclei and includes as well interactions of two 
participants (cascade particles). It uses experimental particle+particle
cross sections at energies below 4.5 GeV/nucleon,
or those calculated by the quark-gluon string model (QGSM)
at higher energies (see, e.g., Ref.\ \cite{Amelin86} and references therein)
to simulate angular and energy distributions of cascade particles, 
and also considers the Pauli exclusion principle.

After the INC, LAQGSM03.03 uses the same preequilibrium, coalescence,
Fermi break-up, and evaporation/fission models as described above
for CEM (with some parameters different from the ones used by CEM,
because the INC of LAQGSM is completely different from
the INC of CEM; see more details in \cite{Trieste08}).  

\begin{figure}[htb!]
\centering
\includegraphics[width=3.2in]{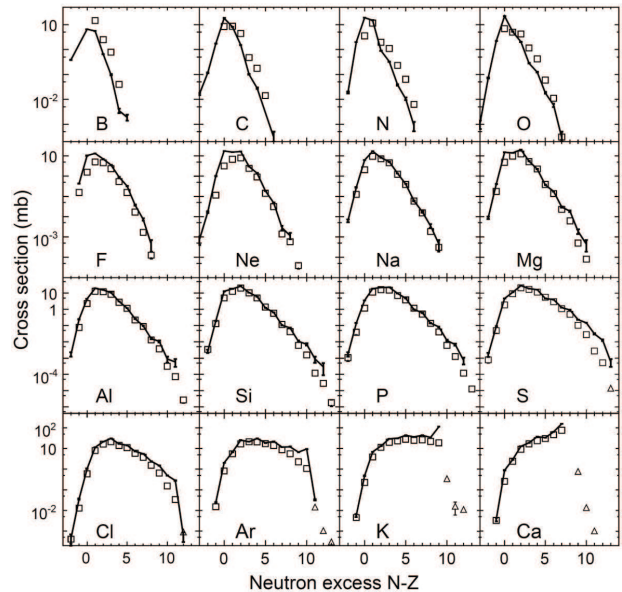}            
\caption[$^{48}$Ca fragmentation on $^9$Be at 140 MeV/A]
 {Measured elemental cross
 sections for $^{48}$Ca fragmentation on $^9$Be at 140
 MeV/nucleon \cite{Mocko} (open symbols) compared to LAQGSM03.03
 predictions (solid lines).}
\label{Ca140Be}
\end{figure}

\begin{figure}[htb!]
\centering
\includegraphics[width=3.in]{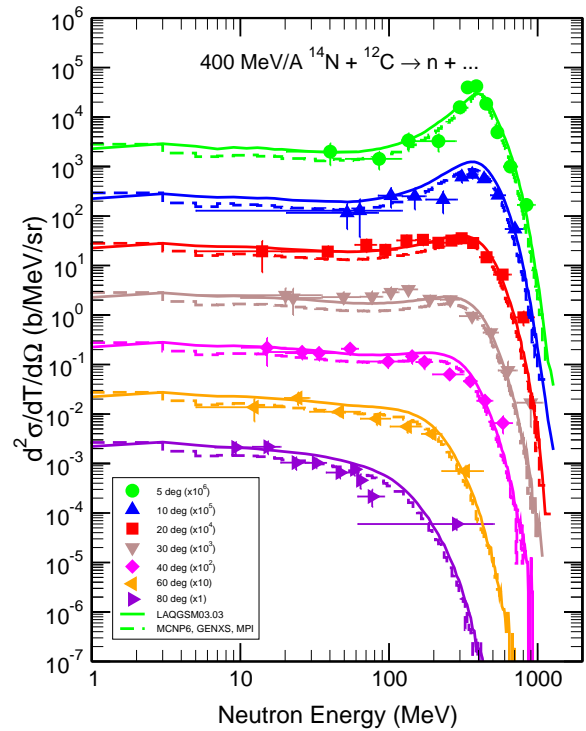}                   
\caption[Neutron spectra from 400 MeV/nucleon $^{14}$N + $^{12}$C]
 {(Color online) Experimental neutron spectra from 400 MeV/nucleon $^{14}$N +
  $^{12}$C \cite{Heilbronn2007} (solid symbols), compared with calculations
  by the production version of MCNP6 (dashed lines) and by the LAQGSM03.03 event
  generator used as a stand-alone code (solid lines).}
\label{N400C}
\end{figure}

\begin{figure}[htb!]
\centering
\includegraphics[width=3.in]{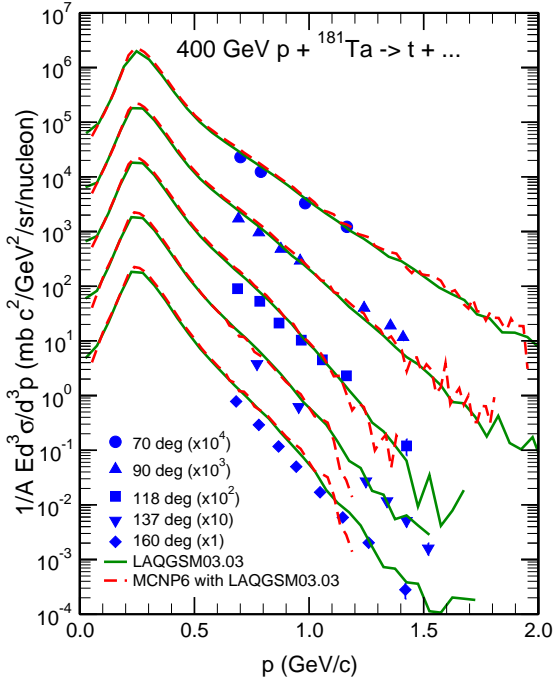}               
\caption{(Color online) Experimental invariant spectra of tritons from the
 reaction 400 GeV $p$ + $^{181}$Ta~\cite{Frankel79} (solid symbols),
 compared with results by LAQGSM03.03 \cite{LAQGSM03.03} used as a stand-alone
 code (solid lines) and by MCNP6 using the LAQGSM03.03 event-generator (dashed
 lines).
}
\label{p400GeV+Ta}
\end{figure}

As examples, Figs.\ \ref{Ca140Be}, \ref{N400C}, and \ref{p400GeV+Ta} show
results for three reactions simulated by LAQGSM compared
with available experimental data, to illustrate the
predictive power of LAQGSM03.03 used as 
the default event generator in MCNP6 for these types of reactions. 

Fig.~\ref{Ca140Be} shows an example of elemental product yields measured
by Mocko et al.\ \cite{Mocko}
from the fragmentation of $^{48}$Ca on $^9$Be at 140 MeV/nucleon.  Many
similar results for other reactions can be found in Ref.~\cite{LAQGSM03.03}.

Fig.~\ref{N400C} displays experimental neutron spectra from 400 MeV/nucleon
$^{14}$N + $^{12}$C \cite{Heilbronn2007}, compared with calculations by 
MCNP6 and the LAQGSM03.03 event generator used as a stand-alone code. Such
data are of significant interest for applications related to cancer treatment
with carbon beams, and most of the neutron spectra from such
reactions were measured at the Heavy-Ion Medical Accelerator in the Chiba
facility (HIMAC) of the Japanese National Institute of Radiological Science
(NIRS). We obtained similar agreement by LAQGSM and by MCNP6 using LAQGSM for
many similar reactions, at different incident energies and for different
projectile-target nuclear combinations (see Ref.\ \cite{Tokyo2014}).

Fig.\ \ref{p400GeV+Ta} shows that LAQGSM predicts well light cluster spectra even
at the ultrarelativistic energies of 400 GeV. Similar results for other
ejectiles from such reactions can be found in Ref.\ \cite{EPJ-Plus2011}. 

In CEM03.03F, we extend the coalescence model to account for heavier fragments
up to $^7$Be. As CEM is restricted to simulate only particle-induced reactions,
and only at energies below about $5$ GeV, such an extension of the coalescence
model may be sufficient. Since LAQGSM is used to calculate also reactions
induced by heavy ions, and at much higher incident energies, where the mean
multiplicities of the secondary nucleons and LF are much higher than for
reactions simulated with CEM, we extend the coalescence model in LAQGSM
to even heavier LF, up to $^{12}$C. Table~\ref{tab:LAQGSM} shows the LF we
produce via the extended coalescence model in LAQGSM, and the real
channels (modes) we consider to form each LF. The values of $p_c$ used in the
extended LAQGSM are also listed; they differ slightly from the ones used by
CEM03.03F.
\begin{table}[htb]
\caption{Coalescence channels (modes) for LF produced in the extended coalescence
 model of LAQGSM03.03F; values of $p_c$ are listed in units of MeV/c/nucleon.}

\centering
\begin{tabular}{ccccccc}
\hline\hline 
LF & $p_c$ & & Channels & (Modes) & \\
\hline
$d$    & 90  & $p+n$ \\
$t$    & 108 & $d+n$ \\
$^3$He & 108 & $d+p$ \\
$^4$He & 115 & $^3$He$+n$ & $t+p$ & $d+d$    \\
$^6$He & 150 & $t+t$  \\
$^6$Li & 150 & $t+^3$He & $^4$He$+d$ \\
$^7$Li & 150 & $t+^4$He & $^6$Li$+n$ \\
$^8$Li & 150 & $^7$Li$+n$ & $^6$He$+d$ \\
$^9$Li & 150 & $^8$Li$+n$ & $^6$He$+t$ \\
$^7$Be & 150 & $^3$He$+^4$He & $^6$Li$+p$ \\
$^9$Be & 150 & $^8$Li$+p$ & $^7$Li$+d$ \\
$^{10}$Be & 150 & $^9$Be$+n$ & $^8$Li$+d$ \\
$^{10}$B &  150 & $^9$Be$+p$ & $^7$Li$+^3$He & $^6$Li$+^4$He \\
$^{11}$B &  150 & $^{10}$B$+n$ & $^9$Be$+d$ & $^7$Li$+^4$He \\
$^{12}$B &  150 & $^{11}$B$+n$ & $^{10}$Be$+d$ & $^8$Li$+^4$He \\
$^{11}$C &  150 & $^{10}$B$+p$ & $^7$Be$+^4$He \\
$^{12}$C &  150 & $^{11}$C$+n$ & $^{11}$B$+p$ & $^{10}$B$+d$ & $^9$Be$+^3$He & $^6$Li$+^6$Li \\
\hline\hline 
\end{tabular}
\label{tab:LAQGSM}
\end{table}

\begin{figure}[htb!]
\centering
\vspace*{5mm}
\includegraphics[width=70mm]{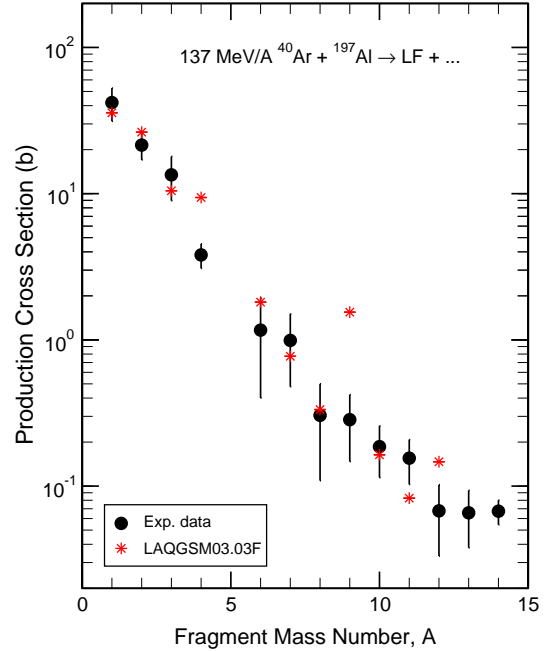}       
\caption[${48}$Ca fragmentation on $^9$Be at 140 MeV/A]
 {(Color online) Measured cross sections for light fragments produced in 137
  MeV/nucleon $^{40}$A + $^{197}$Au reactions \cite{Jacak1987} (circles),
  compared to predictions by LAQGSM03.03 with the extended coalescence model
  (stars).}
\label{Ar137Au}
\end{figure}

Fig.~\ref{Ar137Au} provides an example of preliminary results
for the case of fragment-production cross sections as functions of
mass number, measured by Jacak et al. at the LBL BEVALAC for 137
MeV/nucleon beams of $^{40}$Ar bombarding $^{197}$Au targets \cite{Jacak1987},
compared to LAQGSM03.03F results obtained with the extended coalescence model
(but still using the old preequilibrium model). 
There is reasonably good agreement with experimental
data for mass numbers up to $A=12$, except for $A=9$. 

\begin{figure}[htb!]
\centering
\includegraphics[width=80mm]{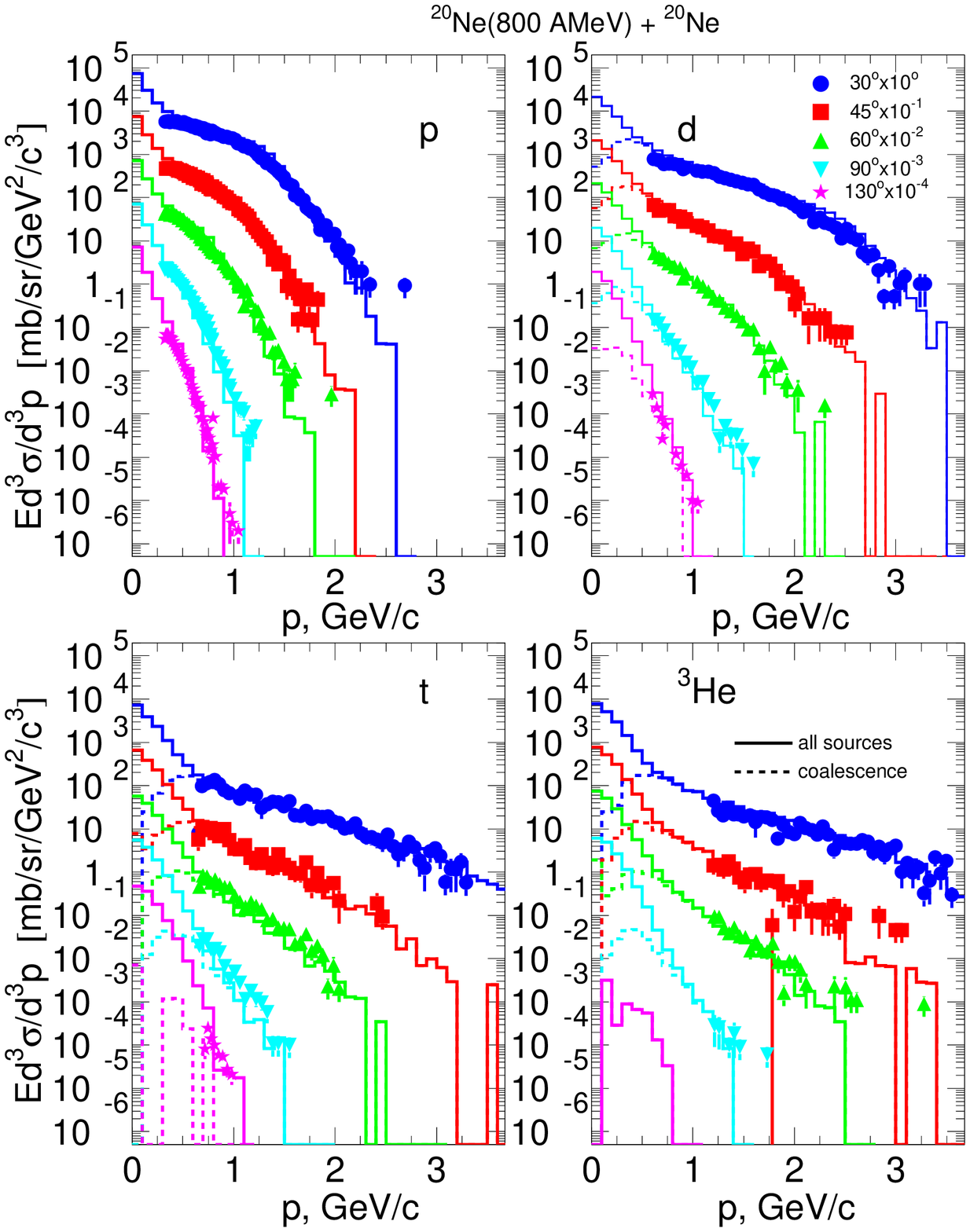}        
\caption{(Color online) Experimental invariant $p$, $d$, $t$, and $^3$He
 spectra at 30, 45, 60, 90, and 130 degrees (symbols) from a thin NaF target
 bombarded with an 800 MeV/nucleon $^{20}$Ne beam \cite{Lemaire78} (symbols),
 compared with results by LAQGSM03.03 using the extended coalescence model, 
 (histograms). The calculations were performed for $^{20}$Ne + $^{20}$Ne.
}
\label{Ne800Ne}
\end{figure}

\begin{figure}[htb!]
\centering
\includegraphics[width=80mm]{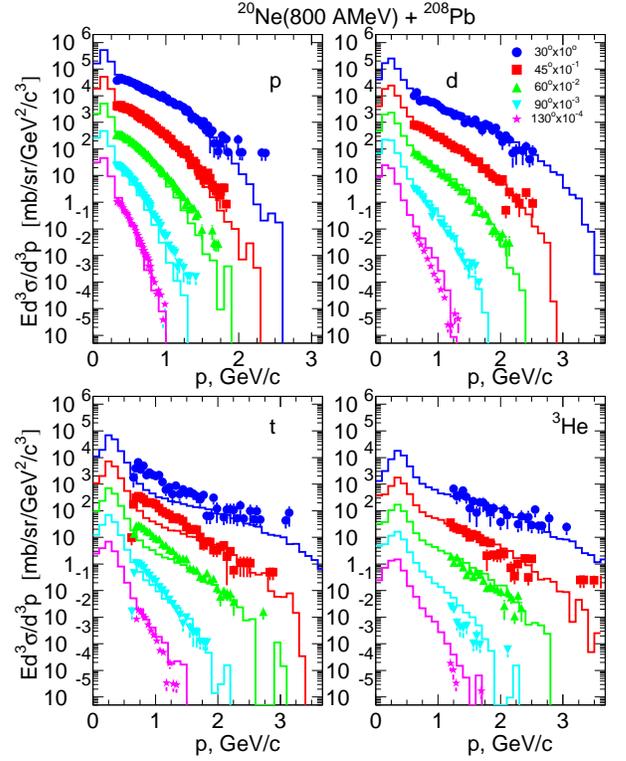}     
\caption{(Color online) Experimental invariant $p$, $d$, $t$, and $^3$He
  spectra at 30, 45, 60, 90, and 130 degrees from a thin Pb target bombarded
  with an 800 MeV/nucleon $^{20}$Ne beam \cite{Lemaire78} (symbols),
  compared with results from LAQGSM03.03 using the extended coalescence model
  (histograms).
}
\label{Ne800Pb}
\end{figure}

Figs.~\ref{Ne800Ne} and \ref{Ne800Pb} show two more examples
of results obtained with the extended coalescence model in LAQGSM,
namely, invariant cross section for the production of $p$, $d$, $t$, and
$^3$He at 30, 45, 60, 90, and 130 deg from 800 MeV/nucleon $^{20}$Ne +
$^{20}$Ne and $^{208}$Pb, respectively. There is a very good agreement of
results by the extended LAQGSM03.03F with these experimental data.
We obtain similar results for several other similar reactions 
measured at Berkeley and published in Ref.\ \cite{Lemaire78}. 

\begin{figure}[htb!]
\centering
\includegraphics[width=70mm]{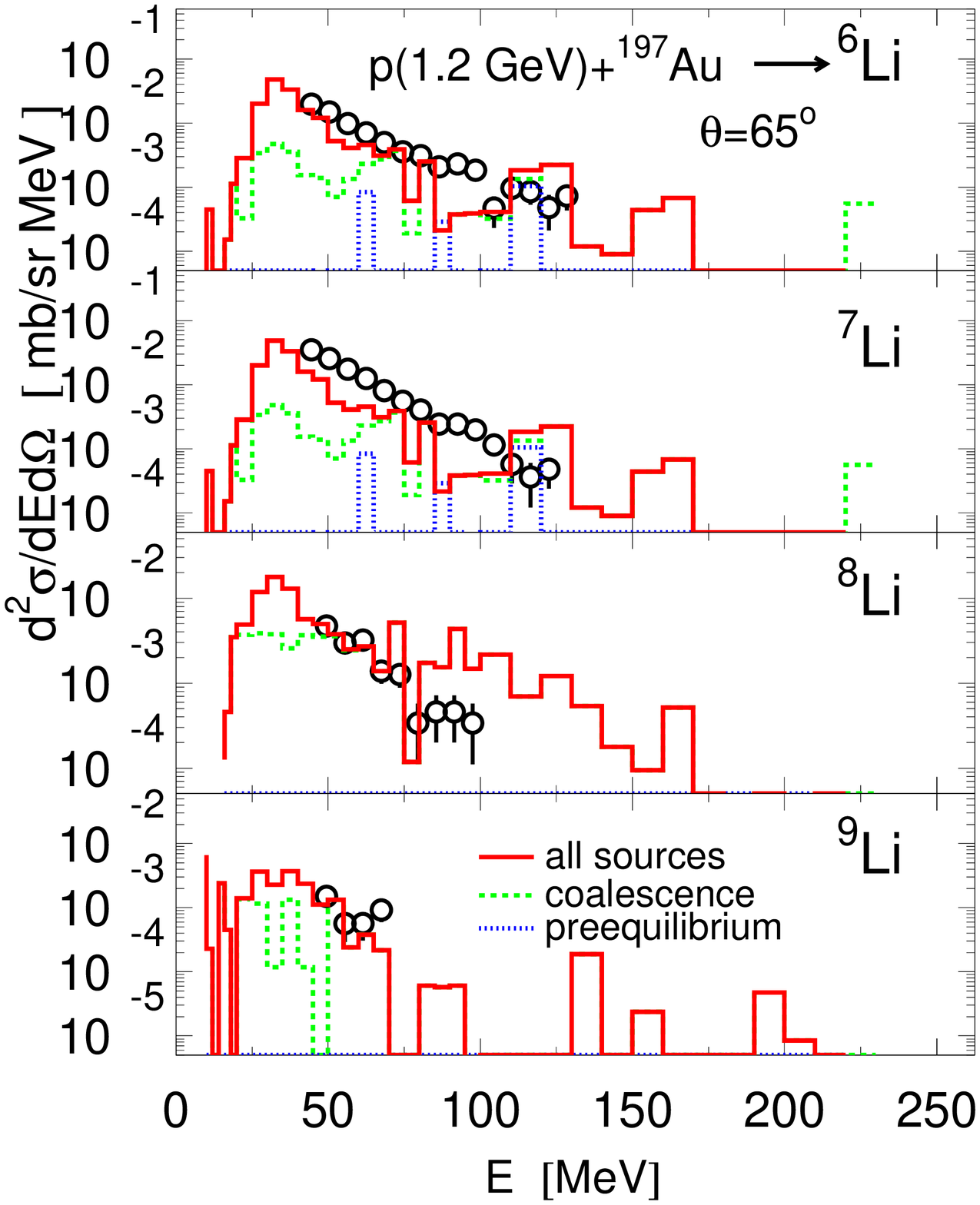}      Fig. 14
\caption{(Color online) Comparison of the experimental data
  for $^{6,7,8,9}$Li spectra at 65${^\circ}$ produced from
  1.2 GeV protons incident on $^{197}$Au \cite{Budzanowski} (open circles),
  compared to results calculated by the preliminary LAQGSM03.03F (histograms).
  The dotted and dashed histograms show the contributions from the preequilibrium
  emission and the extended coalescence model, respectively.
}
\label{1.2GeVpAu}
\end{figure}

\begin{figure}[htb!]
\centering
\includegraphics[width=70mm]{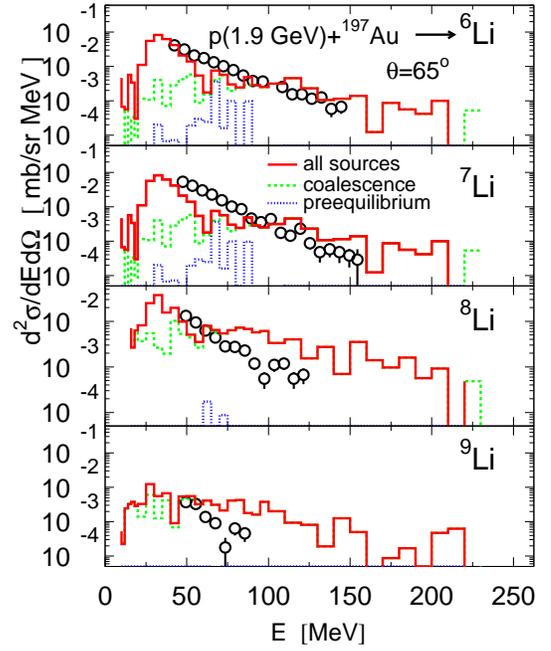}        
\caption{(Color online) Comparison of the experimental data
  for $^{6,7,8,9}$Li spectra at 65${^\circ}$ produced from
  1.9 GeV protons incident on $^{197}$Au \cite{Budzanowski} (open circles),
  compared to results calculated by the preliminary LAQGSM03.03F (histograms).
  The dotted and dashed histograms show the contributions from the preequilibrium
  emission and the extended coalescence model, respectively.
}
\label{1.9GeVpAu}
\end{figure}

The LAQGSM03.03F extension is still a work in progress. We have
extended and frozen its coalescence model, but so far have implemented only
the extended preequilibrium model, exactly as it was developed for CEM03.03F,
with the same parameters. Figs.\ \ref{1.2GeVpAu} and\ \ref{1.9GeVpAu} 
show two examples by this preliminary version of LAQGSM03.03F,
namely, spectra of $^{6,7,8,9}$Li at 65$^\circ$ from proton-gold
interactions at 1.2 and 1.9 GeV, respectively. This
preliminary version of LAQGSM03.03F describes quite well spectra 
of all Li fragments measured for these reactions by the PISA
collaboration and published in Ref.~\cite{Budzanowski}. 
LAQGSM03.03F produces similar results for other LF, from other 
target nuclei, and at other incident energies measured by the PISA
collaboration. However, as Figs.\ \ref{1.2GeVpAu} and \ref{1.9GeVpAu}
indicate, a fine-tuning of several
parameters in the extended preequilibrium model
(and perhaps the values of $p_c$ in the extended coalescence
model) would improve the agreement of the results with the measured
data and would refine the predictive power of LAQGSM03.03F. We hope
to perform such a fine-tuning in the future and to validate LAQGSM03.03F
on as many measured reactions as possible, before implementing
it into MCNP6 to replace the current version of LAQGSM03.03.

\begin{figure*}[]
\centering
\includegraphics[width=6.0in]{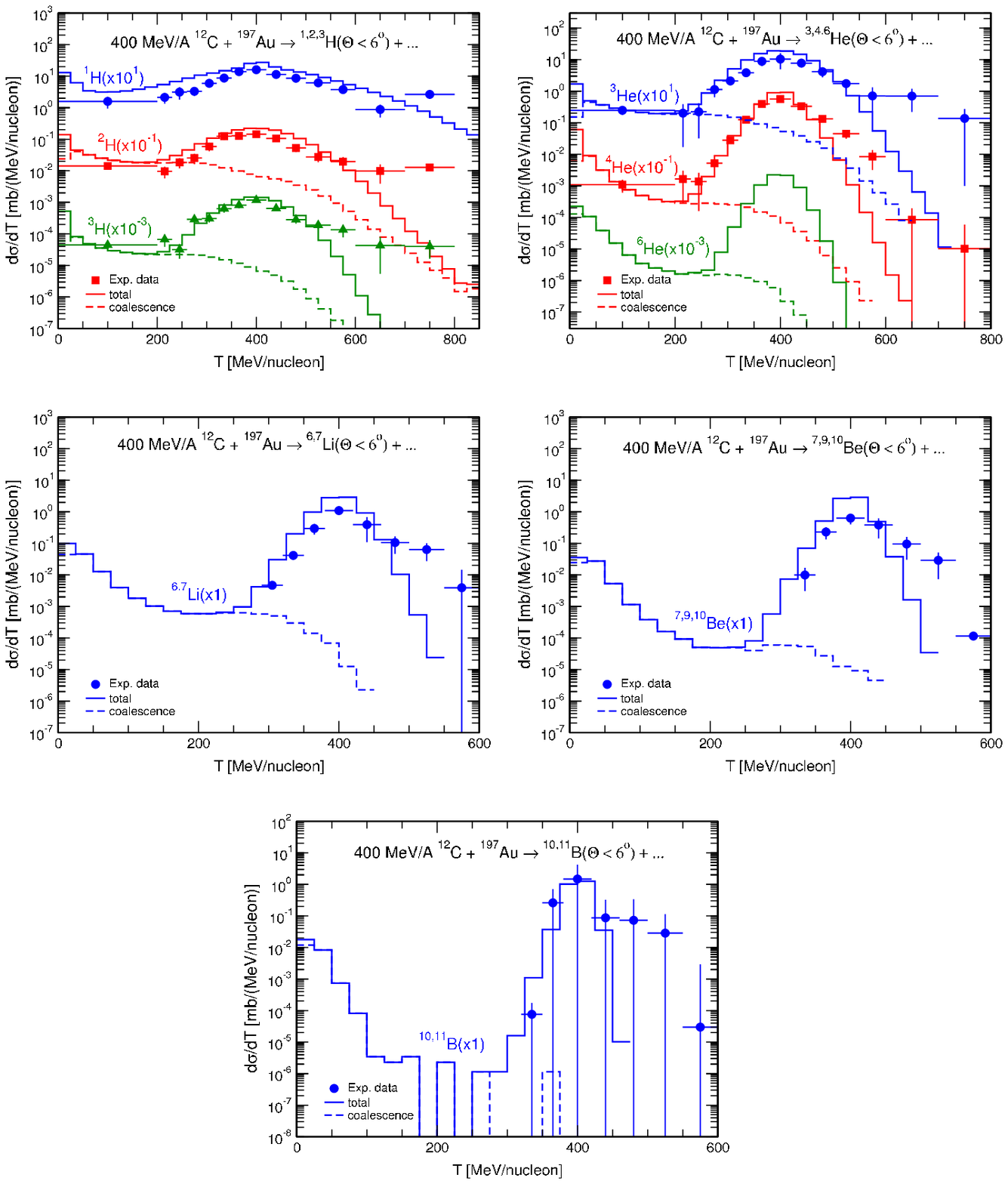}    
\caption{(Color online) Comparison of recently measured product spectra at
  forward lab angles of $\Theta \le 6^\circ $ from the fragmentation of
  $^{12}$C projectiles striking a $^{197}$Au target at 400 MeV/nucleon 
  \cite{Toppi2016} (symbols), compared with results by the previously fixed
  preliminary LAQGSM03.03F code (solid lines). The portion of the cross section
due to coalescence is indicated by dashed lines.
}
\label{C400Au_Z}
\end{figure*}

To demonstrate the reliability of even this non-optimized version of LAQGSM03.03F
for predicting unmeasured reactions, we compare the code predictions to
some recently measured data that were made available only after the
code was put into its current form. We show in Fig.\ \ref{C400Au_Z} the
measured forward-scattered fragmentation products from the interaction of
$^{12}$C nuclei at 400 MeV per nucleon with a $^{197}$Au target.
The measured cross sections are very well predicted, except for the very
highest energies, where the nucleons in these fragments have momenta more than
100 MeV/c above the momentum of the original nucleons
from the $^{12}$C projectiles. This discrepancy may indicate effects of
high-momentum components which are known to exist in real nuclei, and which are
missing from the simple semi-classical nucleon momentum distributions assumed in
the existing Fermi break-up model.

\section{Validation of the extended models}
\label{gamma_j_Chapter}

After extending the preequilibrium and coalescence models in CEM03.03F, we have
analyzed a number of nuclear reactions using different values for $A_{Fermi}$
in the Fermi break-up model discussed in Sec.\ \ref{FBU}, and have concluded
that generally a better agreement with most of the experimental data
so far analyzed is obtained with $A_{Fermi} = 12$, the same value as used
in the original 03.03 versions of CEM and LAQGSM. This value is used
for the extended ``F'' versions of these models.

\begin{figure}[htb!]
\centering
\includegraphics[width=3.5in]{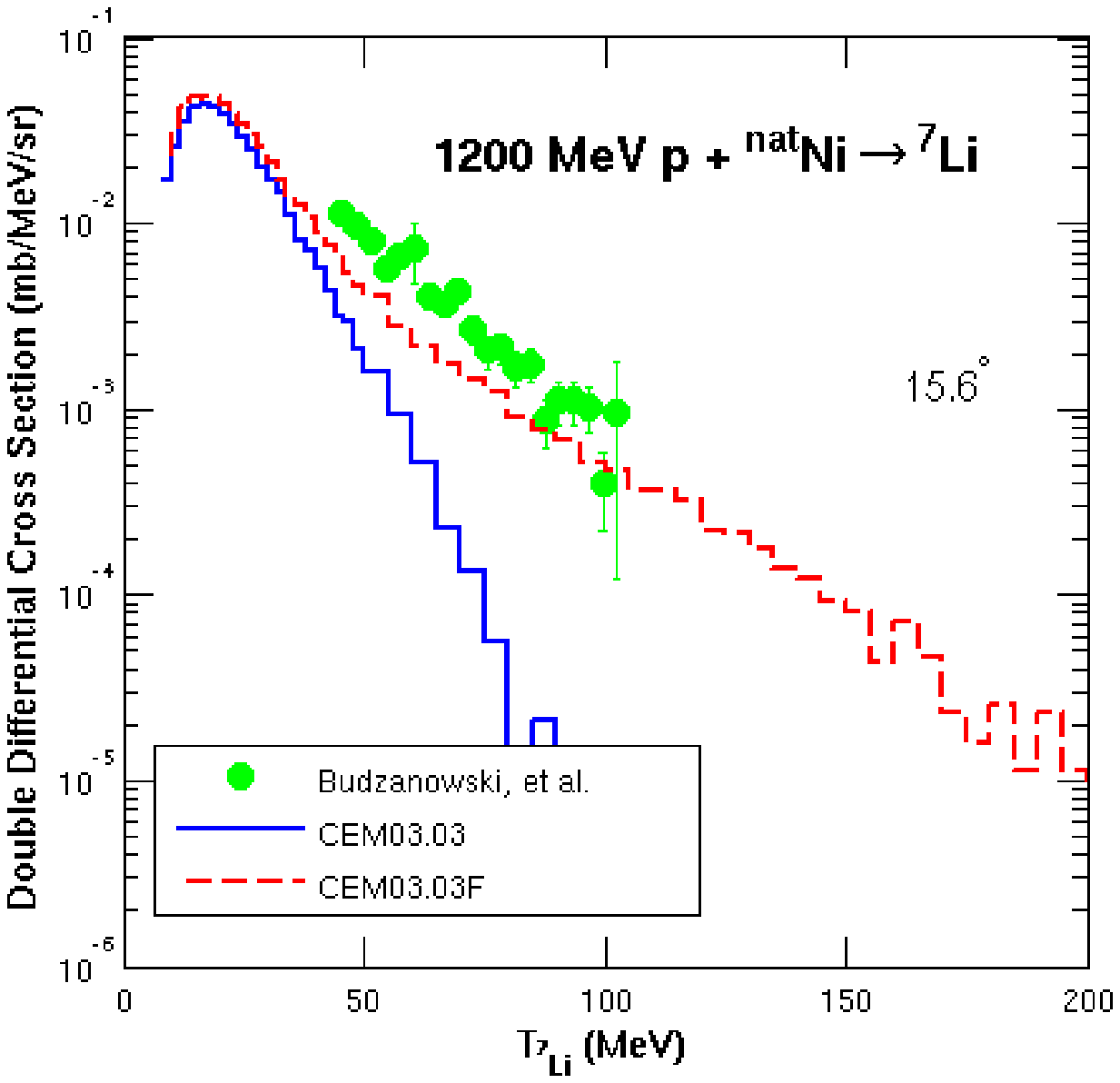}                
\caption{
(Color online) Comparison of experimental data for 1200 MeV $p$ + $^{nat}$Ni
  $\rightarrow$ $^7$Li at 15.6\degree, measured by Budzanowski et 
  al.\ \cite{BudzanowskiNi} (solid circles) to results from the original
  CEM03.03 (solid line) and to those from the improved CEM03.03F (dashed line).}
\label{p1200NiModel2}
\end{figure}

An example of calculations with the final version of
CEM03.03F is shown in Fig.~\ref{p1200NiModel2},
which compares experimental data for the $^7$Li spectrum at 15.6$^\circ$ from
1.2 GeV $p$ + Ni~\cite{BudzanowskiNi}
with results from CEM03.03 and CEM03.03F.
Similar results from MCNP6 are presented below in
Sec.\ \ref{MCNP6}. More extensive results can be found in 
Ref.\ \cite{Leslie-thesis}. CEM03.03F has much improved results compared to
the original CEM03.03, especially for heavy-cluster spectra.

Before implementing the extended ``F'' versions of the models
into the MCNP6 transport code, we have tested that the new models 
do not ``destroy'' the good predictive power and agreement with available
experimental data provided by the original CEM03.03 and LAQGSM03.03 event
generators, considering reactions previously well modeled and not used directly
in the current developments of the preequilibrium and coalescence models. This
is to verify that the extended ``F'' models have similar good predictive powers
established for the original event generators.
We show only a few examples from this extensive effort.

\begin{figure}[]
\centering
\includegraphics[width=3.5in]{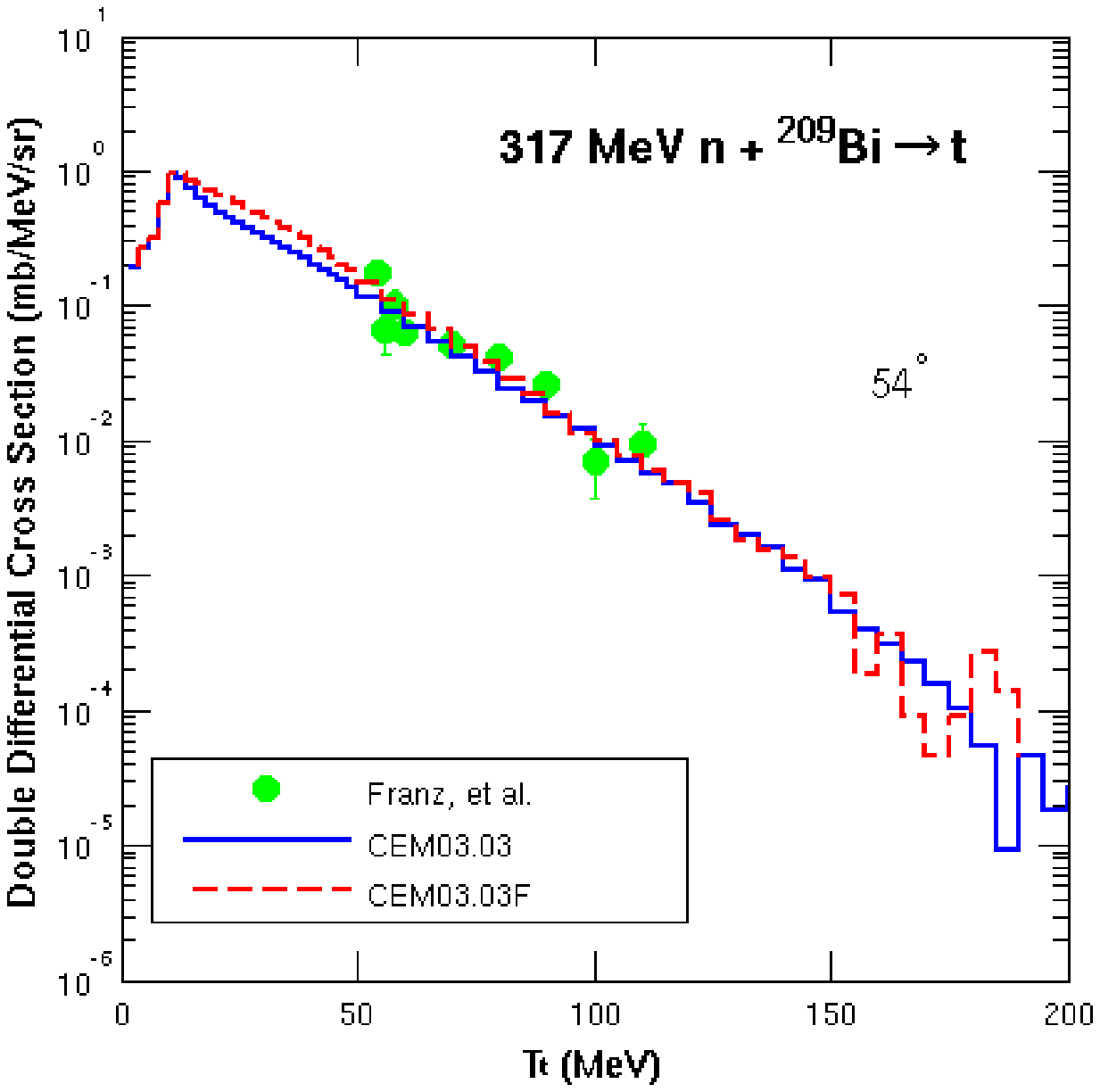}               
\caption{\
(Color online) Comparison of experimental data for 317 MeV
   $n$ + $^{209}$Bi $\rightarrow t$ at 54\degree measured by Franz et 
   al.\ \cite{Franz} (solid circles) to results from the original CEM03.03
   (solid lines) and from the improved CEM03.03F (dashed lines).
  }
\label{n317BiModel}
\end{figure}

Fig.~\ref{n317BiModel} demonstrates this for 317 MeV $n$ + $^{209}$Bi 
$\rightarrow$ $t$
at 54\degree, with experimental data measured by Franz et al.\ \cite{Franz}.
We have obtained similar results for other neutron-induced reactions,
at other incident energies, for other ejectiles and target nuclei.
These results illustrate that the improved production of heavy clusters in
CEM03.03F has not destroyed the spectra of particles and LF of mass 4 and below
from neutron-induced reactions, not considered during this development of the
``F" code versions.

\begin{figure}[]
\centering
\includegraphics[width=3.3in]{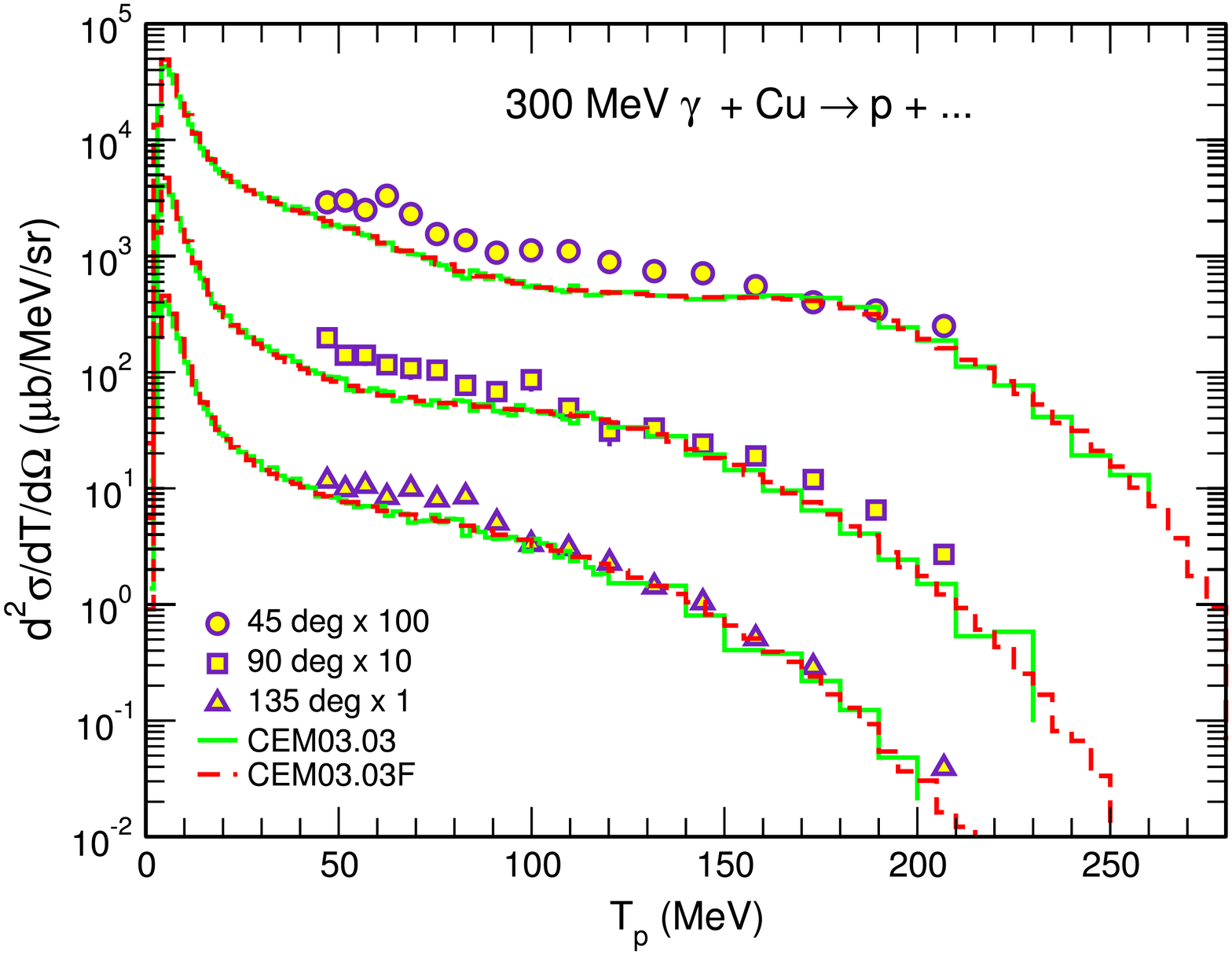}         
\caption[CEM03.03F results for 300 MeV $\gamma$ + $^{nat}$Cu $\rightarrow$ p]
  {(Color online) Experimental data for 300 MeV $\gamma$ + 
  $^{nat}$Cu $\rightarrow p$ at 45\degree, 90\degree, and 135\degree
  from Schumacher et al.\ \cite{Schumacher} (symbols), compared to results
  from the unmodified CEM03.03 (solid lines) and to those from CEM03.03F 
  (dashed lines).
  }
\label{g300CuModel}
\end{figure}

\begin{figure}[]
\centering
\includegraphics[width=3.3in]{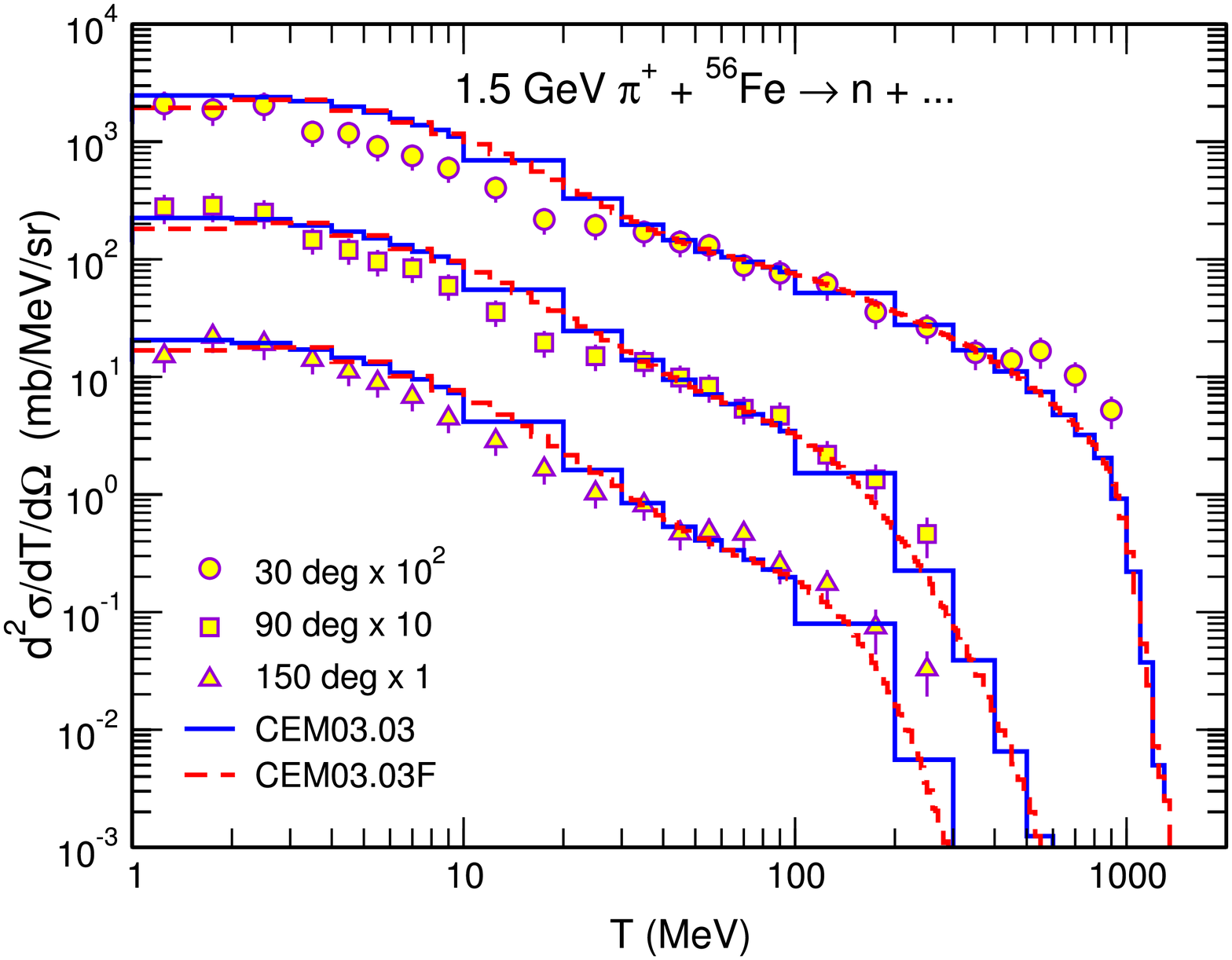}     
\caption{(Color online) Comparison of experimental data for 1500 MeV $\pi^+$ +
  $^{nat}$Fe $\rightarrow n$ at 30\degree, 90\degree, and 150\degree
  from Nakamoto et al.\ \cite{Nakamoto} (symbols), to results from the
  unmodified CEM03.03 (solid lines) and to CEM03.03F (dashed lines).}
\label{pip1500FeModel}
\end{figure}

Figs.\ \ref{g300CuModel} and \ref{pip1500FeModel} compare examples of experimental
data for $\gamma$- and $\pi$-induced reactions to results from CEM03.03 and
CEM03.03F.

Fig.~\ref{g300CuModel} shows the results for 300 MeV $\gamma$ + $^{nat}$Cu
$\rightarrow p$ at 45\degree, 90\degree, and 135\degree compared to experimental
data by Schumacher et al.\ \cite{Schumacher}. 

Fig.~\ref{pip1500FeModel} shows the model results for 1500 MeV $\pi^+$ +
$^{nat}$Fe $\rightarrow n$ at 30\degree, 90\degree, and 150\degree, compared
to experimental data from Nakamoto et al.\ \cite{Nakamoto}. 
The last two figures provide examples of the consistency between CEM03.03F and
CEM03.03 for ejectile spectra from $\gamma$- and $\pi$-induced reactions.

\begin{figure*}[]
\centering
\includegraphics[width=6.0in]{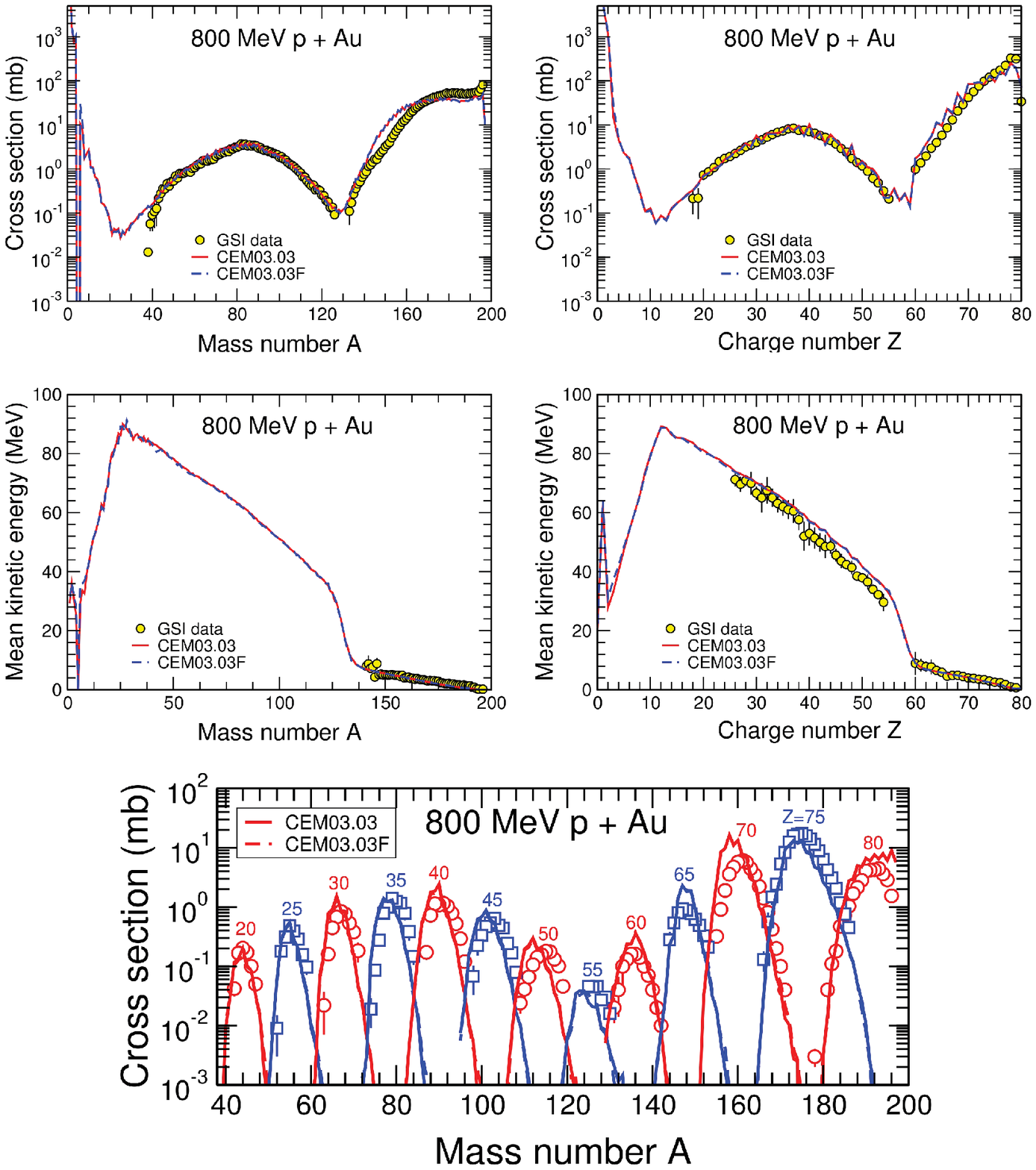}    
\caption[CEM03.03F results for 800 MeV p + $^{197}$Au, product yields]
 {(Color online) Comparison of measured mass and charge distributions
  of the product yields from the reaction 800 MeV $p$ + $^{197}$Au, of the mean
  kinetic energies of these products, and the mass distributions of the cross
  sections for the production of thirteen elements with atomic numbers Z
  ranging from 20 to 80 \cite{Benlliure} (circles), to predicted results
  from the original CEM03.03 (solid lines) and from CEM03.03F (dashed lines).
  }
\label{p800AuModel}
\end{figure*}

Fig.~\ref{p800AuModel} shows the measured~\cite{Benlliure} mass and
charge distributions of the product yields from the reaction 800 MeV $p$ +
$^{197}$Au, of the mean kinetic energy of these products, and the mass
distributions of the cross sections for the production of thirteen elements with
atomic number $Z$ from 20 to 80, compared to predicted results from the
original CEM03.03 and from CEM03.03F. The results are essentially identical
for the two code versions for these observables.

\begin{figure}[t!]
\centering
\includegraphics[width=3.5in]{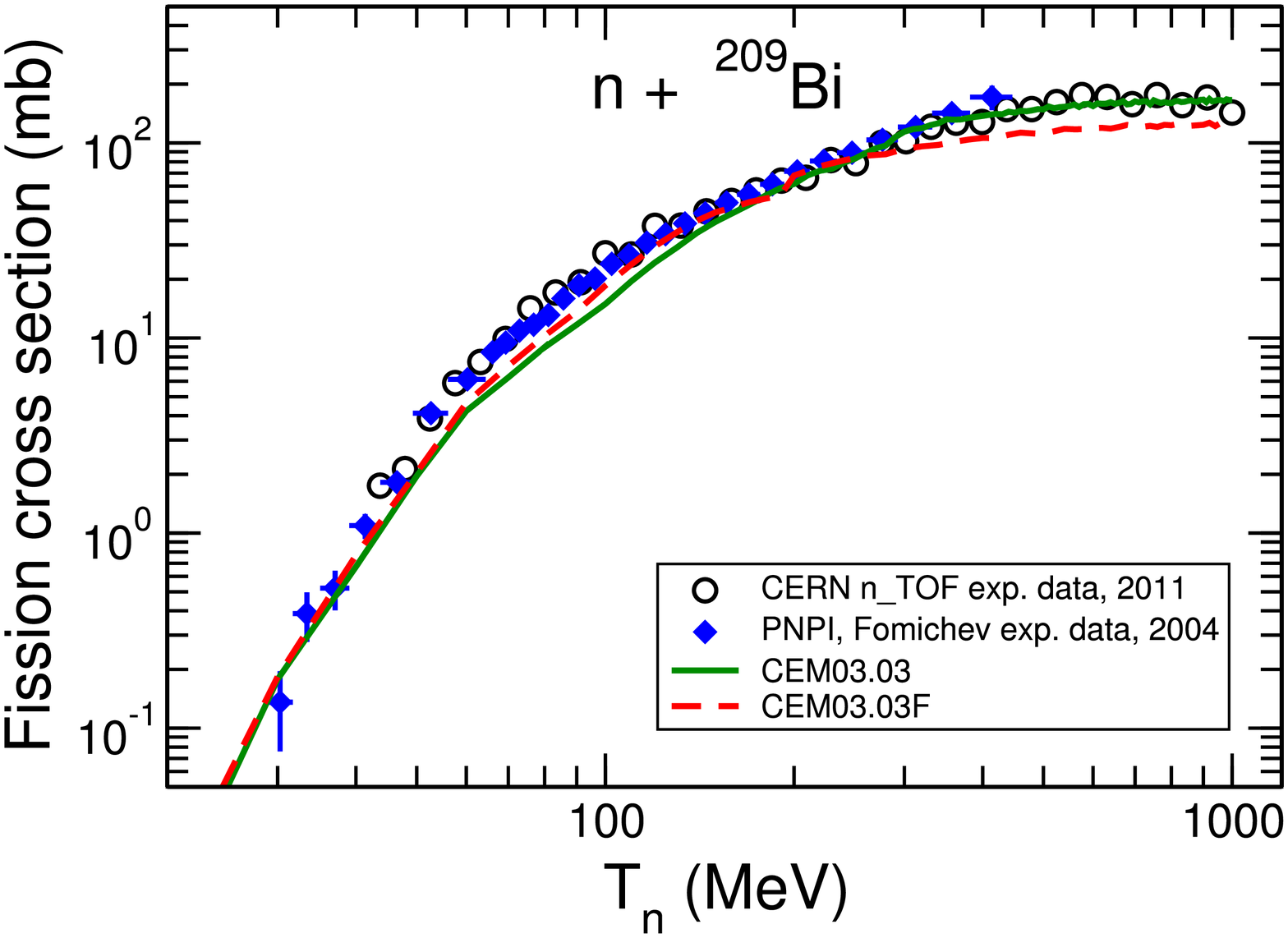}         
\caption{
(Color online) Comparison of measured fission cross sections for
  $n$ + Bi ~\cite{Fomichev} (open circles) and \cite{Tarrio} (solid diamonds)
  to results from the unmodified CEM03.03 (solid lines) and from CEM03.03F
  (dashed lines).}
\label{n+BiFission}
\end{figure}

Fig.~\ref{n+BiFission} shows the measured~\cite{Fomichev, Tarrio} fission
cross sections for $n$ + Bi as a function of neutron energy, compared to results
from CEM03.03 and CEM03.03F. CEM03.03F agrees reasonably well with these new data on
$n$ + Bi fission cross sections, even showing an improvement around energies of
100 MeV, while seeming to slightly underpredict the experiments above about 300 MeV.
But because CEM03.03F considers emission of LF at the preequilibrium
stage, there is some relative depletion from the compound nucleus cross section,
and the mean values of $A$, $Z$, and $E$ of the fissioning nuclei differ 
slightly from the corresponding values in CEM03.03.
 All details of the extended GEM2 code used in CEM
and LAQGSM to calculate $\sigma_f$ can be found in Refs.~\cite{CEM03.03,
Trieste08,GEM2}. In the case of subactinide nuclei, the main parameter that
determines fission cross sections calculated by GEM2 is the level-density
parameter in the fission channel, $a_f$ (or more exactly, the ratio $a_f/a_n$,
where $a_n$ is the level-density parameter for neutron evaporation).  Ideally,
the empirical $a_f/a_n$ parameter in CEM03.03F should be refit to reflect
the changed average properties of the fissioning compound nuclei following
the preequilibrium decay. This effort lies outside the scope of this report,
which is focused on energetic LF emission.

As CEM03.03 is the default event generator within MCNP6 for energies above 150
MeV, its ability to run simulations quickly is important. We tested the impact
of the current improvements on the
computation time with each incremental upgrade, and found either no significant
increase or only a small increase in the computation time. We tested also the
cumulative effect of all of the improvements on the computation time.
Adding all of the upgrades increases the computation time by approximately
one-third, depending upon the incident energy and target nucleus. Considering the
comprehensive nature of the upgrades, and the dramatic improvements made to the
description of heavy cluster production, this seems to be a relatively tolerable
increase.

\section{Implementation into MCNP6}
\label{MCNP6}
 
As mentioned in the Introduction, MCNP6 is a 
general-purpose Monte-Carlo radiation-transport code used by several thousands
of individuals or groups to simulate various nuclear applications. But MCNP6
can be and is actually used also in academic studies, e.g., to simulate 
experimental facitities or only
some of their parts, like target stations, or to estimate some 
unmeasured cross sections. The easiest way to calculate with MCNP6
the absolute values of spectra of ejectiles and/or yields of reaction products
is by using its so-called GENXS option (e.g.\ \cite{MCNP6, GENXS}).

Previously, double differential cross sections
of ejectiles could be calculated by MCNP6 using the GENXS option 
only for elementary particles and very light fragments up
to $^4$He. Thus, a necessary first step in implementing the improved CEM03.03F
into MCNP6 involves extending the ability to output spectra
of heavy clusters. 

We have extended the GENXS option~\cite{GENXS-extend}.  This GENXS upgrade
includes the ability to calculate (or, to ``tally,'' on the language used by
MCNP6) and output double differential cross sections for any fragment or heavy
ion. This upgade also includes the ability to tally and output
angle-integrated cross sections as a function of emitted fragment energy and
energy-integrated cross sections as a function of emitted angle, for any 
products.  More details on using this GENXS
extension can be found in Refs.~\cite{Leslie-thesis, GENXS-extend}. 

After completing and testing the improved CEM03.03F,
and after extending the GENXS option of MCNP6, we inserted CEM03.03F
to replace the older CEM03.03 event generator
into a working test version of MCNP6, called MCNP6-F.
Two of the current improvements are
always implemented in MCNP6-F: the upgraded NASA-Kalbach inverse
cross sections in the preequilibruim stage, and the new energy-dependent
$\gamma_j$ numerical model. The other two improvements (extension of 
preequilibrium emission
to $^{28}$Mg, and the extension of the coalescence model to $^7$Be), both of
which increase slightly the computation time, may be turned off if desired. 
We introduced into MCNP6-F a new input variable to specify the number of types
of the preequilibrium and coalescence fragments to be considered. 
The default of MCNP6-F is to consider the full 
upgrade of CEM03.03F as described above, i.e., up to 66
types of preequilibrium particle and LF and up to $A = 7$ in the 
coalescence model. But if a user wishes to save about 1/3 of
the computing time, this input parameter may be given 
a value of 6, to consider emisssion of only $n$, $p$, $d$, $t$, $^3$He,
and $^4$He, as done in the original CEM03.03.
As mentioned in the previous section, LAQGSM03.03F is still under
development, and is not yet implemented into MCNP6-F; this will be
done in the future, after the completion and validation of LAQGSM03.03F.

We have tested MCNP6-F on a large number of various reactions. A very few
examples from this validation work are presented below.

Fig.~\ref{p1200AuMCNP2} illustrates the results for 1200 MeV $p$ + $^{197}$Au
$\rightarrow$ $^6$Li at 20\degree, with experimental data by  Budzanowski
et al.~\cite{Budzanowski}. This figure provides additional evidence that
MCNP6-F demonstrates increased production of heavy clusters in the mid- and
high-energy regions compared to the original MCNP6. This reaction also
highlights the need to improve the evaporation model used by CEM: The peak 
of the spectrum is too high; such peaks are largely produced by evaporation. 
We expect that this situation might be improved by implementing the improved
inverse cross sections already incorporated into the preequilibrium model into
the evaporation model, and hope to do such work in the future. 
We note there is a recent work on improving
the Li\`{e}ge INC to its INCL4.6 version by A. Boudard et al.\ \cite{Cugnon},
which obtained similar results for heavy-cluster spectra
from this reaction using INCL4.6 + ABLA07.

\begin{figure}[h!]
\centering
\includegraphics[width=3.5in]{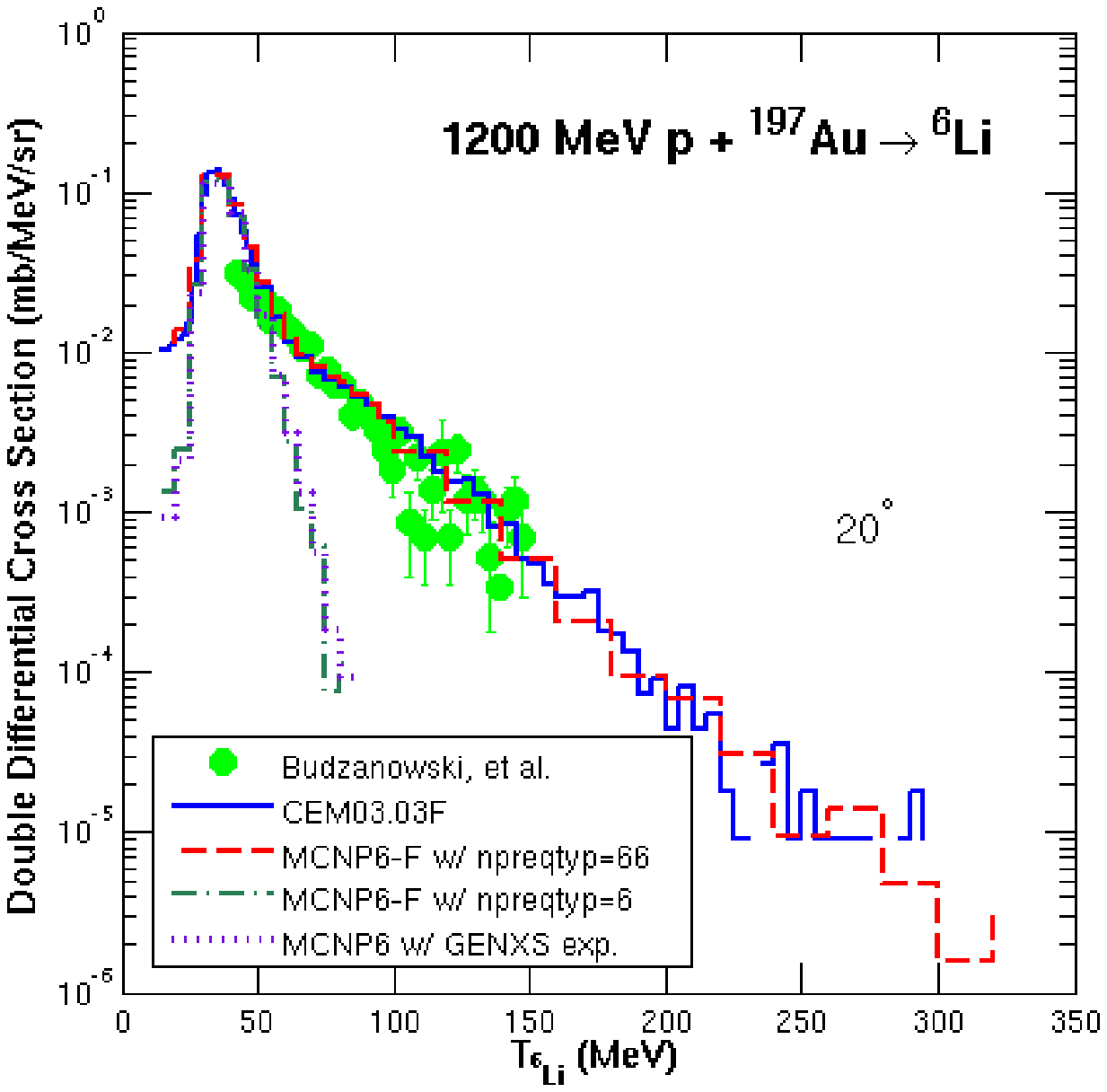}  
\caption{ 
(Color online) Comparison of experimental data on 1200 MeV $p$ +
  $^{197}$Au $\rightarrow$ $^6$Li at 20\degree, measured by Budzanowski
 et al.\ \cite{Budzanowski} (solid circles) to calculated results by CEM03.03F
 (blue solid lines), MCNP6-F with {\sf npreqtyp}=66 (dashed lines), MCNP6-F
 with {\sf npreqtyp}=6 (dash-dotted lines), and the original MCNP6 with the GENXS
 extension only (dotted lines).}
\label{p1200AuMCNP2}
\end{figure}

\begin{figure}[h!]
\centering
\includegraphics[width=3.5in]{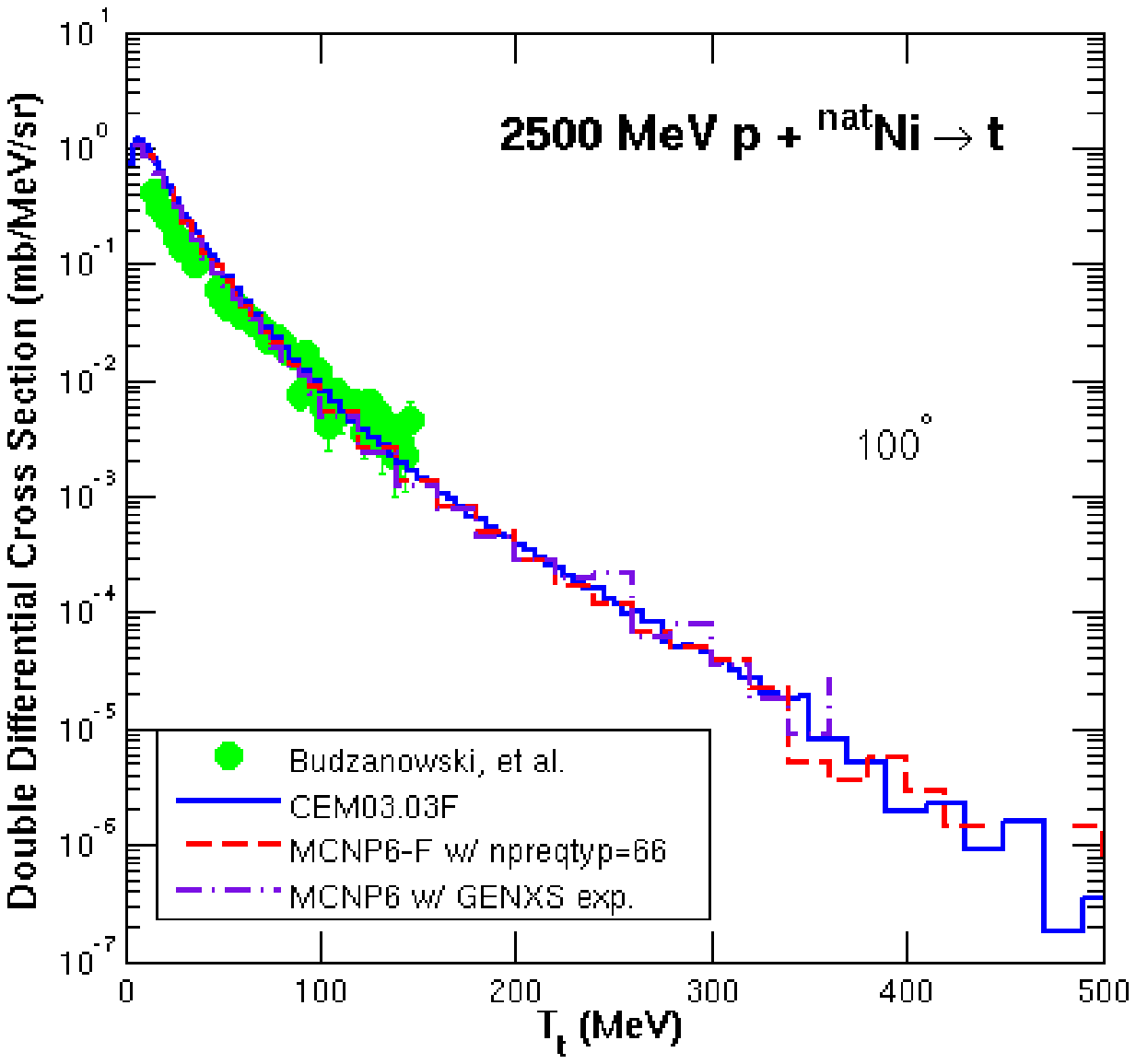}     
\includegraphics[width=3.5in]{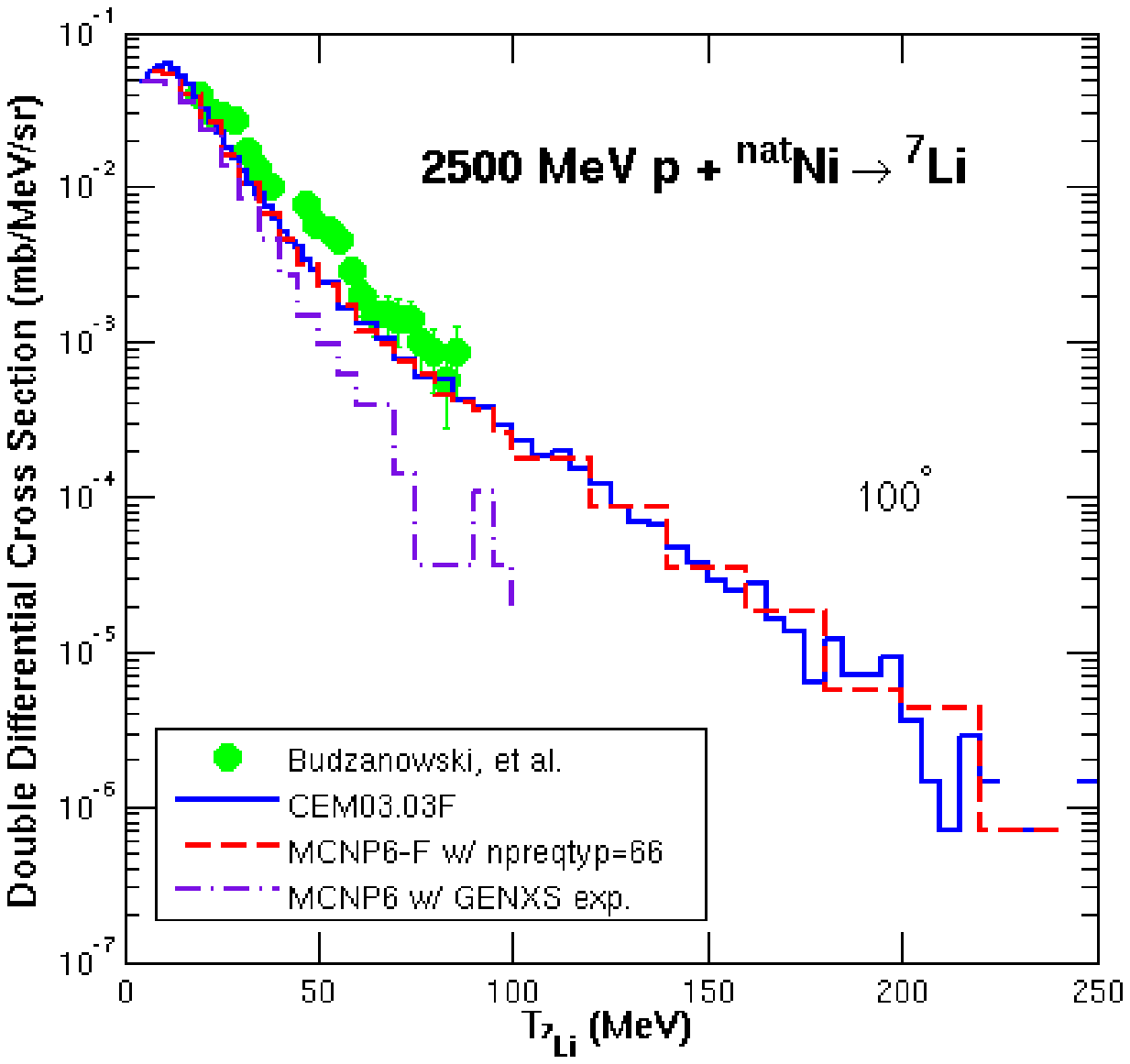}
\caption{
(Color online) Comparison of experimental data for 2500 MeV $p$ + $^{nat}$Ni
   $\rightarrow$ $t$, $^7$Li at 100\degree, measured by Budzanowski et 
   al.\ \cite{BudzanowskiNi} (solid circles) to calculated results from CEM03.03F 
  (solid blue lines), MCNP6-F with {\sf npreqtyp}=66 (dashed red lines), and the
   original MCNP6 with the GENXS extension only (dash-dotted purple lines).}
\label{p2500NiMCNP}
\end{figure}
Fig.~\ref{p2500NiMCNP} demonstrates the results for 2500 MeV $p$ + $^{nat}$Ni
$\rightarrow t$, $^7$Li at 100\degree, compared to experimental data measured by
 Budzanowski et al.\ \cite{BudzanowskiNi}. The triton spectra again illustrate
 that MCNP6-F achieves increased production of heavy clusters without ``destroying''
 the established spectra of nucleons and LF with $A<5$.

Many more similar results on the validation of the extended MCNP6-F for other
reactions can be found in Refs.\ \cite{Leslie-thesis, RMPmanuscript}.

\section{Conclusions}
We have presented the results of our work to improve the description of
energetic light-fragment production by the CEM and LAQGSM models, and by
the Los Alamos MCNP6 transport code from various nuclear 
reactions at energies up to $\sim 1$ TeV/nucleon. 

In these models, energetic LF can be produced via 
Fermi break-up, preequilibrium emission, and coalescence mechanisms.
We extend the modified exciton model used by CEM03.03
to describe emission of preequilibrium particles
to account for a posssiblity of multiple emission of up
to 66 types of particles and LF (up to $^{28}$Mg) at the preequilibrium
stage of reactions. 
For this extension, we had to develop an approximation,
or a ``numerical model,'' to calculate the probability $\gamma_j$
of several excited nucleons to condense into a fragment of the
type ``$j$'' inside the nucleus, that can be emitted at the 
preequilibrium stage of of a reaction.
 
We have also improved the calculation of inverse cross sections 
at the preequilibrium stage of reactions, 
with a new hybrid NASA-Kalbach approach, instead of using the old Dostrovsky
model which was used previously.  This extended version of the MEM is 
implemented into the upgraded CEM, labeled CEM03.03F, as well as into a new
LAQGSM03.03F.

Then, we extend the coalescence models in these
codes to account for coalescence of LF from nucleons emited at the intranuclear
cascade stage of reactions and from lighter clusters, up to fragments with mass
numbers $A \le 7$, in the case of CEM, and $A \le 12$, in the case
of LAQGSM. 
Finally, we study several variations of the Fermi break-up
model and choose the option with the best overall performance to use in
the production versions of the models. 

We have tested the improved versions of CEM and LAQGSM on a
variety on nuclear reactions induced by nucleons, pions,
photons, and heavy ions. On the whole, the 
improved models describe much better than the original
``03.03'' versions production of energetic fragments
heavier than $^4$He, without ``destroying'' the good agreement
provided by the standard versions for the emission of nucleons, 
light complex partiles, and residual nuclei.

Next, we have extended MCNP6 to allow calculation of and outputting
of spectra of fragments and heavier products with arbitrary mass and 
charge numbers.

Last, we implement the improved CEM03.03F 
into MCNP6, producing an upgraded version called MCNP6-F.
LAQGSM03.03F is not completed and will be incorporated into
MCNP6-F at a later time.
We have validated MCNP6-F on a variaty of measured nuclear reactions. 

We conclude that
the improved CEM, LAQGSM, and MCNP6 allow us to describe energetic LF
from particle- and nucleus-induced reactions and provide a
good agreement with available experimental data. They have a
good predictive power 
for various reactions at energies up to $\sim 1$ TeV/nucleon
and can be used as reliable tools in applications
involving such types of nuclear reactions as well as in
scientific studies.

For future work, we hope to complete the development of LAQGSM03.03F and
to implement it into MCNP6. We also hope to develop a better
deexcitation (evaporation/fission + multi-fragmentation) model for 
both the CEM and LAQGSM event generators. 

\section{Acknowledgments}

We are grateful to Drs.\ Christopher Werner and Avneet Sood of the
Los Alamos National Laboratory for encouraging discussions and support.
This study was carried out under the auspices of the National 
Nuclear Security Administration of the U. S. Department of Energy at 
Los Alamos National Laboratory under Contract No. DE-AC52-06NA25396.
This work was supported in part (for LMK) by the M. Hildred Blewett 
Fellowship of the American Physical Society, www.aps.org.



\begin{thebibliography}{166}%
\makeatletter
\providecommand \@ifxundefined [1]{%
 \@ifx{#1\undefined}
}%
\providecommand \@ifnum [1]{%
 \ifnum #1\expandafter \@firstoftwo
 \else \expandafter \@secondoftwo
 \fi
}%
\providecommand \@ifx [1]{%
 \ifx #1\expandafter \@firstoftwo
 \else \expandafter \@secondoftwo
 \fi
}%
\providecommand \natexlab [1]{#1}%
\providecommand \enquote  [1]{``#1''}%
\providecommand \bibnamefont  [1]{#1}%
\providecommand \bibfnamefont [1]{#1}%
\providecommand \citenamefont [1]{#1}%
\providecommand \href@noop [0]{\@secondoftwo}%
\providecommand \href [0]{\begingroup \@sanitize@url \@href}%
\providecommand \@href[1]{\@@startlink{#1}\@@href}%
\providecommand \@@href[1]{\endgroup#1\@@endlink}%
\providecommand \@sanitize@url [0]{\catcode `\\12\catcode `\$12\catcode
  `\&12\catcode `\#12\catcode `\^12\catcode `\_12\catcode `\%12\relax}%
\providecommand \@@startlink[1]{}%
\providecommand \@@endlink[0]{}%
\providecommand \url  [0]{\begingroup\@sanitize@url \@url }%
\providecommand \@url [1]{\endgroup\@href {#1}{\urlprefix }}%
\providecommand \urlprefix  [0]{URL }%
\providecommand \Eprint [0]{\href }%
\providecommand \doibase [0]{http://dx.doi.org/}%
\providecommand \selectlanguage [0]{\@gobble}%
\providecommand \bibinfo  [0]{\@secondoftwo}%
\providecommand \bibfield  [0]{\@secondoftwo}%
\providecommand \translation [1]{[#1]}%
\providecommand \BibitemOpen [0]{}%
\providecommand \bibitemStop [0]{}%
\providecommand \bibitemNoStop [0]{.\EOS\space}%
\providecommand \EOS [0]{\spacefactor3000\relax}%
\providecommand \BibitemShut  [1]{\csname bibitem#1\endcsname}%
\let\auto@bib@innerbib\@empty
\bibitem [{\citenamefont {Amelin}\ \emph {et~al.}(1990)\citenamefont {Amelin},
  \citenamefont {Gudima},\ and\ \citenamefont {Toneev}}]{Amelin}%
  \BibitemOpen
  \bibfield  {author} {\bibinfo {author} {\bibnamefont {Amelin}, \bibfnamefont
  {N.}}, \bibinfo {author} {\bibfnamefont {K.}~\bibnamefont {Gudima}}, \ and\
  \bibinfo {author} {\bibfnamefont {V.}~\bibnamefont {Toneev}}} (\bibinfo
  {year} {1990}),\ \href@noop {} {\bibfield  {journal} {\bibinfo  {journal}
  {Soviet Journal of Nuclear Physics}\ }\textbf {\bibinfo {volume} {51}},\
  \bibinfo {pages} {327}}\BibitemShut {NoStop}%
\bibitem [{\citenamefont {Andersen}\ \emph {et~al.}(2004)\citenamefont
  {Andersen}, \citenamefont {Ballarini}, \citenamefont {Battistoni},
  \citenamefont {Campanella}, \citenamefont {Carboni}, \citenamefont {Cerutti},
  \citenamefont {Empl}, \citenamefont {Fass\`{o}}, \citenamefont {Ferrari},
  \citenamefont {Gadioli}, \citenamefont {Garzelli}, \citenamefont {Lee},
  \citenamefont {Ottolenghi}, \citenamefont {Pelliccioni}, \citenamefont
  {Pinsky}, \citenamefont {Ranft}, \citenamefont {Roesler}, \citenamefont
  {Sala},\ and\ \citenamefont {Wilson}}]{Anderson}%
  \BibitemOpen
  \bibfield  {author} {\bibinfo {author} {\bibnamefont {Andersen},
  \bibfnamefont {V.}}, \bibinfo {author} {\bibfnamefont {F.}~\bibnamefont
  {Ballarini}}, \bibinfo {author} {\bibfnamefont {G.}~\bibnamefont
  {Battistoni}}, \bibinfo {author} {\bibfnamefont {M.}~\bibnamefont
  {Campanella}}, \bibinfo {author} {\bibfnamefont {M.}~\bibnamefont {Carboni}},
  \bibinfo {author} {\bibfnamefont {F.}~\bibnamefont {Cerutti}}, \bibinfo
  {author} {\bibfnamefont {A.}~\bibnamefont {Empl}}, \bibinfo {author}
  {\bibfnamefont {A.}~\bibnamefont {Fass\`{o}}}, \bibinfo {author}
  {\bibfnamefont {A.}~\bibnamefont {Ferrari}}, \bibinfo {author} {\bibfnamefont
  {E.}~\bibnamefont {Gadioli}}, \bibinfo {author} {\bibfnamefont
  {M.}~\bibnamefont {Garzelli}}, \bibinfo {author} {\bibfnamefont
  {K.}~\bibnamefont {Lee}}, \bibinfo {author} {\bibfnamefont {A.}~\bibnamefont
  {Ottolenghi}}, \bibinfo {author} {\bibfnamefont {M.}~\bibnamefont
  {Pelliccioni}}, \bibinfo {author} {\bibfnamefont {L.}~\bibnamefont {Pinsky}},
  \bibinfo {author} {\bibfnamefont {J.}~\bibnamefont {Ranft}}, \bibinfo
  {author} {\bibfnamefont {S.}~\bibnamefont {Roesler}}, \bibinfo {author}
  {\bibfnamefont {P.}~\bibnamefont {Sala}}, \ and\ \bibinfo {author}
  {\bibfnamefont {T.}~\bibnamefont {Wilson}}} (\bibinfo {year} {2004}),\
  \href@noop {} {\bibfield  {journal} {\bibinfo  {journal} {Advances in Space
  Research}\ }\textbf {\bibinfo {volume} {34}},\ \bibinfo {pages}
  {1302}}\BibitemShut {NoStop}%
\bibitem [{\citenamefont {Atchison}(2007)}]{2007Atchison}%
  \BibitemOpen
  \bibfield  {author} {\bibinfo {author} {\bibnamefont {Atchison},
  \bibfnamefont {F.}}} (\bibinfo {year} {2007}),\ \href@noop {} {\bibfield
  {journal} {\bibinfo  {journal} {Nuclear Instruments and Methods in Physics
  Research B}\ }\textbf {\bibinfo {volume} {259}},\ \bibinfo {pages}
  {909}}\BibitemShut {NoStop}%
\bibitem [{\citenamefont {Baktybaev}\ \emph {et~al.}(2003)\citenamefont
  {Baktybaev}, \citenamefont {Duisebaev}, \citenamefont {Duisebaev},
  \citenamefont {Ismailov}, \citenamefont {Itkis}, \citenamefont {Kadyrzhanov},
  \citenamefont {Kalpakchieva}, \citenamefont {Kuznetsov}, \citenamefont
  {Kuterbekov}, \citenamefont {Kukhtina}, \citenamefont {Lukyanov},
  \citenamefont {Mukhamedzhan}, \citenamefont {Penionzhkevich}, \citenamefont
  {Sadykov}, \citenamefont {Sobolev},\ and\ \citenamefont
  {Ugryumov}}]{Baktybaev}%
  \BibitemOpen
  \bibfield  {author} {\bibinfo {author} {\bibnamefont {Baktybaev},
  \bibfnamefont {M.}}, \bibinfo {author} {\bibfnamefont {A.}~\bibnamefont
  {Duisebaev}}, \bibinfo {author} {\bibfnamefont {B.}~\bibnamefont
  {Duisebaev}}, \bibinfo {author} {\bibfnamefont {K.}~\bibnamefont {Ismailov}},
  \bibinfo {author} {\bibfnamefont {M.}~\bibnamefont {Itkis}}, \bibinfo
  {author} {\bibfnamefont {K.}~\bibnamefont {Kadyrzhanov}}, \bibinfo {author}
  {\bibfnamefont {R.}~\bibnamefont {Kalpakchieva}}, \bibinfo {author}
  {\bibfnamefont {I.}~\bibnamefont {Kuznetsov}}, \bibinfo {author}
  {\bibfnamefont {K.}~\bibnamefont {Kuterbekov}}, \bibinfo {author}
  {\bibfnamefont {I.}~\bibnamefont {Kukhtina}}, \bibinfo {author}
  {\bibfnamefont {S.}~\bibnamefont {Lukyanov}}, \bibinfo {author}
  {\bibfnamefont {A.}~\bibnamefont {Mukhamedzhan}}, \bibinfo {author}
  {\bibfnamefont {Y.}~\bibnamefont {Penionzhkevich}}, \bibinfo {author}
  {\bibfnamefont {B.}~\bibnamefont {Sadykov}}, \bibinfo {author} {\bibfnamefont
  {Y.}~\bibnamefont {Sobolev}}, \ and\ \bibinfo {author} {\bibfnamefont
  {V.}~\bibnamefont {Ugryumov}}} (\bibinfo {year} {2003}),\ \href@noop {}
  {\bibfield  {journal} {\bibinfo  {journal} {Physics of Atomic Nuclei}\
  }\textbf {\bibinfo {volume} {66}},\ \bibinfo {pages} {1615}}\BibitemShut
  {NoStop}%
\bibitem [{\citenamefont {Barashenkov}(1993)}]{BarashenkovTables}%
  \BibitemOpen
  \bibfield  {author} {\bibinfo {author} {\bibnamefont {Barashenkov},
  \bibfnamefont {V.}}} (\bibinfo {year} {1993}),\ \href@noop {} {\enquote
  {\bibinfo {title} {Cross sections of interaction of particles and nuclei with
  nuclei (in {Russian})},}\ }\bibinfo {howpublished} {JINR, Dubna, Russia},\
  \bibinfo {note} {tabulated data available at:
  www.nea.fr/html/dbdata/bara.html}\BibitemShut {NoStop}%
\bibitem [{\citenamefont {Barashenkov}\ \emph {et~al.}(1990)\citenamefont
  {Barashenkov}, \citenamefont {Gudowski},\ and\ \citenamefont
  {Polanski}}]{BP-last}%
  \BibitemOpen
  \bibfield  {author} {\bibinfo {author} {\bibnamefont {Barashenkov},
  \bibfnamefont {V.}}, \bibinfo {author} {\bibfnamefont {W.}~\bibnamefont
  {Gudowski}}, \ and\ \bibinfo {author} {\bibfnamefont {A.}~\bibnamefont
  {Polanski}}} (\bibinfo {year} {1990}),\ \href@noop {} {\enquote {\bibinfo
  {title} {Integral high-energy nucleon-nucleus cross sections for mathematical
  experiments with electronuclear facilities},}\ }\bibinfo {howpublished} {JINR
  Communication E2-99-207, JINR, Dubna, Russia},\ \bibinfo {note} {private
  communications from Drs. Alexander Polanski and Dick Prael to Stepan
  Mashnik}\BibitemShut {NoStop}%
\bibitem [{\citenamefont {Barashenkov}\ \emph {et~al.}(1973)\citenamefont
  {Barashenkov}, \citenamefont {Il'inov}, \citenamefont {Sobolevskii},\ and\
  \citenamefont {Toneev}}]{Barashenkov1973}%
  \BibitemOpen
  \bibfield  {author} {\bibinfo {author} {\bibnamefont {Barashenkov},
  \bibfnamefont {V.}}, \bibinfo {author} {\bibfnamefont {A.}~\bibnamefont
  {Il'inov}}, \bibinfo {author} {\bibfnamefont {N.}~\bibnamefont
  {Sobolevskii}}, \ and\ \bibinfo {author} {\bibfnamefont {V.}~\bibnamefont
  {Toneev}}} (\bibinfo {year} {1973}),\ \href@noop {} {\bibfield  {journal}
  {\bibinfo  {journal} {Uspekhi Fiziches-kikh Nauk}\ }\textbf {\bibinfo
  {volume} {109}},\ \bibinfo {pages} {91}}\BibitemShut {NoStop}%
\bibitem [{\citenamefont {Barashenkov}\ and\ \citenamefont
  {Polanski}(1994)}]{BP}%
  \BibitemOpen
  \bibfield  {author} {\bibinfo {author} {\bibnamefont {Barashenkov},
  \bibfnamefont {V.}}, \ and\ \bibinfo {author} {\bibfnamefont
  {A.}~\bibnamefont {Polanski}}} (\bibinfo {year} {1994}),\ \href@noop {}
  {\enquote {\bibinfo {title} {Electronic guide for nuclear cross-sections},}\
  }\bibinfo {howpublished} {Joint Institute for Nuclear Research Communication
  E2-94-417, JINR, Dubna, Russia}\BibitemShut {NoStop}%
\bibitem [{\citenamefont {Barashenkov}\ and\ \citenamefont
  {Toneev}(1972)}]{Barashenkov1972}%
  \BibitemOpen
  \bibfield  {author} {\bibinfo {author} {\bibnamefont {Barashenkov},
  \bibfnamefont {V.}}, \ and\ \bibinfo {author} {\bibfnamefont
  {V.}~\bibnamefont {Toneev}}} (\bibinfo {year} {1972}),\ \href@noop {}
  {\bibinfo  {journal} {Atomizdat}\ }\BibitemShut {NoStop}%
\bibitem [{\citenamefont {Beghian}\ \emph {et~al.}(1966)\citenamefont
  {Beghian}, \citenamefont {Hofman},\ and\ \citenamefont {Wilensky}}]{Beghian}%
  \BibitemOpen
\bibfield  {journal} {  }\bibfield  {author} {\bibinfo {author} {\bibnamefont
  {Beghian}, \bibfnamefont {L.}}, \bibinfo {author} {\bibfnamefont
  {F.}~\bibnamefont {Hofman}}, \ and\ \bibinfo {author} {\bibfnamefont
  {S.}~\bibnamefont {Wilensky}}} (\bibinfo {year} {1966}),\ \href@noop {}
  {\bibfield  {journal} {\bibinfo  {journal} {Neutron Cross-Section Technical
  Conference, Washington}\ }\textbf {\bibinfo {volume} {2}},\ \bibinfo {pages}
  {726}}\BibitemShut {NoStop}%
\bibitem [{\citenamefont {Benecke}\ \emph {et~al.}(1969)\citenamefont
  {Benecke}, \citenamefont {Chou}, \citenamefont {Yang},\ and\ \citenamefont
  {Yen}}]{Benecke}%
  \BibitemOpen
  \bibfield  {author} {\bibinfo {author} {\bibnamefont {Benecke}, \bibfnamefont
  {J.}}, \bibinfo {author} {\bibfnamefont {T.}~\bibnamefont {Chou}}, \bibinfo
  {author} {\bibfnamefont {C.}~\bibnamefont {Yang}}, \ and\ \bibinfo {author}
  {\bibfnamefont {E.}~\bibnamefont {Yen}}} (\bibinfo {year} {1969}),\
  \href@noop {} {\bibfield  {journal} {\bibinfo  {journal} {Physical Review}\
  }\textbf {\bibinfo {volume} {188}},\ \bibinfo {pages} {2159}}\BibitemShut
  {NoStop}%
\bibitem [{\citenamefont {Benlliure}\ \emph {et~al.}(2001)\citenamefont
  {Benlliure}, \citenamefont {Armbruster}, \citenamefont {Bernas},
  \citenamefont {Boudard}, \citenamefont {Dufour}, \citenamefont {Enqvist},
  \citenamefont {Legrain}, \citenamefont {Leray}, \citenamefont {Mustapha},
  \citenamefont {Rejmund}, \citenamefont {Schmidt}, \citenamefont {St\'ephan},
  \citenamefont {Tasaan-Got},\ and\ \citenamefont {Volant}}]{Benlliure}%
  \BibitemOpen
  \bibfield  {author} {\bibinfo {author} {\bibnamefont {Benlliure},
  \bibfnamefont {J.}}, \bibinfo {author} {\bibfnamefont {P.}~\bibnamefont
  {Armbruster}}, \bibinfo {author} {\bibfnamefont {M.}~\bibnamefont {Bernas}},
  \bibinfo {author} {\bibfnamefont {A.}~\bibnamefont {Boudard}}, \bibinfo
  {author} {\bibfnamefont {J.}~\bibnamefont {Dufour}}, \bibinfo {author}
  {\bibfnamefont {T.}~\bibnamefont {Enqvist}}, \bibinfo {author} {\bibfnamefont
  {R.}~\bibnamefont {Legrain}}, \bibinfo {author} {\bibfnamefont
  {S.}~\bibnamefont {Leray}}, \bibinfo {author} {\bibfnamefont
  {B.}~\bibnamefont {Mustapha}}, \bibinfo {author} {\bibfnamefont
  {F.}~\bibnamefont {Rejmund}}, \bibinfo {author} {\bibfnamefont {K.-H.}\
  \bibnamefont {Schmidt}}, \bibinfo {author} {\bibfnamefont {C.}~\bibnamefont
  {St\'ephan}}, \bibinfo {author} {\bibfnamefont {L.}~\bibnamefont
  {Tasaan-Got}}, \ and\ \bibinfo {author} {\bibfnamefont {C.}~\bibnamefont
  {Volant}}} (\bibinfo {year} {2001}),\ \href@noop {} {\bibfield  {journal}
  {\bibinfo  {journal} {Nuclear Physics A}\ }\textbf {\bibinfo {volume}
  {683}},\ \bibinfo {pages} {513}}\BibitemShut {NoStop}%
\bibitem [{\citenamefont {Bernas}\ \emph {et~al.}(2003)\citenamefont {Bernas},
  \citenamefont {Armstrong}, \citenamefont {Benlliure}, \citenamefont
  {Boudard}, \citenamefont {Casarejos}, \citenamefont {Czajkowski},
  \citenamefont {Enqvist}, \citenamefont {Legrain}, \citenamefont {Leray},
  \citenamefont {Mustapha}, \citenamefont {Napolitani}, \citenamefont
  {Pereira}, \citenamefont {Rejmund}, \citenamefont {Ricciardi}, \citenamefont
  {Schmidt}, \citenamefont {St\'ephan}, \citenamefont {Taieb}, \citenamefont
  {Tissan-Got},\ and\ \citenamefont {Volant}}]{p1000U_2}%
  \BibitemOpen
  \bibfield  {author} {\bibinfo {author} {\bibnamefont {Bernas}, \bibfnamefont
  {M.}}, \bibinfo {author} {\bibfnamefont {P.}~\bibnamefont {Armstrong}},
  \bibinfo {author} {\bibfnamefont {J.}~\bibnamefont {Benlliure}}, \bibinfo
  {author} {\bibfnamefont {A.}~\bibnamefont {Boudard}}, \bibinfo {author}
  {\bibfnamefont {E.}~\bibnamefont {Casarejos}}, \bibinfo {author}
  {\bibfnamefont {S.}~\bibnamefont {Czajkowski}}, \bibinfo {author}
  {\bibfnamefont {T.}~\bibnamefont {Enqvist}}, \bibinfo {author} {\bibfnamefont
  {R.}~\bibnamefont {Legrain}}, \bibinfo {author} {\bibfnamefont
  {S.}~\bibnamefont {Leray}}, \bibinfo {author} {\bibfnamefont
  {B.}~\bibnamefont {Mustapha}}, \bibinfo {author} {\bibfnamefont
  {P.}~\bibnamefont {Napolitani}}, \bibinfo {author} {\bibfnamefont
  {J.}~\bibnamefont {Pereira}}, \bibinfo {author} {\bibfnamefont
  {F.}~\bibnamefont {Rejmund}}, \bibinfo {author} {\bibfnamefont {M.-V.}\
  \bibnamefont {Ricciardi}}, \bibinfo {author} {\bibfnamefont {K.-H.}\
  \bibnamefont {Schmidt}}, \bibinfo {author} {\bibfnamefont {C.}~\bibnamefont
  {St\'ephan}}, \bibinfo {author} {\bibfnamefont {J.}~\bibnamefont {Taieb}},
  \bibinfo {author} {\bibfnamefont {L.}~\bibnamefont {Tissan-Got}}, \ and\
  \bibinfo {author} {\bibfnamefont {C.}~\bibnamefont {Volant}}} (\bibinfo
  {year} {2003}),\ \href@noop {} {\bibfield  {journal} {\bibinfo  {journal}
  {Nuclear Physics A}\ }\textbf {\bibinfo {volume} {725}},\ \bibinfo {pages}
  {213}},\ \bibinfo {note} {nucl-ex/0304003}\BibitemShut {NoStop}%
\bibitem [{\citenamefont {Bertini}(1963)}]{Bertini}%
  \BibitemOpen
  \bibfield  {author} {\bibinfo {author} {\bibnamefont {Bertini}, \bibfnamefont
  {H.}}} (\bibinfo {year} {1963}),\ \href@noop {} {\bibfield  {journal}
  {\bibinfo  {journal} {Physical Review}\ }\textbf {\bibinfo {volume} {131}},\
  \bibinfo {pages} {1801}}\BibitemShut {NoStop}%
\bibitem [{\citenamefont {Bertini}(1969)}]{Bertini2}%
  \BibitemOpen
  \bibfield  {author} {\bibinfo {author} {\bibnamefont {Bertini}, \bibfnamefont
  {H.}}} (\bibinfo {year} {1969}),\ \href@noop {} {\bibfield  {journal}
  {\bibinfo  {journal} {Physical Review}\ }\textbf {\bibinfo {volume} {188}},\
  \bibinfo {pages} {1711}}\BibitemShut {NoStop}%
\bibitem [{\citenamefont {Betak}(1976)}]{betak76}%
  \BibitemOpen
  \bibfield  {author} {\bibinfo {author} {\bibnamefont {Betak}, \bibfnamefont
  {E.}}} (\bibinfo {year} {1976}),\ \href@noop {} {\bibfield  {journal}
  {\bibinfo  {journal} {Acta Physica Slovaka}\ }\textbf {\bibinfo {volume}
  {26}},\ \bibinfo {pages} {21}}\BibitemShut {NoStop}%
\bibitem [{\citenamefont {Beyster}\ \emph {et~al.}(1955)\citenamefont
  {Beyster}, \citenamefont {Henkel}, \citenamefont {Nobles},\ and\
  \citenamefont {Kister}}]{Beyster1955}%
  \BibitemOpen
  \bibfield  {author} {\bibinfo {author} {\bibnamefont {Beyster}, \bibfnamefont
  {J.}}, \bibinfo {author} {\bibfnamefont {R.}~\bibnamefont {Henkel}}, \bibinfo
  {author} {\bibfnamefont {R.}~\bibnamefont {Nobles}}, \ and\ \bibinfo {author}
  {\bibfnamefont {J.}~\bibnamefont {Kister}}} (\bibinfo {year} {1955}),\
  \href@noop {} {\bibfield  {journal} {\bibinfo  {journal} {Physical Review}\
  }\textbf {\bibinfo {volume} {98}},\ \bibinfo {pages} {1216}}\BibitemShut
  {NoStop}%
\bibitem [{\citenamefont {Beyster}\ \emph {et~al.}(1956)\citenamefont
  {Beyster}, \citenamefont {Walt},\ and\ \citenamefont {Salmi}}]{Beyster1956}%
  \BibitemOpen
  \bibfield  {author} {\bibinfo {author} {\bibnamefont {Beyster}, \bibfnamefont
  {J.}}, \bibinfo {author} {\bibfnamefont {M.}~\bibnamefont {Walt}}, \ and\
  \bibinfo {author} {\bibfnamefont {E.}~\bibnamefont {Salmi}}} (\bibinfo {year}
  {1956}),\ \href@noop {} {\bibfield  {journal} {\bibinfo  {journal} {Physical
  Review}\ }\textbf {\bibinfo {volume} {104}},\ \bibinfo {pages}
  {1319}}\BibitemShut {NoStop}%
\bibitem [{\citenamefont {Blideanu}\ \emph {et~al.}(2004)\citenamefont
  {Blideanu}, \citenamefont {Lecolley}, \citenamefont {Lecolley}, \citenamefont
  {Lefort}, \citenamefont {Marie}, \citenamefont {{Ata{\c c} }and G.~Ban},
  \citenamefont {Bergenwall}, \citenamefont {Blomgren}, \citenamefont
  {Dangtip}, \citenamefont {Elmgren}, \citenamefont {Eudes}, \citenamefont
  {Foucher}, \citenamefont {Guertin}, \citenamefont {Haddad}, \citenamefont
  {Hildebrand}, \citenamefont {Johansson}, \citenamefont {Jonsson},
  \citenamefont {Kerveno}, \citenamefont {Kirchner}, \citenamefont {Klug},
  \citenamefont {Brun}, \citenamefont {Lebrun}, \citenamefont {Louvel},
  \citenamefont {Nadel-Turonski}, \citenamefont {Nilsson}, \citenamefont
  {Olsson}, \citenamefont {Pomp}, \citenamefont {Prokofiev}, \citenamefont
  {Renberg}, \citenamefont {{Rivi\`ere}}, \citenamefont {Slypen}, \citenamefont
  {{Stuttg\'e}}, \citenamefont {Tippawan},\ and\ \citenamefont
  {{\"Osterlund}}}]{Blideanu}%
  \BibitemOpen
  \bibfield  {author} {\bibinfo {author} {\bibnamefont {Blideanu},
  \bibfnamefont {V.}}, \bibinfo {author} {\bibfnamefont {F.}~\bibnamefont
  {Lecolley}}, \bibinfo {author} {\bibfnamefont {J.}~\bibnamefont {Lecolley}},
  \bibinfo {author} {\bibfnamefont {T.}~\bibnamefont {Lefort}}, \bibinfo
  {author} {\bibfnamefont {N.}~\bibnamefont {Marie}}, \bibinfo {author}
  {\bibfnamefont {A.}~\bibnamefont {{Ata{\c c} }and G.~Ban}}, \bibinfo {author}
  {\bibfnamefont {B.}~\bibnamefont {Bergenwall}}, \bibinfo {author}
  {\bibfnamefont {J.}~\bibnamefont {Blomgren}}, \bibinfo {author}
  {\bibfnamefont {S.}~\bibnamefont {Dangtip}}, \bibinfo {author} {\bibfnamefont
  {K.}~\bibnamefont {Elmgren}}, \bibinfo {author} {\bibfnamefont
  {P.}~\bibnamefont {Eudes}}, \bibinfo {author} {\bibfnamefont
  {Y.}~\bibnamefont {Foucher}}, \bibinfo {author} {\bibfnamefont
  {A.}~\bibnamefont {Guertin}}, \bibinfo {author} {\bibfnamefont
  {F.}~\bibnamefont {Haddad}}, \bibinfo {author} {\bibfnamefont
  {A.}~\bibnamefont {Hildebrand}}, \bibinfo {author} {\bibfnamefont
  {C.}~\bibnamefont {Johansson}}, \bibinfo {author} {\bibfnamefont
  {O.}~\bibnamefont {Jonsson}}, \bibinfo {author} {\bibfnamefont
  {M.}~\bibnamefont {Kerveno}}, \bibinfo {author} {\bibfnamefont
  {T.}~\bibnamefont {Kirchner}}, \bibinfo {author} {\bibfnamefont
  {J.}~\bibnamefont {Klug}}, \bibinfo {author} {\bibfnamefont {C.}~\bibnamefont
  {Brun}}, \bibinfo {author} {\bibfnamefont {C.}~\bibnamefont {Lebrun}},
  \bibinfo {author} {\bibfnamefont {M.}~\bibnamefont {Louvel}}, \bibinfo
  {author} {\bibfnamefont {P.}~\bibnamefont {Nadel-Turonski}}, \bibinfo
  {author} {\bibfnamefont {L.}~\bibnamefont {Nilsson}}, \bibinfo {author}
  {\bibfnamefont {N.}~\bibnamefont {Olsson}}, \bibinfo {author} {\bibfnamefont
  {S.}~\bibnamefont {Pomp}}, \bibinfo {author} {\bibfnamefont {A.}~\bibnamefont
  {Prokofiev}}, \bibinfo {author} {\bibfnamefont {P.-U.}\ \bibnamefont
  {Renberg}}, \bibinfo {author} {\bibfnamefont {G.}~\bibnamefont
  {{Rivi\`ere}}}, \bibinfo {author} {\bibfnamefont {I.}~\bibnamefont {Slypen}},
  \bibinfo {author} {\bibfnamefont {L.}~\bibnamefont {{Stuttg\'e}}}, \bibinfo
  {author} {\bibfnamefont {U.}~\bibnamefont {Tippawan}}, \ and\ \bibinfo
  {author} {\bibfnamefont {M.}~\bibnamefont {{\"Osterlund}}}} (\bibinfo {year}
  {2004}),\ \href@noop {} {\bibfield  {journal} {\bibinfo  {journal} {Physical
  Review C}\ }\textbf {\bibinfo {volume} {70}},\ \bibinfo {pages}
  {014607}}\BibitemShut {NoStop}%
\bibitem [{\citenamefont {Bondorf}\ \emph {et~al.}(1995)\citenamefont
  {Bondorf}, \citenamefont {Botvina}, \citenamefont {Iljinov}, \citenamefont
  {Mishustin},\ and\ \citenamefont {Sneppen}}]{SMM}%
  \BibitemOpen
  \bibfield  {author} {\bibinfo {author} {\bibnamefont {Bondorf}, \bibfnamefont
  {J.}}, \bibinfo {author} {\bibfnamefont {A.}~\bibnamefont {Botvina}},
  \bibinfo {author} {\bibfnamefont {A.}~\bibnamefont {Iljinov}}, \bibinfo
  {author} {\bibfnamefont {I.}~\bibnamefont {Mishustin}}, \ and\ \bibinfo
  {author} {\bibfnamefont {K.}~\bibnamefont {Sneppen}}} (\bibinfo {year}
  {1995}),\ \href@noop {} {\bibfield  {journal} {\bibinfo  {journal} {Physics
  Reports}\ }\textbf {\bibinfo {volume} {257}},\ \bibinfo {pages}
  {133}}\BibitemShut {NoStop}%
\bibitem [{\citenamefont {Bonner}\ and\ \citenamefont
  {Slattery}(1959)}]{Bonner}%
  \BibitemOpen
  \bibfield  {author} {\bibinfo {author} {\bibnamefont {Bonner}, \bibfnamefont
  {T.}}, \ and\ \bibinfo {author} {\bibfnamefont {J.}~\bibnamefont {Slattery}}}
  (\bibinfo {year} {1959}),\ \href@noop {} {\bibfield  {journal} {\bibinfo
  {journal} {Physical Review}\ }\textbf {\bibinfo {volume} {113}},\ \bibinfo
  {pages} {1088}}\BibitemShut {NoStop}%
\bibitem [{\citenamefont {Borning}(1987)}]{Borning}%
  \BibitemOpen
  \bibfield  {author} {\bibinfo {author} {\bibnamefont {Borning}, \bibfnamefont
  {A.}}} (\bibinfo {year} {1987}),\ \href@noop {} {\enquote {\bibinfo {title}
  {Computer system reliability and nuclear war},}\ }\bibinfo {howpublished}
  {Foundation for Global Community}\BibitemShut {NoStop}%
\bibitem [{\citenamefont {Boudard}\ \emph {et~al.}(2013)\citenamefont
  {Boudard}, \citenamefont {Cugnon}, \citenamefont {David}, \citenamefont
  {Leray},\ and\ \citenamefont {Mancusi}}]{INCL4.6}%
  \BibitemOpen
  \bibfield  {author} {\bibinfo {author} {\bibnamefont {Boudard}, \bibfnamefont
  {A.}}, \bibinfo {author} {\bibfnamefont {J.}~\bibnamefont {Cugnon}}, \bibinfo
  {author} {\bibfnamefont {J.-C.}\ \bibnamefont {David}}, \bibinfo {author}
  {\bibfnamefont {S.}~\bibnamefont {Leray}}, \ and\ \bibinfo {author}
  {\bibfnamefont {D.}~\bibnamefont {Mancusi}}} (\bibinfo {year} {2013}),\
  \href@noop {} {\bibfield  {journal} {\bibinfo  {journal} {Physical Review C}\
  }\textbf {\bibinfo {volume} {87}},\ \bibinfo {pages} {014606}}\BibitemShut
  {NoStop}%
\bibitem [{\citenamefont {Boudard}\ \emph {et~al.}(2002)\citenamefont
  {Boudard}, \citenamefont {Cugnon}, \citenamefont {Leray},\ and\ \citenamefont
  {Volant}}]{INCL2}%
  \BibitemOpen
  \bibfield  {author} {\bibinfo {author} {\bibnamefont {Boudard}, \bibfnamefont
  {A.}}, \bibinfo {author} {\bibfnamefont {J.}~\bibnamefont {Cugnon}}, \bibinfo
  {author} {\bibfnamefont {S.}~\bibnamefont {Leray}}, \ and\ \bibinfo {author}
  {\bibfnamefont {C.}~\bibnamefont {Volant}}} (\bibinfo {year} {2002}),\
  \href@noop {} {\bibfield  {journal} {\bibinfo  {journal} {Physical Review C}\
  }\textbf {\bibinfo {volume} {66}},\ \bibinfo {pages} {044615}}\BibitemShut
  {NoStop}%
\bibitem [{\citenamefont {Bubak}\ \emph {et~al.}(2007)\citenamefont {Bubak},
  \citenamefont {Budzanowski}, \citenamefont {Filges}, \citenamefont
  {Goldenbaum}, \citenamefont {Heczko}, \citenamefont {Hodde}, \citenamefont
  {Jarczyk}, \citenamefont {Kamys}, \citenamefont {Kistryn}, \citenamefont
  {Kistryn}, \citenamefont {Kliczewski}, \citenamefont {Kowalczyk},
  \citenamefont {Kozik}, \citenamefont {Kulessa}, \citenamefont {Machner},
  \citenamefont {Magiera}, \citenamefont {Migda{\l}}, \citenamefont {Paul},
  \citenamefont {Piskor-Ignatowicz}, \citenamefont {Pucha{\l}a}, \citenamefont
  {Pysz}, \citenamefont {Rudy}, \citenamefont {Siudak}, \citenamefont
  {Wojciechowski},\ and\ \citenamefont {W{\"u}stner}}]{Bubak}%
  \BibitemOpen
  \bibfield  {author} {\bibinfo {author} {\bibnamefont {Bubak}, \bibfnamefont
  {A.}}, \bibinfo {author} {\bibfnamefont {A.}~\bibnamefont {Budzanowski}},
  \bibinfo {author} {\bibfnamefont {D.}~\bibnamefont {Filges}}, \bibinfo
  {author} {\bibfnamefont {F.}~\bibnamefont {Goldenbaum}}, \bibinfo {author}
  {\bibfnamefont {A.}~\bibnamefont {Heczko}}, \bibinfo {author} {\bibfnamefont
  {H.}~\bibnamefont {Hodde}}, \bibinfo {author} {\bibfnamefont
  {L.}~\bibnamefont {Jarczyk}}, \bibinfo {author} {\bibfnamefont
  {B.}~\bibnamefont {Kamys}}, \bibinfo {author} {\bibfnamefont
  {M.}~\bibnamefont {Kistryn}}, \bibinfo {author} {\bibfnamefont
  {S.}~\bibnamefont {Kistryn}}, \bibinfo {author} {\bibfnamefont
  {S.}~\bibnamefont {Kliczewski}}, \bibinfo {author} {\bibfnamefont
  {A.}~\bibnamefont {Kowalczyk}}, \bibinfo {author} {\bibfnamefont
  {E.}~\bibnamefont {Kozik}}, \bibinfo {author} {\bibfnamefont
  {P.}~\bibnamefont {Kulessa}}, \bibinfo {author} {\bibfnamefont
  {H.}~\bibnamefont {Machner}}, \bibinfo {author} {\bibfnamefont
  {A.}~\bibnamefont {Magiera}}, \bibinfo {author} {\bibfnamefont
  {W.}~\bibnamefont {Migda{\l}}}, \bibinfo {author} {\bibfnamefont
  {N.}~\bibnamefont {Paul}}, \bibinfo {author} {\bibfnamefont {B.}~\bibnamefont
  {Piskor-Ignatowicz}}, \bibinfo {author} {\bibfnamefont {M.}~\bibnamefont
  {Pucha{\l}a}}, \bibinfo {author} {\bibfnamefont {K.}~\bibnamefont {Pysz}},
  \bibinfo {author} {\bibfnamefont {Z.}~\bibnamefont {Rudy}}, \bibinfo {author}
  {\bibfnamefont {R.}~\bibnamefont {Siudak}}, \bibinfo {author} {\bibfnamefont
  {M.}~\bibnamefont {Wojciechowski}}, \ and\ \bibinfo {author} {\bibfnamefont
  {P.}~\bibnamefont {W{\"u}stner}}} (\bibinfo {year} {2007}),\ \href@noop {}
  {\bibfield  {journal} {\bibinfo  {journal} {Physical Review C}\ }\textbf
  {\bibinfo {volume} {76}},\ \bibinfo {pages} {014618}}\BibitemShut {NoStop}%
\bibitem [{\citenamefont {Budzanowski}\ \emph {et~al.}(2008)\citenamefont
  {Budzanowski}, \citenamefont {Fidelus}, \citenamefont {Filges}, \citenamefont
  {Goldenbaum}, \citenamefont {Hodde}, \citenamefont {Jarczyk}, \citenamefont
  {Kamys}, \citenamefont {Kistryn}, \citenamefont {Kistryn}, \citenamefont
  {Kliczewski}, \citenamefont {Kowalczyk}, \citenamefont {Kozik}, \citenamefont
  {Kulessa}, \citenamefont {Machner}, \citenamefont {Magiera}, \citenamefont
  {Piskor-Ignatowicz}, \citenamefont {Pysz}, \citenamefont {Rudy},
  \citenamefont {Siudak},\ and\ \citenamefont {Wojciechowski}}]{Budzanowski}%
  \BibitemOpen
  \bibfield  {author} {\bibinfo {author} {\bibnamefont {Budzanowski},
  \bibfnamefont {A.}}, \bibinfo {author} {\bibfnamefont {M.}~\bibnamefont
  {Fidelus}}, \bibinfo {author} {\bibfnamefont {D.}~\bibnamefont {Filges}},
  \bibinfo {author} {\bibfnamefont {F.}~\bibnamefont {Goldenbaum}}, \bibinfo
  {author} {\bibfnamefont {H.}~\bibnamefont {Hodde}}, \bibinfo {author}
  {\bibfnamefont {L.}~\bibnamefont {Jarczyk}}, \bibinfo {author} {\bibfnamefont
  {B.}~\bibnamefont {Kamys}}, \bibinfo {author} {\bibfnamefont
  {M.}~\bibnamefont {Kistryn}}, \bibinfo {author} {\bibfnamefont
  {S.}~\bibnamefont {Kistryn}}, \bibinfo {author} {\bibfnamefont
  {S.}~\bibnamefont {Kliczewski}}, \bibinfo {author} {\bibfnamefont
  {A.}~\bibnamefont {Kowalczyk}}, \bibinfo {author} {\bibfnamefont
  {E.}~\bibnamefont {Kozik}}, \bibinfo {author} {\bibfnamefont
  {P.}~\bibnamefont {Kulessa}}, \bibinfo {author} {\bibfnamefont
  {H.}~\bibnamefont {Machner}}, \bibinfo {author} {\bibfnamefont
  {A.}~\bibnamefont {Magiera}}, \bibinfo {author} {\bibfnamefont
  {B.}~\bibnamefont {Piskor-Ignatowicz}}, \bibinfo {author} {\bibfnamefont
  {K.}~\bibnamefont {Pysz}}, \bibinfo {author} {\bibfnamefont {Z.}~\bibnamefont
  {Rudy}}, \bibinfo {author} {\bibfnamefont {R.}~\bibnamefont {Siudak}}, \ and\
  \bibinfo {author} {\bibfnamefont {M.}~\bibnamefont {Wojciechowski}}}
  (\bibinfo {year} {2008}),\ \href@noop {} {\bibfield  {journal} {\bibinfo
  {journal} {Physical Review C}\ }\textbf {\bibinfo {volume} {78}},\ \bibinfo
  {pages} {024603}}\BibitemShut {NoStop}%
\bibitem [{\citenamefont {Budzanowski}\ \emph {et~al.}(2010)\citenamefont
  {Budzanowski}, \citenamefont {Fidelus}, \citenamefont {Filges}, \citenamefont
  {Goldenbaum}, \citenamefont {Hodde}, \citenamefont {Jarczyk}, \citenamefont
  {Kamys}, \citenamefont {Kistryn}, \citenamefont {Kistryn}, \citenamefont
  {Kliczewski}, \citenamefont {Kowalczyk}, \citenamefont {Kozik}, \citenamefont
  {Kulessa}, \citenamefont {Machner}, \citenamefont {Magiera}, \citenamefont
  {Piskor-Ignatowicz}, \citenamefont {Pysz}, \citenamefont {Rudy},
  \citenamefont {Siudak},\ and\ \citenamefont {Wojciechowski}}]{BudzanowskiNi}%
  \BibitemOpen
  \bibfield  {author} {\bibinfo {author} {\bibnamefont {Budzanowski},
  \bibfnamefont {A.}}, \bibinfo {author} {\bibfnamefont {M.}~\bibnamefont
  {Fidelus}}, \bibinfo {author} {\bibfnamefont {D.}~\bibnamefont {Filges}},
  \bibinfo {author} {\bibfnamefont {F.}~\bibnamefont {Goldenbaum}}, \bibinfo
  {author} {\bibfnamefont {H.}~\bibnamefont {Hodde}}, \bibinfo {author}
  {\bibfnamefont {L.}~\bibnamefont {Jarczyk}}, \bibinfo {author} {\bibfnamefont
  {B.}~\bibnamefont {Kamys}}, \bibinfo {author} {\bibfnamefont
  {M.}~\bibnamefont {Kistryn}}, \bibinfo {author} {\bibfnamefont
  {S.}~\bibnamefont {Kistryn}}, \bibinfo {author} {\bibfnamefont
  {S.}~\bibnamefont {Kliczewski}}, \bibinfo {author} {\bibfnamefont
  {A.}~\bibnamefont {Kowalczyk}}, \bibinfo {author} {\bibfnamefont
  {E.}~\bibnamefont {Kozik}}, \bibinfo {author} {\bibfnamefont
  {P.}~\bibnamefont {Kulessa}}, \bibinfo {author} {\bibfnamefont
  {H.}~\bibnamefont {Machner}}, \bibinfo {author} {\bibfnamefont
  {A.}~\bibnamefont {Magiera}}, \bibinfo {author} {\bibfnamefont
  {B.}~\bibnamefont {Piskor-Ignatowicz}}, \bibinfo {author} {\bibfnamefont
  {K.}~\bibnamefont {Pysz}}, \bibinfo {author} {\bibfnamefont {Z.}~\bibnamefont
  {Rudy}}, \bibinfo {author} {\bibfnamefont {R.}~\bibnamefont {Siudak}}, \ and\
  \bibinfo {author} {\bibfnamefont {M.}~\bibnamefont {Wojciechowski}}}
  (\bibinfo {year} {2010}),\ \href@noop {} {\bibfield  {journal} {\bibinfo
  {journal} {Physical Review C}\ }\textbf {\bibinfo {volume} {82}},\ \bibinfo
  {pages} {034605}}\BibitemShut {NoStop}%
\bibitem [{\citenamefont {Carlson}(1996)}]{Carlson}%
  \BibitemOpen
  \bibfield  {author} {\bibinfo {author} {\bibnamefont {Carlson}, \bibfnamefont
  {R.}}} (\bibinfo {year} {1996}),\ \href@noop {} {\bibfield  {journal}
  {\bibinfo  {journal} {Atomic Data and Nuclear Data Tables}\ }\textbf
  {\bibinfo {volume} {63}},\ \bibinfo {pages} {93}}\BibitemShut {NoStop}%
\bibitem [{\citenamefont {Charity}\ \emph {et~al.}(2001)\citenamefont
  {Charity}, \citenamefont {Sobotka}, \citenamefont {Cibor}, \citenamefont
  {Hagel}, \citenamefont {Murray}, \citenamefont {Natowitz}, \citenamefont
  {Wada}, \citenamefont {{Y. El Masri}}, \citenamefont {Fabris}, \citenamefont
  {Nebbia}, \citenamefont {Viesti}, \citenamefont {Cinausero}, \citenamefont
  {Fioretto}, \citenamefont {Prete}, \citenamefont {Wagner},\ and\
  \citenamefont {Xu}}]{Charity01}%
  \BibitemOpen
  \bibfield  {author} {\bibinfo {author} {\bibnamefont {Charity}, \bibfnamefont
  {R.}}, \bibinfo {author} {\bibfnamefont {L.}~\bibnamefont {Sobotka}},
  \bibinfo {author} {\bibfnamefont {J.}~\bibnamefont {Cibor}}, \bibinfo
  {author} {\bibfnamefont {K.}~\bibnamefont {Hagel}}, \bibinfo {author}
  {\bibfnamefont {M.}~\bibnamefont {Murray}}, \bibinfo {author} {\bibfnamefont
  {J.}~\bibnamefont {Natowitz}}, \bibinfo {author} {\bibfnamefont
  {R.}~\bibnamefont {Wada}}, \bibinfo {author} {\bibnamefont {{Y. El Masri}}},
  \bibinfo {author} {\bibfnamefont {D.}~\bibnamefont {Fabris}}, \bibinfo
  {author} {\bibfnamefont {G.}~\bibnamefont {Nebbia}}, \bibinfo {author}
  {\bibfnamefont {G.}~\bibnamefont {Viesti}}, \bibinfo {author} {\bibfnamefont
  {M.}~\bibnamefont {Cinausero}}, \bibinfo {author} {\bibfnamefont
  {E.}~\bibnamefont {Fioretto}}, \bibinfo {author} {\bibfnamefont
  {G.}~\bibnamefont {Prete}}, \bibinfo {author} {\bibfnamefont
  {A.}~\bibnamefont {Wagner}}, \ and\ \bibinfo {author} {\bibfnamefont
  {H.}~\bibnamefont {Xu}}} (\bibinfo {year} {2001}),\ \href@noop {} {\bibfield
  {journal} {\bibinfo  {journal} {Physical Review C}\ }\textbf {\bibinfo
  {volume} {63}},\ \bibinfo {pages} {024611}},\ \bibinfo {note}
  {{http://www.chemistry.wustl.edu/~rc/gemini/}}\BibitemShut {NoStop}%
\bibitem [{\citenamefont {Cline}(1972)}]{Cline}%
  \BibitemOpen
  \bibfield  {author} {\bibinfo {author} {\bibnamefont {Cline}, \bibfnamefont
  {C.}}} (\bibinfo {year} {1972}),\ \href@noop {} {\bibfield  {journal}
  {\bibinfo  {journal} {Nuclear Physics A}\ }\textbf {\bibinfo {volume}
  {193}},\ \bibinfo {pages} {417}}\BibitemShut {NoStop}%
\bibitem [{\citenamefont {Cooper}(2012)}]{NeciaGrantCooper}%
  \BibitemOpen
  \bibfield  {author} {\bibinfo {author} {\bibnamefont {Cooper}, \bibfnamefont
  {N.}}} (\bibinfo {year} {2012}),\ \href@noop {} {\bibinfo  {journal}
  {National Security Science}\ ,\ \bibinfo {pages} {12}}\BibitemShut {NoStop}%
\bibitem [{\citenamefont {Cugnon}\ \emph
  {et~al.}(2011{\natexlab{a}})\citenamefont {Cugnon}, \citenamefont {Boudard},
  \citenamefont {David}, \citenamefont {Keli\'c-Heil}, \citenamefont {Leray},
  \citenamefont {Mancusi},\ and\ \citenamefont {Ricciardi}}]{Cugnon2010}%
  \BibitemOpen
\bibfield  {journal} {  }\bibfield  {author} {\bibinfo {author} {\bibnamefont
  {Cugnon}, \bibfnamefont {J.}}, \bibinfo {author} {\bibfnamefont
  {A.}~\bibnamefont {Boudard}}, \bibinfo {author} {\bibfnamefont {J.-C.}\
  \bibnamefont {David}}, \bibinfo {author} {\bibfnamefont {A.}~\bibnamefont
  {Keli\'c-Heil}}, \bibinfo {author} {\bibfnamefont {S.}~\bibnamefont {Leray}},
  \bibinfo {author} {\bibfnamefont {D.}~\bibnamefont {Mancusi}}, \ and\
  \bibinfo {author} {\bibfnamefont {M.~V.}\ \bibnamefont {Ricciardi}}}
  (\bibinfo {year} {2011}{\natexlab{a}}),\ \href@noop {} {\bibfield  {journal}
  {\bibinfo  {journal} {Journal of Physics: Conference Series}\ }\textbf
  {\bibinfo {volume} {312}},\ \bibinfo {pages} {082019}}\BibitemShut {NoStop}%
\bibitem [{\citenamefont {Cugnon}\ \emph
  {et~al.}(2011{\natexlab{b}})\citenamefont {Cugnon}, \citenamefont {Boudard},
  \citenamefont {David}, \citenamefont {Keli\'c-Heil}, \citenamefont {Mancusi},
  \citenamefont {Ricciardi},\ and\ \citenamefont {Leray}}]{Cugnon2011}%
  \BibitemOpen
  \bibfield  {author} {\bibinfo {author} {\bibnamefont {Cugnon}, \bibfnamefont
  {J.}}, \bibinfo {author} {\bibfnamefont {A.}~\bibnamefont {Boudard}},
  \bibinfo {author} {\bibfnamefont {J.-C.}\ \bibnamefont {David}}, \bibinfo
  {author} {\bibfnamefont {A.}~\bibnamefont {Keli\'c-Heil}}, \bibinfo {author}
  {\bibfnamefont {D.}~\bibnamefont {Mancusi}}, \bibinfo {author} {\bibfnamefont
  {V.}~\bibnamefont {Ricciardi}}, \ and\ \bibinfo {author} {\bibfnamefont
  {S.}~\bibnamefont {Leray}}} (\bibinfo {year} {2011}{\natexlab{b}}),\
  \href@noop {} {\bibinfo  {journal} {{Proc. of the Tenth Int. Topical Meeting
  on Nuclear Applications of Accelerators (accApp), April 3-7, 2011, Knoxville,
  USA, 978-0-89448-706-4, 2012}}\ }\BibitemShut {NoStop}%
\bibitem [{\citenamefont {Cugnon}\ \emph {et~al.}(1997)\citenamefont {Cugnon},
  \citenamefont {Volant},\ and\ \citenamefont {Vuillier}}]{INCL}%
  \BibitemOpen
\bibfield  {journal} {  }\bibfield  {author} {\bibinfo {author} {\bibnamefont
  {Cugnon}, \bibfnamefont {J.}}, \bibinfo {author} {\bibfnamefont
  {C.}~\bibnamefont {Volant}}, \ and\ \bibinfo {author} {\bibfnamefont
  {S.}~\bibnamefont {Vuillier}}} (\bibinfo {year} {1997}),\ \href@noop {}
  {\bibfield  {journal} {\bibinfo  {journal} {Nuclear Physics A}\ }\textbf
  {\bibinfo {volume} {620}},\ \bibinfo {pages} {475}}\BibitemShut {NoStop}%
\bibitem [{\citenamefont {David}(2015)}]{David2015}%
  \BibitemOpen
  \bibfield  {author} {\bibinfo {author} {\bibnamefont {David}, \bibfnamefont
  {J.-C.}}} (\bibinfo {year} {2015}),\ \href@noop {} {\bibfield  {journal}
  {\bibinfo  {journal} {European Physical Journal A}\ }\textbf {\bibinfo
  {volume} {51}},\ \bibinfo {pages} {157}},\ \bibinfo {note}
  {arXiv:1505.0382}\BibitemShut {NoStop}%
\bibitem [{\citenamefont {David}\ \emph {et~al.}(2011)\citenamefont {David},
  \citenamefont {Boudard}, \citenamefont {Cugnon}, \citenamefont {Leray},\ and\
  \citenamefont {Mancusi}}]{Cugnon}%
  \BibitemOpen
  \bibfield  {author} {\bibinfo {author} {\bibnamefont {David}, \bibfnamefont
  {J.-C.}}, \bibinfo {author} {\bibfnamefont {A.}~\bibnamefont {Boudard}},
  \bibinfo {author} {\bibfnamefont {J.}~\bibnamefont {Cugnon}}, \bibinfo
  {author} {\bibfnamefont {S.}~\bibnamefont {Leray}}, \ and\ \bibinfo {author}
  {\bibfnamefont {D.}~\bibnamefont {Mancusi}}} (\bibinfo {year} {2011}),\
  \href@noop {} {\bibfield  {journal} {\bibinfo  {journal} {Rapport interne
  IRFU-11-249, {ANDES (Accurate Nuclear Data for nuclear Energy Sustainability)
  EURATOM FP7 grant agreement No. 249671, Task T4.1 - Deliverable D4.1,
  FP7-ANDES - WP 4}}\ }}\bibinfo {note}
  {Http://www-ist.cea.fr/publicea/exl-php/cadcgp.php}\BibitemShut {NoStop}%
\bibitem [{\citenamefont {Degtjarev}(1966)}]{Degtjarev}%
  \BibitemOpen
  \bibfield  {author} {\bibinfo {author} {\bibnamefont {Degtjarev},
  \bibfnamefont {J.}}} (\bibinfo {year} {1966}),\ \href@noop {} {\bibfield
  {journal} {\bibinfo  {journal} {Journal Nuclear Energy, Part A+B (Reactor
  Sci. Techn.)}\ }\textbf {\bibinfo {volume} {20}},\ \bibinfo {pages}
  {818}}\BibitemShut {NoStop}%
\bibitem [{\citenamefont {Dostrovsky}\ \emph {et~al.}(1959)\citenamefont
  {Dostrovsky}, \citenamefont {Fraenkel},\ and\ \citenamefont
  {Friedlander}}]{Dostrovsky}%
  \BibitemOpen
  \bibfield  {author} {\bibinfo {author} {\bibnamefont {Dostrovsky},
  \bibfnamefont {I.}}, \bibinfo {author} {\bibfnamefont {Z.}~\bibnamefont
  {Fraenkel}}, \ and\ \bibinfo {author} {\bibfnamefont {G.}~\bibnamefont
  {Friedlander}}} (\bibinfo {year} {1959}),\ \href@noop {} {\bibfield
  {journal} {\bibinfo  {journal} {Physical Review}\ }\textbf {\bibinfo {volume}
  {116}},\ \bibinfo {pages} {683}}\BibitemShut {NoStop}%
\bibitem [{\citenamefont {El-Nagdy}\ \emph {et~al.}(2013)\citenamefont
  {El-Nagdy}, \citenamefont {Abdelsalam}, \citenamefont {Abou-Moussa},\ and\
  \citenamefont {Badawy}}]{El-Nagdy}%
  \BibitemOpen
  \bibfield  {author} {\bibinfo {author} {\bibnamefont {El-Nagdy},
  \bibfnamefont {M.}}, \bibinfo {author} {\bibfnamefont {A.}~\bibnamefont
  {Abdelsalam}}, \bibinfo {author} {\bibfnamefont {Z.}~\bibnamefont
  {Abou-Moussa}}, \ and\ \bibinfo {author} {\bibfnamefont {B.}~\bibnamefont
  {Badawy}}} (\bibinfo {year} {2013}),\ \href@noop {} {\bibfield  {journal}
  {\bibinfo  {journal} {Canadian Journal of Physics}\ }\textbf {\bibinfo
  {volume} {91}},\ \bibinfo {pages} {737}}\BibitemShut {NoStop}%
\bibitem [{\citenamefont {Ericson}(1960)}]{Ericson}%
  \BibitemOpen
  \bibfield  {author} {\bibinfo {author} {\bibnamefont {Ericson}, \bibfnamefont
  {T.}}} (\bibinfo {year} {1960}),\ \href@noop {} {\bibfield  {journal}
  {\bibinfo  {journal} {Advances in Physics}\ }\textbf {\bibinfo {volume}
  {9}},\ \bibinfo {pages} {425}}\BibitemShut {NoStop}%
\bibitem [{\citenamefont {{F. Williams Jr.}}(1970)}]{Williams}%
  \BibitemOpen
  \bibfield  {author} {\bibinfo {author} {\bibnamefont {{F. Williams Jr.}},}}
  (\bibinfo {year} {1970}),\ \href@noop {} {\bibfield  {journal} {\bibinfo
  {journal} {Physics Letters B}\ }\textbf {\bibinfo {volume} {31}},\ \bibinfo
  {pages} {184}}\BibitemShut {NoStop}%
\bibitem [{\citenamefont {{F. Williams Jr.}}(1971)}]{Williams2}%
  \BibitemOpen
  \bibfield  {author} {\bibinfo {author} {\bibnamefont {{F. Williams Jr.}},}}
  (\bibinfo {year} {1971}),\ \href@noop {} {\bibfield  {journal} {\bibinfo
  {journal} {Nuclear Physics A}\ }\textbf {\bibinfo {volume} {161}},\ \bibinfo
  {pages} {231}}\BibitemShut {NoStop}%
\bibitem [{\citenamefont {Fermi}(1950)}]{Fermi}%
  \BibitemOpen
  \bibfield  {author} {\bibinfo {author} {\bibnamefont {Fermi}, \bibfnamefont
  {E.}}} (\bibinfo {year} {1950}),\ \href@noop {} {\bibfield  {journal}
  {\bibinfo  {journal} {Progress of Theoretical Physics}\ }\textbf {\bibinfo
  {volume} {5}},\ \bibinfo {pages} {570}}\BibitemShut {NoStop}%
\bibitem [{\citenamefont {Fidelus}(2010)}]{Fidelus}%
  \BibitemOpen
  \bibfield  {author} {\bibinfo {author} {\bibnamefont {Fidelus}, \bibfnamefont
  {M.}}} (\bibinfo {year} {2010}),\ \href@noop {} {\enquote {\bibinfo {title}
  {Model description of proton induced fragmentation of atomic nuclei},}\
  }\bibinfo {howpublished} {Ph.D. Thesis, Cracow}\BibitemShut {NoStop}%
\bibitem [{\citenamefont {Filges}\ and\ \citenamefont
  {Goldenbaum}(2009)}]{HandbookSR}%
  \BibitemOpen
  \bibfield  {author} {\bibinfo {author} {\bibnamefont {Filges}, \bibfnamefont
  {D.}}, \ and\ \bibinfo {author} {\bibfnamefont {F.}~\bibnamefont
  {Goldenbaum}}} (\bibinfo {year} {2009}),\ \href@noop {} {\emph {\bibinfo
  {title} {Handbook of Spallation Research: Theory, Experiments and
  Applications}}}\ (\bibinfo  {publisher} {WILEY-VCH Verlag GmbH \& Co.})\
  \bibinfo {note} {{ISBN:} 978-3-527-40714-9}\BibitemShut {NoStop}%
\bibitem [{\citenamefont {Fomichev}\ \emph {et~al.}(2005)\citenamefont
  {Fomichev}, \citenamefont {Dushin}, \citenamefont {Soloviev}, \citenamefont
  {Fomichev},\ and\ \citenamefont {Mashnik}}]{n+BiFomichev}%
  \BibitemOpen
  \bibfield  {author} {\bibinfo {author} {\bibnamefont {Fomichev},
  \bibfnamefont {A.}}, \bibinfo {author} {\bibfnamefont {V.}~\bibnamefont
  {Dushin}}, \bibinfo {author} {\bibfnamefont {S.}~\bibnamefont {Soloviev}},
  \bibinfo {author} {\bibfnamefont {A.}~\bibnamefont {Fomichev}}, \ and\
  \bibinfo {author} {\bibfnamefont {S.}~\bibnamefont {Mashnik}}} (\bibinfo
  {year} {2005}),\ \href@noop {} {\enquote {\bibinfo {title} {Fission cross
  sections for {$^{240}$Pu, $^{243}$Am, $^{209}$Bi, $^{nat}$W} induced by
  neutrons up to 500 {MeV }measured relative to $^{235}${U}},}\ }\bibinfo
  {howpublished} {LANL Report LA-UR-05-1533; V. G. Khlopin Radium Institute
  Preprint RI-262, St. Petersburg, Russia}\BibitemShut {NoStop}%
\bibitem [{\citenamefont {Franz}\ \emph {et~al.}(1990)\citenamefont {Franz},
  \citenamefont {Koncz}, \citenamefont {Roessle}, \citenamefont {Sauerwein},
  \citenamefont {Schmitt}, \citenamefont {Schmoll}, \citenamefont {Eroe},
  \citenamefont {Fodor}, \citenamefont {Kecskemeti}, \citenamefont {Kovacs},\
  and\ \citenamefont {Seres}}]{Franz}%
  \BibitemOpen
  \bibfield  {author} {\bibinfo {author} {\bibnamefont {Franz}, \bibfnamefont
  {J.}}, \bibinfo {author} {\bibfnamefont {P.}~\bibnamefont {Koncz}}, \bibinfo
  {author} {\bibfnamefont {E.}~\bibnamefont {Roessle}}, \bibinfo {author}
  {\bibfnamefont {C.}~\bibnamefont {Sauerwein}}, \bibinfo {author}
  {\bibfnamefont {H.}~\bibnamefont {Schmitt}}, \bibinfo {author} {\bibfnamefont
  {K.}~\bibnamefont {Schmoll}}, \bibinfo {author} {\bibfnamefont
  {J.}~\bibnamefont {Eroe}}, \bibinfo {author} {\bibfnamefont {Z.}~\bibnamefont
  {Fodor}}, \bibinfo {author} {\bibfnamefont {J.}~\bibnamefont {Kecskemeti}},
  \bibinfo {author} {\bibfnamefont {Z.}~\bibnamefont {Kovacs}}, \ and\ \bibinfo
  {author} {\bibfnamefont {Z.}~\bibnamefont {Seres}}} (\bibinfo {year}
  {1990}),\ \href@noop {} {\bibfield  {journal} {\bibinfo  {journal} {Nuclear
  Physics A}\ }\textbf {\bibinfo {volume} {510}},\ \bibinfo {pages}
  {774}}\BibitemShut {NoStop}%
\bibitem [{\citenamefont {Furihata}(2000)}]{2000Furihata}%
  \BibitemOpen
  \bibfield  {author} {\bibinfo {author} {\bibnamefont {Furihata},
  \bibfnamefont {S.}}} (\bibinfo {year} {2000}),\ \href@noop {} {\bibfield
  {journal} {\bibinfo  {journal} {Nuclear Instruments and Methods in Physics
  Research B}\ }\textbf {\bibinfo {volume} {171}},\ \bibinfo {pages}
  {252}}\BibitemShut {NoStop}%
\bibitem [{\citenamefont {Furihata}(2003)}]{GEM2thesis}%
  \BibitemOpen
  \bibfield  {author} {\bibinfo {author} {\bibnamefont {Furihata},
  \bibfnamefont {S.}}} (\bibinfo {year} {2003}),\ \href@noop {} {\enquote
  {\bibinfo {title} {Development of a generalized evaporation model and study
  of residual nuclei prodution},}\ }\bibinfo {howpublished} {Ph.D. thesis,
  Tohoku University}\BibitemShut {NoStop}%
\bibitem [{\citenamefont {Furihata}\ \emph {et~al.}(2001)\citenamefont
  {Furihata}, \citenamefont {Nita}, \citenamefont {Meigo}, \citenamefont
  {Ikeda},\ and\ \citenamefont {Maekawa}}]{GEM2}%
  \BibitemOpen
  \bibfield  {author} {\bibinfo {author} {\bibnamefont {Furihata},
  \bibfnamefont {S.}}, \bibinfo {author} {\bibfnamefont {K.}~\bibnamefont
  {Nita}}, \bibinfo {author} {\bibfnamefont {S.}~\bibnamefont {Meigo}},
  \bibinfo {author} {\bibfnamefont {Y.}~\bibnamefont {Ikeda}}, \ and\ \bibinfo
  {author} {\bibfnamefont {F.}~\bibnamefont {Maekawa}}} (\bibinfo {year}
  {2001}),\ \href@noop {} {\bibinfo  {journal} {Japan Atomic Energy Research
  Institute}\ }\BibitemShut {NoStop}%
\bibitem [{\citenamefont {Golovchenko}\ \emph {et~al.}(2002)\citenamefont
  {Golovchenko}, \citenamefont {{Skvar\v{c}}}, \citenamefont {Yasuda},
  \citenamefont {Giacomelli}, \citenamefont {Tretyakova}, \citenamefont
  {{R.~Ili\'{c}}}, \citenamefont {Bimbot}, \citenamefont {Toulemonde},\ and\
  \citenamefont {Murakami}}]{Golovchenko}%
  \BibitemOpen
\bibfield  {journal} {  }\bibfield  {author} {\bibinfo {author} {\bibnamefont
  {Golovchenko}, \bibfnamefont {A.}}, \bibinfo {author} {\bibfnamefont
  {J.}~\bibnamefont {{Skvar\v{c}}}}, \bibinfo {author} {\bibfnamefont
  {N.}~\bibnamefont {Yasuda}}, \bibinfo {author} {\bibfnamefont
  {M.}~\bibnamefont {Giacomelli}}, \bibinfo {author} {\bibfnamefont
  {S.}~\bibnamefont {Tretyakova}}, \bibinfo {author} {\bibnamefont
  {{R.~Ili\'{c}}}}, \bibinfo {author} {\bibfnamefont {R.}~\bibnamefont
  {Bimbot}}, \bibinfo {author} {\bibfnamefont {M.}~\bibnamefont {Toulemonde}},
  \ and\ \bibinfo {author} {\bibfnamefont {T.}~\bibnamefont {Murakami}}}
  (\bibinfo {year} {2002}),\ \href@noop {} {\bibfield  {journal} {\bibinfo
  {journal} {Physical Review C}\ }\textbf {\bibinfo {volume} {66}},\ \bibinfo
  {pages} {014609}}\BibitemShut {NoStop}%
\bibitem [{\citenamefont {Goorley}\ \emph {et~al.}(2012)\citenamefont
  {Goorley}, \citenamefont {James}, \citenamefont {Booth}, \citenamefont
  {Brown}, \citenamefont {Bull}, \citenamefont {Cox}, \citenamefont {Durkee},
  \citenamefont {Elson}, \citenamefont {Fensin}, \citenamefont {Forster},
  \citenamefont {Hendricks}, \citenamefont {Hughes}, \citenamefont {Johns},
  \citenamefont {Kiedrowski}, \citenamefont {Martz}, \citenamefont {Mashnik},
  \citenamefont {Mc{K}inney}, \citenamefont {Pelowitz}, \citenamefont {Prael},
  \citenamefont {Sweezy}, \citenamefont {Waters}, \citenamefont {Wilcox},\ and\
  \citenamefont {Zukaitis}}]{MCNP6}%
  \BibitemOpen
  \bibfield  {author} {\bibinfo {author} {\bibnamefont {Goorley}, \bibfnamefont
  {T.}}, \bibinfo {author} {\bibfnamefont {M.}~\bibnamefont {James}}, \bibinfo
  {author} {\bibfnamefont {T.}~\bibnamefont {Booth}}, \bibinfo {author}
  {\bibfnamefont {F.}~\bibnamefont {Brown}}, \bibinfo {author} {\bibfnamefont
  {J.}~\bibnamefont {Bull}}, \bibinfo {author} {\bibfnamefont {L.}~\bibnamefont
  {Cox}}, \bibinfo {author} {\bibfnamefont {J.}~\bibnamefont {Durkee}},
  \bibinfo {author} {\bibfnamefont {J.}~\bibnamefont {Elson}}, \bibinfo
  {author} {\bibfnamefont {M.}~\bibnamefont {Fensin}}, \bibinfo {author}
  {\bibfnamefont {R.}~\bibnamefont {Forster}}, \bibinfo {author} {\bibfnamefont
  {J.}~\bibnamefont {Hendricks}}, \bibinfo {author} {\bibfnamefont
  {G.}~\bibnamefont {Hughes}}, \bibinfo {author} {\bibfnamefont
  {R.}~\bibnamefont {Johns}}, \bibinfo {author} {\bibfnamefont
  {B.}~\bibnamefont {Kiedrowski}}, \bibinfo {author} {\bibfnamefont
  {R.}~\bibnamefont {Martz}}, \bibinfo {author} {\bibfnamefont
  {S.}~\bibnamefont {Mashnik}}, \bibinfo {author} {\bibfnamefont
  {G.}~\bibnamefont {Mc{K}inney}}, \bibinfo {author} {\bibfnamefont
  {D.}~\bibnamefont {Pelowitz}}, \bibinfo {author} {\bibfnamefont
  {R.}~\bibnamefont {Prael}}, \bibinfo {author} {\bibfnamefont
  {J.}~\bibnamefont {Sweezy}}, \bibinfo {author} {\bibfnamefont
  {L.}~\bibnamefont {Waters}}, \bibinfo {author} {\bibfnamefont
  {T.}~\bibnamefont {Wilcox}}, \ and\ \bibinfo {author} {\bibfnamefont
  {T.}~\bibnamefont {Zukaitis}}} (\bibinfo {year} {2012}),\ \href@noop {}
  {\bibfield  {journal} {\bibinfo  {journal} {Nuclear Technology}\ }\textbf
  {\bibinfo {volume} {180}},\ \bibinfo {pages} {298}}\BibitemShut {NoStop}%
\bibitem [{\citenamefont {Green}\ \emph {et~al.}(1987)\citenamefont {Green},
  \citenamefont {Korteling}, \citenamefont {{J. D'Auria}}, \citenamefont
  {Jackson},\ and\ \citenamefont {Helmer}}]{Green}%
  \BibitemOpen
  \bibfield  {author} {\bibinfo {author} {\bibnamefont {Green}, \bibfnamefont
  {R.}}, \bibinfo {author} {\bibfnamefont {R.}~\bibnamefont {Korteling}},
  \bibinfo {author} {\bibnamefont {{J. D'Auria}}}, \bibinfo {author}
  {\bibfnamefont {K.}~\bibnamefont {Jackson}}, \ and\ \bibinfo {author}
  {\bibfnamefont {R.}~\bibnamefont {Helmer}}} (\bibinfo {year} {1987}),\
  \href@noop {} {\bibfield  {journal} {\bibinfo  {journal} {Physical Review C}\
  }\textbf {\bibinfo {volume} {35}},\ \bibinfo {pages} {1341}}\BibitemShut
  {NoStop}%
\bibitem [{\citenamefont {Green}\ \emph {et~al.}(1984)\citenamefont {Green},
  \citenamefont {Korteling},\ and\ \citenamefont {Jackson}}]{Green480}%
  \BibitemOpen
  \bibfield  {author} {\bibinfo {author} {\bibnamefont {Green}, \bibfnamefont
  {R.}}, \bibinfo {author} {\bibfnamefont {R.}~\bibnamefont {Korteling}}, \
  and\ \bibinfo {author} {\bibfnamefont {K.}~\bibnamefont {Jackson}}} (\bibinfo
  {year} {1984}),\ \href@noop {} {\bibfield  {journal} {\bibinfo  {journal}
  {Physical Review C}\ }\textbf {\bibinfo {volume} {29}},\ \bibinfo {pages}
  {1806}}\BibitemShut {NoStop}%
\bibitem [{\citenamefont {Gudima}\ \emph {et~al.}(2015)\citenamefont {Gudima},
  \citenamefont {Mashnik},\ and\ \citenamefont {Kerby}}]{NUFRA2015slides}%
  \BibitemOpen
  \bibfield  {author} {\bibinfo {author} {\bibnamefont {Gudima}, \bibfnamefont
  {K.}}, \bibinfo {author} {\bibfnamefont {S.}~\bibnamefont {Mashnik}}, \ and\
  \bibinfo {author} {\bibfnamefont {L.}~\bibnamefont {Kerby}}} (\bibinfo {year}
  {2015}),\ \href@noop {} {\enquote {\bibinfo {title} {Fragmentation of light
  nuclei at intermediate energies simulated with {MCNP6}},}\ }\bibinfo
  {howpublished} {LANL Report, LA-UR-15-27417, presented at the Fifth
  International Conference on Nuclear Fragmentation From Basic Research to
  Applications (NUFRA2015), 4 -- 11 October 2015, Kemer (Antalya),
  Turkey}\BibitemShut {NoStop}%
\bibitem [{\citenamefont {Gudima}\ \emph {et~al.}(2001)\citenamefont {Gudima},
  \citenamefont {Mashnik},\ and\ \citenamefont {Sierk}}]{LAQGSM}%
  \BibitemOpen
  \bibfield  {author} {\bibinfo {author} {\bibnamefont {Gudima}, \bibfnamefont
  {K.}}, \bibinfo {author} {\bibfnamefont {S.}~\bibnamefont {Mashnik}}, \ and\
  \bibinfo {author} {\bibfnamefont {A.}~\bibnamefont {Sierk}}} (\bibinfo {year}
  {2001}),\ \href@noop {} {\enquote {\bibinfo {title} {User manual for the code
  {LAQGSM}},}\ }\bibinfo {note} {{LANL Report LA-UR-01-6804;
  http://lib-www.lanl.gov/lapubs/00818645.pdf}}\BibitemShut {NoStop}%
\bibitem [{\citenamefont {Gudima}\ \emph {et~al.}(1983)\citenamefont {Gudima},
  \citenamefont {Mashnik},\ and\ \citenamefont {Toneev}}]{CEMModel}%
  \BibitemOpen
  \bibfield  {author} {\bibinfo {author} {\bibnamefont {Gudima}, \bibfnamefont
  {K.}}, \bibinfo {author} {\bibfnamefont {S.}~\bibnamefont {Mashnik}}, \ and\
  \bibinfo {author} {\bibfnamefont {V.}~\bibnamefont {Toneev}}} (\bibinfo
  {year} {1983}),\ \href@noop {} {\bibfield  {journal} {\bibinfo  {journal}
  {Nuclear Physics A}\ }\textbf {\bibinfo {volume} {401}},\ \bibinfo {pages}
  {329}}\BibitemShut {NoStop}%
\bibitem [{\citenamefont {Gudima}\ \emph {et~al.}(1975)\citenamefont {Gudima},
  \citenamefont {Ososkov},\ and\ \citenamefont {Toneev}}]{Gudima}%
  \BibitemOpen
  \bibfield  {author} {\bibinfo {author} {\bibnamefont {Gudima}, \bibfnamefont
  {K.}}, \bibinfo {author} {\bibfnamefont {G.}~\bibnamefont {Ososkov}}, \ and\
  \bibinfo {author} {\bibfnamefont {V.}~\bibnamefont {Toneev}}} (\bibinfo
  {year} {1975}),\ \href@noop {} {\bibfield  {journal} {\bibinfo  {journal}
  {Yadernaya Fizika}\ }\textbf {\bibinfo {volume} {21}}},\ \bibinfo {note}
  {[Soviet Journal of Nuclear Physics 21 (1975) 139-143]}\BibitemShut {NoStop}%
\bibitem [{\citenamefont {Hansen}\ \emph {et~al.}(2012)\citenamefont {Hansen},
  \citenamefont {{A. L\"uhr}}, \citenamefont {Sobolevsky},\ and\ \citenamefont
  {Bassler}}]{SHIELD2}%
  \BibitemOpen
  \bibfield  {author} {\bibinfo {author} {\bibnamefont {Hansen}, \bibfnamefont
  {D.}}, \bibinfo {author} {\bibnamefont {{A. L\"uhr}}}, \bibinfo {author}
  {\bibfnamefont {N.}~\bibnamefont {Sobolevsky}}, \ and\ \bibinfo {author}
  {\bibfnamefont {N.}~\bibnamefont {Bassler}}} (\bibinfo {year} {2012}),\
  \href@noop {} {\bibfield  {journal} {\bibinfo  {journal} {Physics in Medicine
  \& Biology}\ }\textbf {\bibinfo {volume} {57}},\ \bibinfo {pages}
  {2393}}\BibitemShut {NoStop}%
\bibitem [{\citenamefont {Heilbronn}\ \emph {et~al.}(2007)\citenamefont
  {Heilbronn}, \citenamefont {Zeitlin}, \citenamefont {Iwata}, \citenamefont
  {Murakami}, \citenamefont {Iwase}, \citenamefont {Nakamura}, \citenamefont
  {Nunomiya}, \citenamefont {Sato}, \citenamefont {Yashima}, \citenamefont
  {Ronningen},\ and\ \citenamefont {Ieki}}]{Heilbronn2007}%
  \BibitemOpen
  \bibfield  {author} {\bibinfo {author} {\bibnamefont {Heilbronn},
  \bibfnamefont {L.}}, \bibinfo {author} {\bibfnamefont {C.}~\bibnamefont
  {Zeitlin}}, \bibinfo {author} {\bibfnamefont {Y.}~\bibnamefont {Iwata}},
  \bibinfo {author} {\bibfnamefont {T.}~\bibnamefont {Murakami}}, \bibinfo
  {author} {\bibfnamefont {H.}~\bibnamefont {Iwase}}, \bibinfo {author}
  {\bibfnamefont {T.}~\bibnamefont {Nakamura}}, \bibinfo {author}
  {\bibfnamefont {T.}~\bibnamefont {Nunomiya}}, \bibinfo {author}
  {\bibfnamefont {H.}~\bibnamefont {Sato}}, \bibinfo {author} {\bibfnamefont
  {H.}~\bibnamefont {Yashima}}, \bibinfo {author} {\bibfnamefont {R.~M.}\
  \bibnamefont {Ronningen}}, \ and\ \bibinfo {author} {\bibfnamefont
  {K.}~\bibnamefont {Ieki}}} (\bibinfo {year} {2007}),\ \href@noop {}
  {\bibfield  {journal} {\bibinfo  {journal} {Nuclear Science and Engineering}\
  }\textbf {\bibinfo {volume} {157}},\ \bibinfo {pages} {142}}\BibitemShut
  {NoStop}%
\bibitem [{\citenamefont {Hultqvist}\ \emph {et~al.}(2012)\citenamefont
  {Hultqvist}, \citenamefont {Lazzeroni}, \citenamefont {Botvina},
  \citenamefont {Gudowska}, \citenamefont {Sobolevsky},\ and\ \citenamefont
  {Brahme}}]{SHIELD}%
  \BibitemOpen
  \bibfield  {author} {\bibinfo {author} {\bibnamefont {Hultqvist},
  \bibfnamefont {M.}}, \bibinfo {author} {\bibfnamefont {M.}~\bibnamefont
  {Lazzeroni}}, \bibinfo {author} {\bibfnamefont {A.}~\bibnamefont {Botvina}},
  \bibinfo {author} {\bibfnamefont {I.}~\bibnamefont {Gudowska}}, \bibinfo
  {author} {\bibfnamefont {N.}~\bibnamefont {Sobolevsky}}, \ and\ \bibinfo
  {author} {\bibfnamefont {A.}~\bibnamefont {Brahme}}} (\bibinfo {year}
  {2012}),\ \href@noop {} {\bibfield  {journal} {\bibinfo  {journal} {Physics
  in Medicine \& Biology}\ }\textbf {\bibinfo {volume} {57}},\ \bibinfo {pages}
  {4369}}\BibitemShut {NoStop}%
\bibitem [{\citenamefont {Iida}\ \emph {et~al.}(2007)\citenamefont {Iida},
  \citenamefont {Kohama},\ and\ \citenamefont {Oyamatsu}}]{Iida}%
  \BibitemOpen
  \bibfield  {author} {\bibinfo {author} {\bibnamefont {Iida}, \bibfnamefont
  {K.}}, \bibinfo {author} {\bibfnamefont {A.}~\bibnamefont {Kohama}}, \ and\
  \bibinfo {author} {\bibfnamefont {K.}~\bibnamefont {Oyamatsu}}} (\bibinfo
  {year} {2007}),\ \href@noop {} {\bibfield  {journal} {\bibinfo  {journal}
  {Journal of the Physical Society of Japan}\ }\textbf {\bibinfo {volume}
  {76}},\ \bibinfo {pages} {044201}},\ \bibinfo {note}
  {arXiv:nucl-th/0601039}\BibitemShut {NoStop}%
\bibitem [{\citenamefont {Ingemarsson}\ and\ \citenamefont
  {Lantz}(2003)}]{Ingemarsson}%
  \BibitemOpen
  \bibfield  {author} {\bibinfo {author} {\bibnamefont {Ingemarsson},
  \bibfnamefont {A.}}, \ and\ \bibinfo {author} {\bibfnamefont
  {M.}~\bibnamefont {Lantz}}} (\bibinfo {year} {2003}),\ \href@noop {}
  {\bibfield  {journal} {\bibinfo  {journal} {Physical Review C}\ }\textbf
  {\bibinfo {volume} {67}},\ \bibinfo {pages} {064605}}\BibitemShut {NoStop}%
\bibitem [{\citenamefont {Ingemarsson}\ \emph {et~al.}(2000)\citenamefont
  {Ingemarsson}, \citenamefont {Nyberg}, \citenamefont {Renberg}, \citenamefont
  {Sundberg}, \citenamefont {Carlson}, \citenamefont {Cox}, \citenamefont
  {Auce}, \citenamefont {Johansson}, \citenamefont {Tibell}, \citenamefont
  {Khoa},\ and\ \citenamefont {Warner}}]{Ingemarsson2000}%
  \BibitemOpen
  \bibfield  {author} {\bibinfo {author} {\bibnamefont {Ingemarsson},
  \bibfnamefont {A.}}, \bibinfo {author} {\bibfnamefont {J.}~\bibnamefont
  {Nyberg}}, \bibinfo {author} {\bibfnamefont {P.}~\bibnamefont {Renberg}},
  \bibinfo {author} {\bibfnamefont {O.}~\bibnamefont {Sundberg}}, \bibinfo
  {author} {\bibfnamefont {R.}~\bibnamefont {Carlson}}, \bibinfo {author}
  {\bibfnamefont {A.}~\bibnamefont {Cox}}, \bibinfo {author} {\bibfnamefont
  {A.}~\bibnamefont {Auce}}, \bibinfo {author} {\bibfnamefont {R.}~\bibnamefont
  {Johansson}}, \bibinfo {author} {\bibfnamefont {G.}~\bibnamefont {Tibell}},
  \bibinfo {author} {\bibfnamefont {D.}~\bibnamefont {Khoa}}, \ and\ \bibinfo
  {author} {\bibfnamefont {R.}~\bibnamefont {Warner}}} (\bibinfo {year}
  {2000}),\ \href@noop {} {\bibfield  {journal} {\bibinfo  {journal} {Nuclear
  Physics A}\ }\textbf {\bibinfo {volume} {676}},\ \bibinfo {pages}
  {3}}\BibitemShut {NoStop}%
\bibitem [{\citenamefont {Jacak}\ \emph {et~al.}(1987)\citenamefont {Jacak},
  \citenamefont {Westfall}, \citenamefont {Crawley}, \citenamefont {Fox},
  \citenamefont {Gelbke},\ and\ \citenamefont {Harwood}}]{Jacak1987}%
  \BibitemOpen
  \bibfield  {author} {\bibinfo {author} {\bibnamefont {Jacak}, \bibfnamefont
  {B.}}, \bibinfo {author} {\bibfnamefont {G.}~\bibnamefont {Westfall}},
  \bibinfo {author} {\bibfnamefont {G.}~\bibnamefont {Crawley}}, \bibinfo
  {author} {\bibfnamefont {D.}~\bibnamefont {Fox}}, \bibinfo {author}
  {\bibfnamefont {C.}~\bibnamefont {Gelbke}}, \ and\ \bibinfo {author}
  {\bibfnamefont {L.}~\bibnamefont {Harwood}}} (\bibinfo {year} {1987}),\
  \href@noop {} {\bibfield  {journal} {\bibinfo  {journal} {Physical Review C}\
  }\textbf {\bibinfo {volume} {35}},\ \bibinfo {pages} {1751}}\BibitemShut
  {NoStop}%
\bibitem [{\citenamefont {Junghans}\ \emph {et~al.}(1998)\citenamefont
  {Junghans}, \citenamefont {de~Jong}, \citenamefont {Clerc}, \citenamefont
  {Ignatyuk}, \citenamefont {Kudyaev},\ and\ \citenamefont {Schmidt}}]{ABLA}%
  \BibitemOpen
  \bibfield  {author} {\bibinfo {author} {\bibnamefont {Junghans},
  \bibfnamefont {A.}}, \bibinfo {author} {\bibfnamefont {M.}~\bibnamefont
  {de~Jong}}, \bibinfo {author} {\bibfnamefont {H.-G.}\ \bibnamefont {Clerc}},
  \bibinfo {author} {\bibfnamefont {A.}~\bibnamefont {Ignatyuk}}, \bibinfo
  {author} {\bibfnamefont {G.}~\bibnamefont {Kudyaev}}, \ and\ \bibinfo
  {author} {\bibfnamefont {K.-H.}\ \bibnamefont {Schmidt}}} (\bibinfo {year}
  {1998}),\ \href@noop {} {\bibfield  {journal} {\bibinfo  {journal} {Nuclear
  Physics A}\ }\textbf {\bibinfo {volume} {629}},\ \bibinfo {pages}
  {635}}\BibitemShut {NoStop}%
\bibitem [{\citenamefont {Kalbach}(1988)}]{Kalbach1988}%
  \BibitemOpen
  \bibfield  {author} {\bibinfo {author} {\bibnamefont {Kalbach}, \bibfnamefont
  {C.}}} (\bibinfo {year} {1988}),\ \href@noop {} {\bibfield  {journal}
  {\bibinfo  {journal} {Physical Review C}\ }\textbf {\bibinfo {volume} {37}},\
  \bibinfo {pages} {2350}}\BibitemShut {NoStop}%
\bibitem [{\citenamefont {Kalbach}(1998)}]{Kalbach}%
  \BibitemOpen
  \bibfield  {author} {\bibinfo {author} {\bibnamefont {Kalbach}, \bibfnamefont
  {C.}}} (\bibinfo {year} {1998}),\ \href@noop {} {\bibfield  {journal}
  {\bibinfo  {journal} {Journal of Physics G: Nuclear and Particle Physics}\
  }\textbf {\bibinfo {volume} {24}},\ \bibinfo {pages} {847}}\BibitemShut
  {NoStop}%
\bibitem [{\citenamefont {Kerby}(2015{\natexlab{a}})}]{CPC2015}%
  \BibitemOpen
  \bibfield  {author} {\bibinfo {author} {\bibnamefont {Kerby}, \bibfnamefont
  {L.}}} (\bibinfo {year} {2015}{\natexlab{a}}),\ \href@noop {} {\enquote
  {\bibinfo {title} {An energy-dependent numerical model for the condensation
  probability, $\gamma_j$},}\ }\bibinfo {howpublished} {LANL Report,
  LA-UR-15-26648},\ \bibinfo {note} {submitted to Computer Physics
  Communications}\BibitemShut {NoStop}%
\bibitem [{\citenamefont {Kerby}(2015{\natexlab{b}})}]{KerbyThesis}%
  \BibitemOpen
  \bibfield  {author} {\bibinfo {author} {\bibnamefont {Kerby}, \bibfnamefont
  {L.}}} (\bibinfo {year} {2015}{\natexlab{b}}),\ \href@noop {} {\enquote
  {\bibinfo {title} {Precompound emission of energetic light fragments in
  spallation reactions},}\ }\bibinfo {howpublished} {Ph. D. Thesis, University
  of Idaho}\BibitemShut {NoStop}%
\bibitem [{\citenamefont {Kerby}\ and\ \citenamefont {Mashnik}(2014)}]{FY2014}%
  \BibitemOpen
  \bibfield  {author} {\bibinfo {author} {\bibnamefont {Kerby}, \bibfnamefont
  {L.}}, \ and\ \bibinfo {author} {\bibfnamefont {S.}~\bibnamefont {Mashnik}}}
  (\bibinfo {year} {2014}),\ \href@noop {} {\enquote {\bibinfo {title} {{LANL}
  fiscal year 2014 report},}\ }\bibinfo {howpublished} {LANL Report
  LA-UR-14-27533},\ \bibinfo {note}
  {www.osti.gov/scitech/biblio/1162152}\BibitemShut {NoStop}%
\bibitem [{\citenamefont {Kerby}\ and\ \citenamefont
  {Mashnik}(2015{\natexlab{a}})}]{Coal2015}%
  \BibitemOpen
  \bibfield  {author} {\bibinfo {author} {\bibnamefont {Kerby}, \bibfnamefont
  {L.}}, \ and\ \bibinfo {author} {\bibfnamefont {S.}~\bibnamefont {Mashnik}}}
  (\bibinfo {year} {2015}{\natexlab{a}}),\ \href@noop {} {\enquote {\bibinfo
  {title} {An expanded coalescence model within the intranuclear cascade of
  {CEM}},}\ }\bibinfo {howpublished} {LANL Report, LA-UR-15-20322}\BibitemShut
  {NoStop}%
\bibitem [{\citenamefont {Kerby}\ and\ \citenamefont
  {Mashnik}(2015{\natexlab{b}})}]{gamma2015}%
  \BibitemOpen
  \bibfield  {author} {\bibinfo {author} {\bibnamefont {Kerby}, \bibfnamefont
  {L.}}, \ and\ \bibinfo {author} {\bibfnamefont {S.}~\bibnamefont {Mashnik}}}
  (\bibinfo {year} {2015}{\natexlab{b}}),\ \href@noop {} {\enquote {\bibinfo
  {title} {A new model for the condensation probability, $\gamma_j$, in
  {CEM}},}\ }\bibinfo {howpublished} {LANL Report, LA-UR-15-22370}\BibitemShut
  {NoStop}%
\bibitem [{\citenamefont {Kerby}\ and\ \citenamefont
  {Mashnik}(2015{\natexlab{c}})}]{ANS2015}%
  \BibitemOpen
  \bibfield  {author} {\bibinfo {author} {\bibnamefont {Kerby}, \bibfnamefont
  {L.}}, \ and\ \bibinfo {author} {\bibfnamefont {S.}~\bibnamefont {Mashnik}}}
  (\bibinfo {year} {2015}{\natexlab{c}}),\ \href@noop {} {\bibfield  {journal}
  {\bibinfo  {journal} {Transactions of the American Nuclear Society}\ }\textbf
  {\bibinfo {volume} {112}},\ \bibinfo {pages} {577}}\BibitemShut {NoStop}%
\bibitem [{\citenamefont {Kerby}\ and\ \citenamefont
  {Mashnik}(2015{\natexlab{d}})}]{NIMB2015}%
  \BibitemOpen
  \bibfield  {author} {\bibinfo {author} {\bibnamefont {Kerby}, \bibfnamefont
  {L.}}, \ and\ \bibinfo {author} {\bibfnamefont {S.}~\bibnamefont {Mashnik}}}
  (\bibinfo {year} {2015}{\natexlab{d}}),\ \href@noop {} {\bibfield  {journal}
  {\bibinfo  {journal} {Nuclear Instruments and Methods B}\ }\textbf {\bibinfo
  {volume} {356-357}},\ \bibinfo {pages} {135}}\BibitemShut {NoStop}%
\bibitem [{\citenamefont {Kerby}\ \emph {et~al.}(2015)\citenamefont {Kerby},
  \citenamefont {Mashnik},\ and\ \citenamefont {Bull}}]{PHYSOR2016}%
  \BibitemOpen
  \bibfield  {author} {\bibinfo {author} {\bibnamefont {Kerby}, \bibfnamefont
  {L.}}, \bibinfo {author} {\bibfnamefont {S.}~\bibnamefont {Mashnik}}, \ and\
  \bibinfo {author} {\bibfnamefont {J.}~\bibnamefont {Bull}}} (\bibinfo {year}
  {2015}),\ \href@noop {} {\enquote {\bibinfo {title} {{MCNP6 GENXS} option
  expansion to include fragment spectra of heavy ions},}\ }\bibinfo
  {howpublished} {LANL Report, LA-UR-15-27858},\ \bibinfo {note} {summary
  accepted for presentation at PHYSOR 2016, Sun Valley, Idaho, USA, May 1-5,
  2016}\BibitemShut {NoStop}%
\bibitem [{\citenamefont {Kerby}\ \emph {et~al.}(2012)\citenamefont {Kerby},
  \citenamefont {Mashnik},\ and\ \citenamefont {Sierk}}]{Summer2012}%
  \BibitemOpen
  \bibfield  {author} {\bibinfo {author} {\bibnamefont {Kerby}, \bibfnamefont
  {L.}}, \bibinfo {author} {\bibfnamefont {S.}~\bibnamefont {Mashnik}}, \ and\
  \bibinfo {author} {\bibfnamefont {A.}~\bibnamefont {Sierk}}} (\bibinfo {year}
  {2012}),\ \href@noop {} {\enquote {\bibinfo {title} {Preliminary results of
  investigating precompound emission of light fragments in spallation
  reactions, summer 2012},}\ }\bibinfo {howpublished} {LANL Report,
  LA-UR-12-24190}\BibitemShut {NoStop}%
\bibitem [{\citenamefont {Kerby}\ \emph
  {et~al.}(2014{\natexlab{a}})\citenamefont {Kerby}, \citenamefont {Mashnik},\
  and\ \citenamefont {Sierk}}]{ND2013}%
  \BibitemOpen
  \bibfield  {author} {\bibinfo {author} {\bibnamefont {Kerby}, \bibfnamefont
  {L.}}, \bibinfo {author} {\bibfnamefont {S.}~\bibnamefont {Mashnik}}, \ and\
  \bibinfo {author} {\bibfnamefont {A.}~\bibnamefont {Sierk}}} (\bibinfo {year}
  {2014}{\natexlab{a}}),\ \href@noop {} {\bibfield  {journal} {\bibinfo
  {journal} {Nuclear Data Sheets}\ }\textbf {\bibinfo {volume} {118}},\
  \bibinfo {pages} {316}},\ \bibinfo {note} {arXiv:1303.4311}\BibitemShut
  {NoStop}%
\bibitem [{\citenamefont {Kerby}\ \emph
  {et~al.}(2014{\natexlab{b}})\citenamefont {Kerby}, \citenamefont {Mashnik},\
  and\ \citenamefont {Tokuhiro}}]{ANS2014}%
  \BibitemOpen
  \bibfield  {author} {\bibinfo {author} {\bibnamefont {Kerby}, \bibfnamefont
  {L.}}, \bibinfo {author} {\bibfnamefont {S.}~\bibnamefont {Mashnik}}, \ and\
  \bibinfo {author} {\bibfnamefont {A.}~\bibnamefont {Tokuhiro}}} (\bibinfo
  {year} {2014}{\natexlab{b}}),\ \href@noop {} {\bibfield  {journal} {\bibinfo
  {journal} {Transactions of the American Nuclear Society}\ }\textbf {\bibinfo
  {volume} {110}},\ \bibinfo {pages} {465}}\BibitemShut {NoStop}%
\bibitem [{\citenamefont {Koi}\ and\ \citenamefont
  {Wright}(2013)}]{GEANTManual}%
  \BibitemOpen
  \bibfield  {author} {\bibinfo {author} {\bibnamefont {Koi}, \bibfnamefont
  {T.}}, \ and\ \bibinfo {author} {\bibfnamefont {D.}~\bibnamefont {Wright}}}
  (\bibinfo {year} {2013}),\ \href@noop {} {\enquote {\bibinfo {title}
  {{GEANT4} physics reference manual chapter 23 total reaction cross section in
  nucleus-nucleus reactions},}\ }\bibinfo {howpublished}
  {gentoo.osuosl.org/distfiles/PhysicsReferenceManual-4.10.0.pdf}\BibitemShut
  {NoStop}%
\bibitem [{\citenamefont {Koning}\ \emph {et~al.}(2004)\citenamefont {Koning},
  \citenamefont {Hilaire},\ and\ \citenamefont {Duijvestijn}}]{TALYS}%
  \BibitemOpen
  \bibfield  {author} {\bibinfo {author} {\bibnamefont {Koning}, \bibfnamefont
  {A.}}, \bibinfo {author} {\bibfnamefont {S.}~\bibnamefont {Hilaire}}, \ and\
  \bibinfo {author} {\bibfnamefont {M.}~\bibnamefont {Duijvestijn}}} (\bibinfo
  {year} {2004}),\ \href@noop {} {\enquote {\bibinfo {title} {{TALYS:
  C}omprehensive nuclear reaction modeling},}\ }\bibinfo {howpublished} {Proc.
  Int. Conf. on Nuclear Data for Sci. \& Techn. (ND2004), September 26 -
  October 1, 2004, Santa Fe, NM, USA},\ \bibinfo {note} {edited by R. Haight,
  M. Chadwick, T. Kawano, and P. Talou, (AIP Conference Proceedings, Volume
  769, Melville, New York, 2005), pp. 1154--1159}\BibitemShut {NoStop}%
\bibitem [{\citenamefont {Konobeyev}\ and\ \citenamefont
  {Fischer}(2014)}]{KonFisch2014}%
  \BibitemOpen
  \bibfield  {author} {\bibinfo {author} {\bibnamefont {Konobeyev},
  \bibfnamefont {A.~Y.}}, \ and\ \bibinfo {author} {\bibfnamefont
  {U.}~\bibnamefont {Fischer}}} (\bibinfo {year} {2014}),\ \href@noop {}
  {\enquote {\bibinfo {title} {Status of evaluation of {$^9$Be DPA and Gas}
  production cross-sections at neutron incident energies up to 200 {MeV}},}\
  }\bibinfo {howpublished} {presentation at the Fall 2014 Nuclear Data Week,
  24-28 November 2014, NEA, Issy-le-Moulineaux, France;
  www.oecd-nea.org/dbdata/meetings/nov2014/}\BibitemShut {NoStop}%
\bibitem [{\citenamefont {Konobeyev}\ and\ \citenamefont
  {Korovin}(1995)}]{Konobeyev}%
  \BibitemOpen
  \bibfield  {author} {\bibinfo {author} {\bibnamefont {Konobeyev},
  \bibfnamefont {A.~Y.}}, \ and\ \bibinfo {author} {\bibfnamefont {Y.~A.}\
  \bibnamefont {Korovin}}} (\bibinfo {year} {1995}),\ \href@noop {} {\bibfield
  {journal} {\bibinfo  {journal} {Kerntechnik}\ }\textbf {\bibinfo {volume}
  {60}},\ \bibinfo {pages} {147}}\BibitemShut {NoStop}%
\bibitem [{\citenamefont {Kox}\ \emph {et~al.}(1987)\citenamefont {Kox},
  \citenamefont {Gamp}, \citenamefont {Perrin}, \citenamefont {Arvieux},
  \citenamefont {Bertholet}, \citenamefont {Bruandet}, \citenamefont {Buenerd},
  \citenamefont {Cherkaoui}, \citenamefont {Cole}, \citenamefont {El-Masri},
  \citenamefont {Longequeue}, \citenamefont {Menet}, \citenamefont {Merchez},\
  and\ \citenamefont {Viano}}]{Kox}%
  \BibitemOpen
  \bibfield  {author} {\bibinfo {author} {\bibnamefont {Kox}, \bibfnamefont
  {S.}}, \bibinfo {author} {\bibfnamefont {A.}~\bibnamefont {Gamp}}, \bibinfo
  {author} {\bibfnamefont {C.}~\bibnamefont {Perrin}}, \bibinfo {author}
  {\bibfnamefont {J.}~\bibnamefont {Arvieux}}, \bibinfo {author} {\bibfnamefont
  {R.}~\bibnamefont {Bertholet}}, \bibinfo {author} {\bibfnamefont
  {J.}~\bibnamefont {Bruandet}}, \bibinfo {author} {\bibfnamefont
  {M.}~\bibnamefont {Buenerd}}, \bibinfo {author} {\bibfnamefont
  {R.}~\bibnamefont {Cherkaoui}}, \bibinfo {author} {\bibfnamefont
  {A.}~\bibnamefont {Cole}}, \bibinfo {author} {\bibfnamefont {Y.}~\bibnamefont
  {El-Masri}}, \bibinfo {author} {\bibfnamefont {N.}~\bibnamefont
  {Longequeue}}, \bibinfo {author} {\bibfnamefont {J.}~\bibnamefont {Menet}},
  \bibinfo {author} {\bibfnamefont {F.}~\bibnamefont {Merchez}}, \ and\
  \bibinfo {author} {\bibfnamefont {J.}~\bibnamefont {Viano}}} (\bibinfo {year}
  {1987}),\ \href@noop {} {\bibfield  {journal} {\bibinfo  {journal} {Physical
  Review C}\ }\textbf {\bibinfo {volume} {35}},\ \bibinfo {pages}
  {1678}}\BibitemShut {NoStop}%
\bibitem [{\citenamefont {Krylov}\ \emph {et~al.}(2014)\citenamefont {Krylov},
  \citenamefont {Paraipan}, \citenamefont {Sobolevsky}, \citenamefont
  {Timoshenko},\ and\ \citenamefont {Tret'yakov}}]{Krylov}%
  \BibitemOpen
  \bibfield  {author} {\bibinfo {author} {\bibnamefont {Krylov}, \bibfnamefont
  {A.}}, \bibinfo {author} {\bibfnamefont {M.}~\bibnamefont {Paraipan}},
  \bibinfo {author} {\bibfnamefont {N.}~\bibnamefont {Sobolevsky}}, \bibinfo
  {author} {\bibfnamefont {G.}~\bibnamefont {Timoshenko}}, \ and\ \bibinfo
  {author} {\bibfnamefont {V.}~\bibnamefont {Tret'yakov}}} (\bibinfo {year}
  {2014}),\ \href@noop {} {\bibfield  {journal} {\bibinfo  {journal} {Physics
  of Particles and Nuclei Letters}\ }\textbf {\bibinfo {volume} {11}},\
  \bibinfo {pages} {549}}\BibitemShut {NoStop}%
\bibitem [{\citenamefont {Leray}\ \emph {et~al.}(2014)\citenamefont {Leray},
  \citenamefont {Boudard}, \citenamefont {Braunn}, \citenamefont {Cugnon},
  \citenamefont {David}, \citenamefont {Leprince},\ and\ \citenamefont
  {Mancusi}}]{Leray2014}%
  \BibitemOpen
  \bibfield  {author} {\bibinfo {author} {\bibnamefont {Leray}, \bibfnamefont
  {S.}}, \bibinfo {author} {\bibfnamefont {A.}~\bibnamefont {Boudard}},
  \bibinfo {author} {\bibfnamefont {B.}~\bibnamefont {Braunn}}, \bibinfo
  {author} {\bibfnamefont {J.}~\bibnamefont {Cugnon}}, \bibinfo {author}
  {\bibfnamefont {J.-C.}\ \bibnamefont {David}}, \bibinfo {author}
  {\bibfnamefont {A.}~\bibnamefont {Leprince}}, \ and\ \bibinfo {author}
  {\bibfnamefont {D.}~\bibnamefont {Mancusi}}} (\bibinfo {year} {2014}),\
  \href@noop {} {\bibfield  {journal} {\bibinfo  {journal} {Nuclear Data
  Sheets}\ }\textbf {\bibinfo {volume} {118}},\ \bibinfo {pages}
  {312}}\BibitemShut {NoStop}%
\bibitem [{\citenamefont {Leray}\ \emph {et~al.}(2011)\citenamefont {Leray},
  \citenamefont {David}, \citenamefont {Khandaker}, \citenamefont {Mank},
  \citenamefont {Mengoni}, \citenamefont {Otsuka}, \citenamefont {Filges},
  \citenamefont {Gallmeier}, \citenamefont {Konobeyev},\ and\ \citenamefont
  {Michel}}]{IAEABenchmark}%
  \BibitemOpen
  \bibfield  {author} {\bibinfo {author} {\bibnamefont {Leray}, \bibfnamefont
  {S.}}, \bibinfo {author} {\bibfnamefont {J.-C.}\ \bibnamefont {David}},
  \bibinfo {author} {\bibfnamefont {M.}~\bibnamefont {Khandaker}}, \bibinfo
  {author} {\bibfnamefont {G.}~\bibnamefont {Mank}}, \bibinfo {author}
  {\bibfnamefont {A.}~\bibnamefont {Mengoni}}, \bibinfo {author} {\bibfnamefont
  {N.}~\bibnamefont {Otsuka}}, \bibinfo {author} {\bibfnamefont
  {D.}~\bibnamefont {Filges}}, \bibinfo {author} {\bibfnamefont
  {F.}~\bibnamefont {Gallmeier}}, \bibinfo {author} {\bibfnamefont
  {A.}~\bibnamefont {Konobeyev}}, \ and\ \bibinfo {author} {\bibfnamefont
  {R.}~\bibnamefont {Michel}}} (\bibinfo {year} {2011}),\ \href@noop {}
  {\bibfield  {journal} {\bibinfo  {journal} {Journal of the Korean Physical
  Society}\ }\textbf {\bibinfo {volume} {59}}~(\bibinfo {number} {2}),\
  \bibinfo {pages} {791}}\BibitemShut {NoStop}%
\bibitem [{\citenamefont {{L\"uhr}}\ \emph {et~al.}(2012)\citenamefont
  {{L\"uhr}}, \citenamefont {Hansen}, \citenamefont {Teiwes}, \citenamefont
  {Sobolevsky}, \citenamefont {{O. J\"akel}},\ and\ \citenamefont
  {Bassler}}]{SHIELD3}%
  \BibitemOpen
  \bibfield  {author} {\bibinfo {author} {\bibnamefont {{L\"uhr}},
  \bibfnamefont {A.}}, \bibinfo {author} {\bibfnamefont {D.}~\bibnamefont
  {Hansen}}, \bibinfo {author} {\bibfnamefont {R.}~\bibnamefont {Teiwes}},
  \bibinfo {author} {\bibfnamefont {N.}~\bibnamefont {Sobolevsky}}, \bibinfo
  {author} {\bibnamefont {{O. J\"akel}}}, \ and\ \bibinfo {author}
  {\bibfnamefont {N.}~\bibnamefont {Bassler}}} (\bibinfo {year} {2012}),\
  \href@noop {} {\bibfield  {journal} {\bibinfo  {journal} {Physics in Medicine
  \& Biology}\ }\textbf {\bibinfo {volume} {57}},\ \bibinfo {pages}
  {5169}}\BibitemShut {NoStop}%
\bibitem [{\citenamefont {Machner}\ \emph {et~al.}(2006)\citenamefont
  {Machner}, \citenamefont {Aschman}, \citenamefont {Baruth-Ram}, \citenamefont
  {Carter}, \citenamefont {Cowley}, \citenamefont {Goldenbaum}, \citenamefont
  {Nangu}, \citenamefont {Pilcher}, \citenamefont {Sideras-Haddad},
  \citenamefont {Sellschop}, \citenamefont {Smit}, \citenamefont {Spoelstra},\
  and\ \citenamefont {Steyn}}]{Machner}%
  \BibitemOpen
  \bibfield  {author} {\bibinfo {author} {\bibnamefont {Machner}, \bibfnamefont
  {H.}}, \bibinfo {author} {\bibfnamefont {D.}~\bibnamefont {Aschman}},
  \bibinfo {author} {\bibfnamefont {K.}~\bibnamefont {Baruth-Ram}}, \bibinfo
  {author} {\bibfnamefont {J.}~\bibnamefont {Carter}}, \bibinfo {author}
  {\bibfnamefont {A.}~\bibnamefont {Cowley}}, \bibinfo {author} {\bibfnamefont
  {F.}~\bibnamefont {Goldenbaum}}, \bibinfo {author} {\bibfnamefont
  {B.}~\bibnamefont {Nangu}}, \bibinfo {author} {\bibfnamefont
  {J.}~\bibnamefont {Pilcher}}, \bibinfo {author} {\bibfnamefont
  {E.}~\bibnamefont {Sideras-Haddad}}, \bibinfo {author} {\bibfnamefont
  {J.}~\bibnamefont {Sellschop}}, \bibinfo {author} {\bibfnamefont
  {F.}~\bibnamefont {Smit}}, \bibinfo {author} {\bibfnamefont {B.}~\bibnamefont
  {Spoelstra}}, \ and\ \bibinfo {author} {\bibfnamefont {D.}~\bibnamefont
  {Steyn}}} (\bibinfo {year} {2006}),\ \href@noop {} {\bibfield  {journal}
  {\bibinfo  {journal} {Physical Review C}\ }\textbf {\bibinfo {volume} {73}},\
  \bibinfo {pages} {044606}}\BibitemShut {NoStop}%
\bibitem [{\citenamefont {Mac{R}eady}(2012)}]{Protons}%
  \BibitemOpen
  \bibfield  {author} {\bibinfo {author} {\bibnamefont {Mac{R}eady},
  \bibfnamefont {N.}}} (\bibinfo {year} {2012}),\ \href@noop {} {\bibfield
  {journal} {\bibinfo  {journal} {Journal of the National Cancer Institute}\
  }\textbf {\bibinfo {volume} {104}}~(\bibinfo {number} {9})}\BibitemShut
  {NoStop}%
\bibitem [{\citenamefont {Mancusi}\ \emph {et~al.}(2014)\citenamefont
  {Mancusi}, \citenamefont {Boudard}, \citenamefont {Cugnon}, \citenamefont
  {David}, \citenamefont {Kaitaniemi},\ and\ \citenamefont {Leray}}]{Mancusi}%
  \BibitemOpen
  \bibfield  {author} {\bibinfo {author} {\bibnamefont {Mancusi}, \bibfnamefont
  {D.}}, \bibinfo {author} {\bibfnamefont {A.}~\bibnamefont {Boudard}},
  \bibinfo {author} {\bibfnamefont {J.}~\bibnamefont {Cugnon}}, \bibinfo
  {author} {\bibfnamefont {J.-C.}\ \bibnamefont {David}}, \bibinfo {author}
  {\bibfnamefont {P.}~\bibnamefont {Kaitaniemi}}, \ and\ \bibinfo {author}
  {\bibfnamefont {S.}~\bibnamefont {Leray}}} (\bibinfo {year} {2014}),\
  \href@noop {} {\bibfield  {journal} {\bibinfo  {journal} {Physical Review C}\
  }\textbf {\bibinfo {volume} {90}},\ \bibinfo {pages} {054602}},\ \bibinfo
  {note} {arXiv:1407.7755v2}\BibitemShut {NoStop}%
\bibitem [{\citenamefont {Mantzouranis}\ \emph {et~al.}(1976)\citenamefont
  {Mantzouranis}, \citenamefont {Weidenm{\"u}ller},\ and\ \citenamefont
  {Agassi}}]{Mantzouranis}%
  \BibitemOpen
  \bibfield  {author} {\bibinfo {author} {\bibnamefont {Mantzouranis},
  \bibfnamefont {G.}}, \bibinfo {author} {\bibfnamefont {H.}~\bibnamefont
  {Weidenm{\"u}ller}}, \ and\ \bibinfo {author} {\bibfnamefont
  {D.}~\bibnamefont {Agassi}}} (\bibinfo {year} {1976}),\ \href@noop {}
  {\bibfield  {journal} {\bibinfo  {journal} {Z. Phys. A}\ }\textbf {\bibinfo
  {volume} {276}},\ \bibinfo {pages} {145}}\BibitemShut {NoStop}%
\bibitem [{\citenamefont {Mashnik}(2011{\natexlab{a}})}]{VVCEM}%
  \BibitemOpen
  \bibfield  {author} {\bibinfo {author} {\bibnamefont {Mashnik}, \bibfnamefont
  {S.}}} (\bibinfo {year} {2011}{\natexlab{a}}),\ \href@noop {} {\enquote
  {\bibinfo {title} {Validation and verification of {MCNP6} against high-energy
  experimental data and calculations by other codes. {I. The CEM} testing
  primer},}\ }\bibinfo {howpublished} {LANL Report LA-UR-11-05129},\ \bibinfo
  {note} {mcnp.lanl.gov}\BibitemShut {NoStop}%
\bibitem [{\citenamefont {Mashnik}(2011{\natexlab{b}})}]{VVLAQGSM}%
  \BibitemOpen
  \bibfield  {author} {\bibinfo {author} {\bibnamefont {Mashnik}, \bibfnamefont
  {S.}}} (\bibinfo {year} {2011}{\natexlab{b}}),\ \href@noop {} {\enquote
  {\bibinfo {title} {Validation and verification of {MCNP6} against high-energy
  experimental data and calculations by other codes. {II. The LAQGSM} testing
  primer},}\ }\bibinfo {howpublished} {LANL Report LA-UR-11-05627},\ \bibinfo
  {note} {mcnp.lanl.gov}\BibitemShut {NoStop}%
\bibitem [{\citenamefont {Mashnik}(2011{\natexlab{c}})}]{VV}%
  \BibitemOpen
  \bibfield  {author} {\bibinfo {author} {\bibnamefont {Mashnik}, \bibfnamefont
  {S.}}} (\bibinfo {year} {2011}{\natexlab{c}}),\ \href@noop {} {\bibfield
  {journal} {\bibinfo  {journal} {European Physical Journal - Plus}\ }\textbf
  {\bibinfo {volume} {126}},\ \bibinfo {pages} {49}},\ \bibinfo {note}
  {arXiv:1011.4978}\BibitemShut {NoStop}%
\bibitem [{\citenamefont {Mashnik}(2013)}]{VVMPI}%
  \BibitemOpen
  \bibfield  {author} {\bibinfo {author} {\bibnamefont {Mashnik}, \bibfnamefont
  {S.}}} (\bibinfo {year} {2013}),\ \href@noop {} {\enquote {\bibinfo {title}
  {Validation and verification of {MCNP6} against high-energy experimental data
  and calculations by other codes. {III. The MPI} testing primer},}\ }\bibinfo
  {howpublished} {LANL Report LA-UR-13-26944},\ \bibinfo {note}
  {mcnp.lanl.gov}\BibitemShut {NoStop}%
\bibitem [{\citenamefont {Mashnik}(2014)}]{Mashnik2014FRIB}%
  \BibitemOpen
  \bibfield  {author} {\bibinfo {author} {\bibnamefont {Mashnik}, \bibfnamefont
  {S.}}} (\bibinfo {year} {2014}),\ \href@noop {} {\enquote {\bibinfo {title}
  {{MCNP6} simulation of reactions of interest to {FRIB}, medical, and space
  applications},}\ }\bibinfo {howpublished} {LANL Report LA-UR-14-25095,
  arXiv:1407.2832}\BibitemShut {NoStop}%
\bibitem [{\citenamefont {Mashnik}\ \emph
  {et~al.}(2005{\natexlab{a}})\citenamefont {Mashnik}, \citenamefont {Gudima},
  \citenamefont {Baznat}, \citenamefont {Sierk}, \citenamefont {Prael},\ and\
  \citenamefont {Mokhov}}]{CEMLAQ}%
  \BibitemOpen
  \bibfield  {author} {\bibinfo {author} {\bibnamefont {Mashnik}, \bibfnamefont
  {S.}}, \bibinfo {author} {\bibfnamefont {K.}~\bibnamefont {Gudima}}, \bibinfo
  {author} {\bibfnamefont {M.}~\bibnamefont {Baznat}}, \bibinfo {author}
  {\bibfnamefont {A.}~\bibnamefont {Sierk}}, \bibinfo {author} {\bibfnamefont
  {R.}~\bibnamefont {Prael}}, \ and\ \bibinfo {author} {\bibfnamefont
  {N.}~\bibnamefont {Mokhov}}} (\bibinfo {year} {2005}{\natexlab{a}}),\
  \href@noop {} {\enquote {\bibinfo {title} {{CEM03.01 and LAQGSM03.01}
  versions of the improved {Cascade-Exciton Model (CEM) and Los Alamos
  Quark-Gluon String Model (LAQGSM)} codes},}\ }\bibinfo {howpublished} {LANL
  Report LA-UR-05-2686},\ \bibinfo {note} {mcnp.lanl.gov}\BibitemShut {NoStop}%
\bibitem [{\citenamefont {Mashnik}\ \emph
  {et~al.}(2007{\natexlab{a}})\citenamefont {Mashnik}, \citenamefont {Gudima},
  \citenamefont {Mokhov},\ and\ \citenamefont {Prael}}]{LAQGSM03}%
  \BibitemOpen
  \bibfield  {author} {\bibinfo {author} {\bibnamefont {Mashnik}, \bibfnamefont
  {S.}}, \bibinfo {author} {\bibfnamefont {K.}~\bibnamefont {Gudima}}, \bibinfo
  {author} {\bibfnamefont {N.}~\bibnamefont {Mokhov}}, \ and\ \bibinfo {author}
  {\bibfnamefont {R.}~\bibnamefont {Prael}}} (\bibinfo {year}
  {2007}{\natexlab{a}}),\ \href@noop {} {\enquote {\bibinfo {title}
  {{LAQGSM03.03} upgrade and its validation},}\ }\bibinfo {howpublished} {LANL
  Report LA-UR-07-6198},\ \bibinfo {note} {arXiv:0709.1736}\BibitemShut
  {NoStop}%
\bibitem [{\citenamefont {Mashnik}\ \emph {et~al.}(2008)\citenamefont
  {Mashnik}, \citenamefont {Gudima}, \citenamefont {Prael}, \citenamefont
  {Sierk}, \citenamefont {Baznat},\ and\ \citenamefont
  {Mokhov}}]{ICTP-IAEAWorkshop}%
  \BibitemOpen
  \bibfield  {author} {\bibinfo {author} {\bibnamefont {Mashnik}, \bibfnamefont
  {S.}}, \bibinfo {author} {\bibfnamefont {K.}~\bibnamefont {Gudima}}, \bibinfo
  {author} {\bibfnamefont {R.}~\bibnamefont {Prael}}, \bibinfo {author}
  {\bibfnamefont {A.}~\bibnamefont {Sierk}}, \bibinfo {author} {\bibfnamefont
  {M.}~\bibnamefont {Baznat}}, \ and\ \bibinfo {author} {\bibfnamefont
  {N.}~\bibnamefont {Mokhov}}} (\bibinfo {year} {2008}),\ \href@noop {}
  {\enquote {\bibinfo {title} {{CEM03.03 and LAQGSM03.03} event generators for
  the {MCNP6, MCNPX, and MARS15} transport codes},}\ }\bibinfo {howpublished}
  {Joint {ICTP-IAEA} Advanced Workshop on Model Codes for Spallation Reactions.
  Trieste, Italy},\ \bibinfo {note} {{LANL} Report LA-UR-08-2931,
  arXiv:0812.1820}\BibitemShut {NoStop}%
\bibitem [{\citenamefont {Mashnik}\ \emph
  {et~al.}(2005{\natexlab{b}})\citenamefont {Mashnik}, \citenamefont {Gudima},
  \citenamefont {Sierk}, \citenamefont {Baznat},\ and\ \citenamefont
  {Mokhov}}]{CEM03.01}%
  \BibitemOpen
  \bibfield  {author} {\bibinfo {author} {\bibnamefont {Mashnik}, \bibfnamefont
  {S.}}, \bibinfo {author} {\bibfnamefont {K.}~\bibnamefont {Gudima}}, \bibinfo
  {author} {\bibfnamefont {A.}~\bibnamefont {Sierk}}, \bibinfo {author}
  {\bibfnamefont {M.}~\bibnamefont {Baznat}}, \ and\ \bibinfo {author}
  {\bibfnamefont {N.}~\bibnamefont {Mokhov}}} (\bibinfo {year}
  {2005}{\natexlab{b}}),\ \href@noop {} {\enquote {\bibinfo {title} {{CEM03.01}
  user manual},}\ }\bibinfo {howpublished} {LANL Report LA-UR-05-7321},\
  \bibinfo {note} {mcnp.lanl.gov}\BibitemShut {NoStop}%
\bibitem [{\citenamefont {Mashnik}\ and\ \citenamefont
  {Kerby}(2014)}]{NIMA2014}%
  \BibitemOpen
  \bibfield  {author} {\bibinfo {author} {\bibnamefont {Mashnik}, \bibfnamefont
  {S.}}, \ and\ \bibinfo {author} {\bibfnamefont {L.}~\bibnamefont {Kerby}}}
  (\bibinfo {year} {2014}),\ \href@noop {} {\bibfield  {journal} {\bibinfo
  {journal} {Nuclear Instruments and Methods in Physics Research A}\ }\textbf
  {\bibinfo {volume} {764}},\ \bibinfo {pages} {59}},\ \bibinfo {note}
  {arXiv:1404.7820}\BibitemShut {NoStop}%
\bibitem [{\citenamefont {Mashnik}\ and\ \citenamefont
  {Kerby}(2015)}]{NN2015slides}%
  \BibitemOpen
  \bibfield  {author} {\bibinfo {author} {\bibnamefont {Mashnik}, \bibfnamefont
  {S.}}, \ and\ \bibinfo {author} {\bibfnamefont {L.}~\bibnamefont {Kerby}}}
  (\bibinfo {year} {2015}),\ \href@noop {} {\enquote {\bibinfo {title} {{MCNP6}
  simulation of light and medium nuclei fragmentation at intermediate
  energies},}\ }\bibinfo {howpublished} {LANL Report, LA-UR-15-22811, presented
  at the 12th International Conference on Nucleus-Nucleus Collisions (NN2015),
  June 21-26, 2015, Catania, Italy}\BibitemShut {NoStop}%
\bibitem [{\citenamefont {Mashnik}\ \emph
  {et~al.}(2007{\natexlab{b}})\citenamefont {Mashnik}, \citenamefont {Prael},\
  and\ \citenamefont {Gudima}}]{ResNote2006}%
  \BibitemOpen
  \bibfield  {author} {\bibinfo {author} {\bibnamefont {Mashnik}, \bibfnamefont
  {S.}}, \bibinfo {author} {\bibfnamefont {R.}~\bibnamefont {Prael}}, \ and\
  \bibinfo {author} {\bibfnamefont {K.}~\bibnamefont {Gudima}}} (\bibinfo
  {year} {2007}{\natexlab{b}}),\ \href@noop {} {\enquote {\bibinfo {title}
  {Implementation of {CEM03.01 into MCNP6} and its verification and validation
  running through {MCNP6. CEM03.02} upgrade},}\ }\bibinfo {howpublished} {LANL
  Research Note X-3-RN(U)-07-03, LANL Report LA-UR-06-8652},\ \bibinfo {note}
  {mcnp.lanl.gov}\BibitemShut {NoStop}%
\bibitem [{\citenamefont {Mashnik}\ and\ \citenamefont {Sierk}(2012)}]{CEM03}%
  \BibitemOpen
  \bibfield  {author} {\bibinfo {author} {\bibnamefont {Mashnik}, \bibfnamefont
  {S.}}, \ and\ \bibinfo {author} {\bibfnamefont {A.}~\bibnamefont {Sierk}}}
  (\bibinfo {year} {2012}),\ \href@noop {} {\enquote {\bibinfo {title}
  {{CEM03.03} user manual},}\ }\bibinfo {howpublished} {LANL Report
  LA-UR-12-01364},\ \bibinfo {note} {mcnp.lanl.gov}\BibitemShut {NoStop}%
\bibitem [{\citenamefont {Mashnik}\ \emph
  {et~al.}(1998{\natexlab{a}})\citenamefont {Mashnik}, \citenamefont {Sierk},
  \citenamefont {Bersillon},\ and\ \citenamefont {Gabriel}}]{Mashnik1998}%
  \BibitemOpen
  \bibfield  {author} {\bibinfo {author} {\bibnamefont {Mashnik}, \bibfnamefont
  {S.}}, \bibinfo {author} {\bibfnamefont {A.}~\bibnamefont {Sierk}}, \bibinfo
  {author} {\bibfnamefont {O.}~\bibnamefont {Bersillon}}, \ and\ \bibinfo
  {author} {\bibfnamefont {T.}~\bibnamefont {Gabriel}}} (\bibinfo {year}
  {1998}{\natexlab{a}}),\ \href@noop {} {\bibfield  {journal} {\bibinfo
  {journal} {Nuclear Instruments and Methods in Physics Research A}\ }\textbf
  {\bibinfo {volume} {414}},\ \bibinfo {pages} {68}},\ \bibinfo {note} {{LANL
  Report LA-UR-97-97-2905, mcnp.lanl.gov}}\BibitemShut {NoStop}%
\bibitem [{\citenamefont {Mashnik}\ \emph {et~al.}(2002)\citenamefont
  {Mashnik}, \citenamefont {Sierk},\ and\ \citenamefont
  {Gudima}}]{SantaFe2002}%
  \BibitemOpen
  \bibfield  {author} {\bibinfo {author} {\bibnamefont {Mashnik}, \bibfnamefont
  {S.}}, \bibinfo {author} {\bibfnamefont {A.}~\bibnamefont {Sierk}}, \ and\
  \bibinfo {author} {\bibfnamefont {K.}~\bibnamefont {Gudima}}} (\bibinfo
  {year} {2002}),\ \href@noop {} {\enquote {\bibinfo {title} {Complex particle
  and light fragment emission in the {Cascade-Exciton Model} of nuclear
  reactions},}\ }\bibinfo {howpublished} {LANL Report LA-UR-02-5185},\ \bibinfo
  {note} {invited talk presented at the 12th Biennial Topical Meeting of the
  Radiation Protection and Shielding Division (RPSD) of the American Nuclear
  Society, April 14-17, 2002, Santa Fe, NM; E-print:
  nucl-th/0208048}\BibitemShut {NoStop}%
\bibitem [{\citenamefont {Mashnik}\ \emph {et~al.}(2010)\citenamefont
  {Mashnik}, \citenamefont {Sierk}, \citenamefont {Gudima},\ and\ \citenamefont
  {Baznat}}]{SecondAdvancedWorkshop}%
  \BibitemOpen
  \bibfield  {author} {\bibinfo {author} {\bibnamefont {Mashnik}, \bibfnamefont
  {S.}}, \bibinfo {author} {\bibfnamefont {A.}~\bibnamefont {Sierk}}, \bibinfo
  {author} {\bibfnamefont {K.}~\bibnamefont {Gudima}}, \ and\ \bibinfo {author}
  {\bibfnamefont {M.}~\bibnamefont {Baznat}}} (\bibinfo {year} {2010}),\
  \href@noop {} {\bibinfo  {journal} {{LANL Report LA-UR-10-00510, viewgraphs
  of the invited talk presented at the Second Advanced Workshop on Model Codes
  for Spallation Reactions, 8-11 February 2010, CEA-Saclay, France,
  http://nds121.iaea.org/alberto/mediawiki-1.6.10/index.php/Benchmark:2ndWorkP%
rog}}\ }\BibitemShut {NoStop}%
\bibitem [{\citenamefont {Mashnik}\ \emph
  {et~al.}(1998{\natexlab{b}})\citenamefont {Mashnik}, \citenamefont {Sierk},
  \citenamefont {Riper},\ and\ \citenamefont {Wilson}}]{T16}%
  \BibitemOpen
\bibfield  {journal} {  }\bibfield  {author} {\bibinfo {author} {\bibnamefont
  {Mashnik}, \bibfnamefont {S.}}, \bibinfo {author} {\bibfnamefont
  {A.}~\bibnamefont {Sierk}}, \bibinfo {author} {\bibfnamefont
  {K.}~\bibnamefont {Riper}}, \ and\ \bibinfo {author} {\bibfnamefont
  {W.}~\bibnamefont {Wilson}}} (\bibinfo {year} {1998}{\natexlab{b}}),\
  \href@noop {} {\enquote {\bibinfo {title} {Production and validation of
  isotope production cross section libraries for neutrons and protons to 1.7
  {GeV}},}\ }\bibinfo {howpublished} {LANL Report LA-UR-98-6000},\ \bibinfo
  {note} {in: Proc. SARE-4, Knoxville, TN, September 13-16, 1998 (ORNL, 1999,
  pp. 151-162); arXiv:nucl-th/9812071; our T-16 Library ``T-16 Lib'' is updated
  permanently when new experimental data became available to us}\BibitemShut
  {NoStop}%
\bibitem [{\citenamefont {Mashnik}\ and\ \citenamefont
  {Smolyansky}(1996)}]{Mashnik1994}%
  \BibitemOpen
  \bibfield  {author} {\bibinfo {author} {\bibnamefont {Mashnik}, \bibfnamefont
  {S.}}, \ and\ \bibinfo {author} {\bibfnamefont {S.}~\bibnamefont
  {Smolyansky}}} (\bibinfo {year} {1996}),\ \href@noop {} {\bibinfo  {journal}
  {{JINR Preprint E2-94-353, Dubna, 1994, 24 pp.; Proc. Int. Study Center in
  Nonlinear Science} {\em Dynamics of Transport in Fluids, Plasmas and Charged
  Beams}, {Villa Gualino, Torino, Italy, June-September, 1994, Eds. G. Maino
  and M. Ottaviani, Singapore: World Scientific, 1996, pp. 137--159}}\
  }\BibitemShut {NoStop}%
\bibitem [{\citenamefont {Mashnik}\ and\ \citenamefont {Toneev}(1974)}]{MODEX}%
  \BibitemOpen
\bibfield  {journal} {  }\bibfield  {author} {\bibinfo {author} {\bibnamefont
  {Mashnik}, \bibfnamefont {S.}}, \ and\ \bibinfo {author} {\bibfnamefont
  {V.}~\bibnamefont {Toneev}}} (\bibinfo {year} {1974}),\ \href@noop {}
  {\bibfield  {journal} {\bibinfo  {journal} {JINR Communication}\ }\textbf
  {\bibinfo {volume} {P4-8417}}}\BibitemShut {NoStop}%
\bibitem [{\citenamefont {Mazarakis}\ and\ \citenamefont
  {Stephens}(1973)}]{Mazarakis}%
  \BibitemOpen
  \bibfield  {author} {\bibinfo {author} {\bibnamefont {Mazarakis},
  \bibfnamefont {M.}}, \ and\ \bibinfo {author} {\bibfnamefont
  {W.}~\bibnamefont {Stephens}}} (\bibinfo {year} {1973}),\ \href@noop {}
  {\bibfield  {journal} {\bibinfo  {journal} {Physical Review C}\ }\textbf
  {\bibinfo {volume} {7}},\ \bibinfo {pages} {1280}}\BibitemShut {NoStop}%
\bibitem [{\citenamefont {Meier}\ \emph {et~al.}(1992)\citenamefont {Meier},
  \citenamefont {Amian}, \citenamefont {Goulding}, \citenamefont {Morgan},\
  and\ \citenamefont {Moss}}]{Meier1992}%
  \BibitemOpen
  \bibfield  {author} {\bibinfo {author} {\bibnamefont {Meier}, \bibfnamefont
  {M.}}, \bibinfo {author} {\bibfnamefont {W.}~\bibnamefont {Amian}}, \bibinfo
  {author} {\bibfnamefont {C.}~\bibnamefont {Goulding}}, \bibinfo {author}
  {\bibfnamefont {G.}~\bibnamefont {Morgan}}, \ and\ \bibinfo {author}
  {\bibfnamefont {C.}~\bibnamefont {Moss}}} (\bibinfo {year} {1992}),\
  \href@noop {} {\bibfield  {journal} {\bibinfo  {journal} {Nuclear Science and
  Engineering}\ }\textbf {\bibinfo {volume} {110}},\ \bibinfo {pages}
  {289}}\BibitemShut {NoStop}%
\bibitem [{\citenamefont {Meier}\ \emph {et~al.}(1989)\citenamefont {Meier},
  \citenamefont {Clark}, \citenamefont {Goulding}, \citenamefont {McClelland},
  \citenamefont {Morgan}, \citenamefont {Moss},\ and\ \citenamefont
  {Mian}}]{Meier1989}%
  \BibitemOpen
  \bibfield  {author} {\bibinfo {author} {\bibnamefont {Meier}, \bibfnamefont
  {M.}}, \bibinfo {author} {\bibfnamefont {D.}~\bibnamefont {Clark}}, \bibinfo
  {author} {\bibfnamefont {C.}~\bibnamefont {Goulding}}, \bibinfo {author}
  {\bibfnamefont {J.}~\bibnamefont {McClelland}}, \bibinfo {author}
  {\bibfnamefont {G.}~\bibnamefont {Morgan}}, \bibinfo {author} {\bibfnamefont
  {C.}~\bibnamefont {Moss}}, \ and\ \bibinfo {author} {\bibfnamefont
  {W.}~\bibnamefont {Mian}}} (\bibinfo {year} {1989}),\ \href@noop {}
  {\bibfield  {journal} {\bibinfo  {journal} {Nuclear Science and Engineering}\
  }\textbf {\bibinfo {volume} {102}},\ \bibinfo {pages} {310}}\BibitemShut
  {NoStop}%
\bibitem [{\citenamefont {Mocko}(2006)}]{MockoThesis}%
  \BibitemOpen
  \bibfield  {author} {\bibinfo {author} {\bibnamefont {Mocko}, \bibfnamefont
  {M.}}} (\bibinfo {year} {2006}),\ \href@noop {} {\enquote {\bibinfo {title}
  {Rare isotope production},}\ }\bibinfo {howpublished} {Ph. D. thesis,
  Michigan State University}\BibitemShut {NoStop}%
\bibitem [{\citenamefont {Mocko}\ \emph {et~al.}(2006)\citenamefont {Mocko},
  \citenamefont {Tsang}, \citenamefont {Andronenko}, \citenamefont
  {Andronenko}, \citenamefont {Delaunay}, \citenamefont {Famiano},
  \citenamefont {Ginter}, \citenamefont {Henzl}, \citenamefont {Henzlov\'a},
  \citenamefont {Hua}, \citenamefont {Lukyanov}, \citenamefont {Lynch},
  \citenamefont {Rogers}, \citenamefont {Steiner}, \citenamefont {Stolz},
  \citenamefont {Tarasov}, \citenamefont {van Goethem}, \citenamefont {Verde},
  \citenamefont {Wallace},\ and\ \citenamefont {Zalessov}}]{Mocko2006}%
  \BibitemOpen
  \bibfield  {author} {\bibinfo {author} {\bibnamefont {Mocko}, \bibfnamefont
  {M.}}, \bibinfo {author} {\bibfnamefont {B.}~\bibnamefont {Tsang}}, \bibinfo
  {author} {\bibfnamefont {L.}~\bibnamefont {Andronenko}}, \bibinfo {author}
  {\bibfnamefont {M.}~\bibnamefont {Andronenko}}, \bibinfo {author}
  {\bibfnamefont {F.}~\bibnamefont {Delaunay}}, \bibinfo {author}
  {\bibfnamefont {M.}~\bibnamefont {Famiano}}, \bibinfo {author} {\bibfnamefont
  {T.}~\bibnamefont {Ginter}}, \bibinfo {author} {\bibfnamefont
  {V.}~\bibnamefont {Henzl}}, \bibinfo {author} {\bibfnamefont
  {D.}~\bibnamefont {Henzlov\'a}}, \bibinfo {author} {\bibfnamefont
  {H.}~\bibnamefont {Hua}}, \bibinfo {author} {\bibfnamefont {S.}~\bibnamefont
  {Lukyanov}}, \bibinfo {author} {\bibfnamefont {W.}~\bibnamefont {Lynch}},
  \bibinfo {author} {\bibfnamefont {A.}~\bibnamefont {Rogers}}, \bibinfo
  {author} {\bibfnamefont {M.}~\bibnamefont {Steiner}}, \bibinfo {author}
  {\bibfnamefont {A.}~\bibnamefont {Stolz}}, \bibinfo {author} {\bibfnamefont
  {O.}~\bibnamefont {Tarasov}}, \bibinfo {author} {\bibfnamefont {M.-J.}\
  \bibnamefont {van Goethem}}, \bibinfo {author} {\bibfnamefont
  {G.}~\bibnamefont {Verde}}, \bibinfo {author} {\bibfnamefont
  {W.}~\bibnamefont {Wallace}}, \ and\ \bibinfo {author} {\bibfnamefont
  {A.}~\bibnamefont {Zalessov}}} (\bibinfo {year} {2006}),\ \href@noop {}
  {\bibfield  {journal} {\bibinfo  {journal} {Physical Review C}\ }\textbf
  {\bibinfo {volume} {74}},\ \bibinfo {pages} {054612}}\BibitemShut {NoStop}%
\bibitem [{\citenamefont {Morrison}(1956)}]{Morrison}%
  \BibitemOpen
  \bibfield  {author} {\bibinfo {author} {\bibnamefont {Morrison},
  \bibfnamefont {G.}}} (\bibinfo {year} {1956}),\ \href@noop {} {\bibfield
  {journal} {\bibinfo  {journal} {Physica (Utrecht)}\ }\textbf {\bibinfo
  {volume} {22}},\ \bibinfo {pages} {1135}}\BibitemShut {NoStop}%
\bibitem [{\citenamefont {Nakamoto}\ \emph {et~al.}(1997)\citenamefont
  {Nakamoto}, \citenamefont {Ishibashi}, \citenamefont {Matsufuji},
  \citenamefont {Shigyo}, \citenamefont {Maehata}, \citenamefont {Arima},
  \citenamefont {Meigo}, \citenamefont {Takada}, \citenamefont {Chiba},\ and\
  \citenamefont {Numajiri}}]{Nakamoto}%
  \BibitemOpen
  \bibfield  {author} {\bibinfo {author} {\bibnamefont {Nakamoto},
  \bibfnamefont {T.}}, \bibinfo {author} {\bibfnamefont {K.}~\bibnamefont
  {Ishibashi}}, \bibinfo {author} {\bibfnamefont {N.}~\bibnamefont
  {Matsufuji}}, \bibinfo {author} {\bibfnamefont {N.}~\bibnamefont {Shigyo}},
  \bibinfo {author} {\bibfnamefont {K.}~\bibnamefont {Maehata}}, \bibinfo
  {author} {\bibfnamefont {H.}~\bibnamefont {Arima}}, \bibinfo {author}
  {\bibfnamefont {S.}~\bibnamefont {Meigo}}, \bibinfo {author} {\bibfnamefont
  {H.}~\bibnamefont {Takada}}, \bibinfo {author} {\bibfnamefont
  {S.}~\bibnamefont {Chiba}}, \ and\ \bibinfo {author} {\bibfnamefont
  {M.}~\bibnamefont {Numajiri}}} (\bibinfo {year} {1997}),\ \href@noop {}
  {\bibfield  {journal} {\bibinfo  {journal} {Journal of Nuclear Science and
  Technology}\ }\textbf {\bibinfo {volume} {34}},\ \bibinfo {pages}
  {860}}\BibitemShut {NoStop}%
\bibitem [{\citenamefont {Nakamura}\ and\ \citenamefont
  {Heilbronn}(2006)}]{Nakamura}%
  \BibitemOpen
  \bibfield  {author} {\bibinfo {author} {\bibnamefont {Nakamura},
  \bibfnamefont {T.}}, \ and\ \bibinfo {author} {\bibfnamefont
  {L.}~\bibnamefont {Heilbronn}}} (\bibinfo {year} {2006}),\ \href@noop {}
  {\emph {\bibinfo {title} {Handbook on Secondary Particle Production and
  Transport by High-Energy Heavy Ions}}}\ (\bibinfo  {publisher} {World
  Scientific, Singapore})\BibitemShut {NoStop}%
\bibitem [{\citenamefont {Ogawa}\ \emph {et~al.}(2013)\citenamefont {Ogawa},
  \citenamefont {Sato}, \citenamefont {Hashimoto},\ and\ \citenamefont
  {Niita}}]{PHITS}%
  \BibitemOpen
  \bibfield  {author} {\bibinfo {author} {\bibnamefont {Ogawa}, \bibfnamefont
  {T.}}, \bibinfo {author} {\bibfnamefont {T.}~\bibnamefont {Sato}}, \bibinfo
  {author} {\bibfnamefont {S.}~\bibnamefont {Hashimoto}}, \ and\ \bibinfo
  {author} {\bibfnamefont {K.}~\bibnamefont {Niita}}} (\bibinfo {year}
  {2013}),\ \href@noop {} {\bibfield  {journal} {\bibinfo  {journal} {Nuclear
  Instruments and Methods in Physics Research A}\ }\textbf {\bibinfo {volume}
  {723}},\ \bibinfo {pages} {36}}\BibitemShut {NoStop}%
\bibitem [{\citenamefont {Pasechnik}(1955)}]{Pasechnik}%
  \BibitemOpen
  \bibfield  {author} {\bibinfo {author} {\bibnamefont {Pasechnik},
  \bibfnamefont {M.}}} (\bibinfo {year} {1955}),\ \href@noop {} {\bibfield
  {journal} {\bibinfo  {journal} {1st UN Conf. Peaceful Uses Atomic Energy{,}
  Geneva}\ }\textbf {\bibinfo {volume} {2}},\ \bibinfo {pages} {3}}\BibitemShut
  {NoStop}%
\bibitem [{\citenamefont {Polf}\ and\ \citenamefont {Parodi}(2015)}]{Polf}%
  \BibitemOpen
  \bibfield  {author} {\bibinfo {author} {\bibnamefont {Polf}, \bibfnamefont
  {J.}}, \ and\ \bibinfo {author} {\bibfnamefont {K.}~\bibnamefont {Parodi}}}
  (\bibinfo {year} {2015}),\ \href@noop {} {\bibfield  {journal} {\bibinfo
  {journal} {Physics Today}\ }\textbf {\bibinfo {volume} {68}},\ \bibinfo
  {pages} {28}}\BibitemShut {NoStop}%
\bibitem [{\citenamefont {Poze}\ and\ \citenamefont {Glazkov}(1956)}]{Poze}%
  \BibitemOpen
  \bibfield  {author} {\bibinfo {author} {\bibnamefont {Poze}, \bibfnamefont
  {K.}}, \ and\ \bibinfo {author} {\bibfnamefont {N.}~\bibnamefont {Glazkov}}}
  (\bibinfo {year} {1956}),\ \href@noop {} {\bibfield  {journal} {\bibinfo
  {journal} {Soviet Physics - JETP}\ }\textbf {\bibinfo {volume} {3}},\
  \bibinfo {pages} {745}}\BibitemShut {NoStop}%
\bibitem [{\citenamefont {Prael}(2011)}]{GENXS_Prael}%
  \BibitemOpen
  \bibfield  {author} {\bibinfo {author} {\bibnamefont {Prael}, \bibfnamefont
  {R.}}} (\bibinfo {year} {2011}),\ \href@noop {} {\enquote {\bibinfo {title}
  {Tally edits for the {MCNP6 GENXS} option},}\ }\bibinfo {howpublished} {LANL
  Report, LA-UR-11-02146},\ \bibinfo {note} {mcnp.lanl.gov}\BibitemShut
  {NoStop}%
\bibitem [{\citenamefont {Prael}\ \emph
  {et~al.}(1998{\natexlab{a}})\citenamefont {Prael}, \citenamefont {Ferrari},
  \citenamefont {Tripathi},\ and\ \citenamefont {Polanski}}]{SARE4REP}%
  \BibitemOpen
  \bibfield  {author} {\bibinfo {author} {\bibnamefont {Prael}, \bibfnamefont
  {R.}}, \bibinfo {author} {\bibfnamefont {A.}~\bibnamefont {Ferrari}},
  \bibinfo {author} {\bibfnamefont {R.}~\bibnamefont {Tripathi}}, \ and\
  \bibinfo {author} {\bibfnamefont {A.}~\bibnamefont {Polanski}}} (\bibinfo
  {year} {1998}{\natexlab{a}}),\ \href@noop {} {\enquote {\bibinfo {title}
  {Comparison of nucleon cross section parametrization methods for medium and
  high energies},}\ }\bibinfo {howpublished} {LANL Report LA-UR-98-5813},\
  \bibinfo {note} {proc. Forth Int. Workshop on Simulating Accelerator
  Radiation Environments (SARE-4), Hyatt Regency, Knoxville, TN, September
  13-16, 1998, edited by Tony A. Gabriel, Oak Ridge National Laboratory (1999)
  p. 171}\BibitemShut {NoStop}%
\bibitem [{\citenamefont {Prael}\ \emph
  {et~al.}(1998{\natexlab{b}})\citenamefont {Prael}, \citenamefont {Ferrari},
  \citenamefont {Tripathi},\ and\ \citenamefont {Polanski}}]{SARE4REP-suppl}%
  \BibitemOpen
  \bibfield  {author} {\bibinfo {author} {\bibnamefont {Prael}, \bibfnamefont
  {R.}}, \bibinfo {author} {\bibfnamefont {A.}~\bibnamefont {Ferrari}},
  \bibinfo {author} {\bibfnamefont {R.}~\bibnamefont {Tripathi}}, \ and\
  \bibinfo {author} {\bibfnamefont {A.}~\bibnamefont {Polanski}}} (\bibinfo
  {year} {1998}{\natexlab{b}}),\ \href@noop {} {\enquote {\bibinfo {title}
  {Plots supplemental to: Comparison of nucleon cross section parametrization
  methods for medium and high energies},}\ }\bibinfo {howpublished} {LANL
  Report LA-UR-98-5843}\BibitemShut {NoStop}%
\bibitem [{\citenamefont {Press}\ \emph {et~al.}(1992)\citenamefont {Press},
  \citenamefont {Vetterling}, \citenamefont {Teukolsky},\ and\ \citenamefont
  {Flannery}}]{NumericalRecipes}%
  \BibitemOpen
  \bibfield  {author} {\bibinfo {author} {\bibnamefont {Press}, \bibfnamefont
  {W.}}, \bibinfo {author} {\bibfnamefont {W.}~\bibnamefont {Vetterling}},
  \bibinfo {author} {\bibfnamefont {S.}~\bibnamefont {Teukolsky}}, \ and\
  \bibinfo {author} {\bibfnamefont {B.}~\bibnamefont {Flannery}}} (\bibinfo
  {year} {1992}),\ \href@noop {} {\emph {\bibinfo {title} {Numerical Recipes in
  C: The Art of Scientific Computing}}}\ (\bibinfo  {publisher} {Cambridge
  University Press})\BibitemShut {NoStop}%
\bibitem [{\citenamefont {Pshenichnov}\ \emph {et~al.}(2010)\citenamefont
  {Pshenichnov}, \citenamefont {Botvina}, \citenamefont {Mishustin},\ and\
  \citenamefont {Grainer}}]{GEANT4}%
  \BibitemOpen
  \bibfield  {author} {\bibinfo {author} {\bibnamefont {Pshenichnov},
  \bibfnamefont {I.}}, \bibinfo {author} {\bibfnamefont {A.}~\bibnamefont
  {Botvina}}, \bibinfo {author} {\bibfnamefont {I.}~\bibnamefont {Mishustin}},
  \ and\ \bibinfo {author} {\bibfnamefont {W.}~\bibnamefont {Grainer}}}
  (\bibinfo {year} {2010}),\ \href@noop {} {\bibfield  {journal} {\bibinfo
  {journal} {Nuclear Instruments and Methods in Physics Research B}\ }\textbf
  {\bibinfo {volume} {268}},\ \bibinfo {pages} {604}}\BibitemShut {NoStop}%
\bibitem [{\citenamefont {Rejmund}\ \emph {et~al.}(2001)\citenamefont
  {Rejmund}, \citenamefont {Mustapha}, \citenamefont {Armbruster},
  \citenamefont {Benlliure}, \citenamefont {Bernas}, \citenamefont {Boudard},
  \citenamefont {Dufour}, \citenamefont {Enqvist}, \citenamefont {Legrain},
  \citenamefont {Leray}, \citenamefont {Schmidt}, \citenamefont {St\'ephan},
  \citenamefont {Taieb}, \citenamefont {Tassan-Got},\ and\ \citenamefont
  {Volant}}]{Rejmund}%
  \BibitemOpen
  \bibfield  {author} {\bibinfo {author} {\bibnamefont {Rejmund}, \bibfnamefont
  {F.}}, \bibinfo {author} {\bibfnamefont {B.}~\bibnamefont {Mustapha}},
  \bibinfo {author} {\bibfnamefont {P.}~\bibnamefont {Armbruster}}, \bibinfo
  {author} {\bibfnamefont {J.}~\bibnamefont {Benlliure}}, \bibinfo {author}
  {\bibfnamefont {M.}~\bibnamefont {Bernas}}, \bibinfo {author} {\bibfnamefont
  {A.}~\bibnamefont {Boudard}}, \bibinfo {author} {\bibfnamefont
  {J.}~\bibnamefont {Dufour}}, \bibinfo {author} {\bibfnamefont
  {T.}~\bibnamefont {Enqvist}}, \bibinfo {author} {\bibfnamefont
  {R.}~\bibnamefont {Legrain}}, \bibinfo {author} {\bibfnamefont
  {S.}~\bibnamefont {Leray}}, \bibinfo {author} {\bibfnamefont {K.-H.}\
  \bibnamefont {Schmidt}}, \bibinfo {author} {\bibfnamefont {C.}~\bibnamefont
  {St\'ephan}}, \bibinfo {author} {\bibfnamefont {J.}~\bibnamefont {Taieb}},
  \bibinfo {author} {\bibfnamefont {L.}~\bibnamefont {Tassan-Got}}, \ and\
  \bibinfo {author} {\bibfnamefont {C.}~\bibnamefont {Volant}}} (\bibinfo
  {year} {2001}),\ \href@noop {} {\bibfield  {journal} {\bibinfo  {journal}
  {Nuclear Physics A}\ }\textbf {\bibinfo {volume} {683}},\ \bibinfo {pages}
  {540}}\BibitemShut {NoStop}%
\bibitem [{\citenamefont {Ribansky}\ and\ \citenamefont
  {Oblozinsky}(1973)}]{RibanskyGamma}%
  \BibitemOpen
  \bibfield  {author} {\bibinfo {author} {\bibnamefont {Ribansky},
  \bibfnamefont {I.}}, \ and\ \bibinfo {author} {\bibfnamefont
  {P.}~\bibnamefont {Oblozinsky}}} (\bibinfo {year} {1973}),\ \href@noop {}
  {\bibfield  {journal} {\bibinfo  {journal} {Physics Letters}\ }\textbf
  {\bibinfo {volume} {45B}}}\BibitemShut {NoStop}%
\bibitem [{\citenamefont {Ribansky}\ \emph {et~al.}(1973)\citenamefont
  {Ribansky}, \citenamefont {Oblozinsky},\ and\ \citenamefont
  {Betak}}]{Ribansky}%
  \BibitemOpen
  \bibfield  {author} {\bibinfo {author} {\bibnamefont {Ribansky},
  \bibfnamefont {I.}}, \bibinfo {author} {\bibfnamefont {P.}~\bibnamefont
  {Oblozinsky}}, \ and\ \bibinfo {author} {\bibfnamefont {E.}~\bibnamefont
  {Betak}}} (\bibinfo {year} {1973}),\ \href@noop {} {\bibfield  {journal}
  {\bibinfo  {journal} {Nuclear Physics A}\ }\textbf {\bibinfo {volume}
  {205}},\ \bibinfo {pages} {545}}\BibitemShut {NoStop}%
\bibitem [{\citenamefont {Ronen}(2012)}]{Ronen}%
  \BibitemOpen
  \bibfield  {author} {\bibinfo {author} {\bibnamefont {Ronen}, \bibfnamefont
  {Y.}}} (\bibinfo {year} {2012}),\ \href@noop {} {\bibfield  {journal}
  {\bibinfo  {journal} {Physica Scripta}\ }\textbf {\bibinfo {volume} {86}},\
  \bibinfo {pages} {065203}}\BibitemShut {NoStop}%
\bibitem [{\citenamefont {Sato}\ \emph {et~al.}(2013)\citenamefont {Sato},
  \citenamefont {Niita}, \citenamefont {Matsuda}, \citenamefont {Hashimoto},
  \citenamefont {Iwamoto}, \citenamefont {Noda}, \citenamefont {Ogawa},
  \citenamefont {Iwase}, \citenamefont {Nakashima}, \citenamefont {Fukahori},
  \citenamefont {Okumura}, \citenamefont {Kai}, \citenamefont {Chiba},
  \citenamefont {Furuta},\ and\ \citenamefont {Sihver}}]{PHITS2}%
  \BibitemOpen
  \bibfield  {author} {\bibinfo {author} {\bibnamefont {Sato}, \bibfnamefont
  {T.}}, \bibinfo {author} {\bibfnamefont {K.}~\bibnamefont {Niita}}, \bibinfo
  {author} {\bibfnamefont {N.}~\bibnamefont {Matsuda}}, \bibinfo {author}
  {\bibfnamefont {S.}~\bibnamefont {Hashimoto}}, \bibinfo {author}
  {\bibfnamefont {Y.}~\bibnamefont {Iwamoto}}, \bibinfo {author} {\bibfnamefont
  {S.}~\bibnamefont {Noda}}, \bibinfo {author} {\bibfnamefont {T.}~\bibnamefont
  {Ogawa}}, \bibinfo {author} {\bibfnamefont {H.}~\bibnamefont {Iwase}},
  \bibinfo {author} {\bibfnamefont {H.}~\bibnamefont {Nakashima}}, \bibinfo
  {author} {\bibfnamefont {T.}~\bibnamefont {Fukahori}}, \bibinfo {author}
  {\bibfnamefont {K.}~\bibnamefont {Okumura}}, \bibinfo {author} {\bibfnamefont
  {T.}~\bibnamefont {Kai}}, \bibinfo {author} {\bibfnamefont {S.}~\bibnamefont
  {Chiba}}, \bibinfo {author} {\bibfnamefont {T.}~\bibnamefont {Furuta}}, \
  and\ \bibinfo {author} {\bibfnamefont {L.}~\bibnamefont {Sihver}}} (\bibinfo
  {year} {2013}),\ \href@noop {} {\bibfield  {journal} {\bibinfo  {journal}
  {Journal of Nuclear Science and Technology}\ }\textbf {\bibinfo {volume}
  {50}},\ \bibinfo {pages} {913}}\BibitemShut {NoStop}%
\bibitem [{\citenamefont {Schulz}\ \emph {et~al.}(1983)\citenamefont {Schulz},
  \citenamefont {{R\"{o}pke }and K.~Gudima},\ and\ \citenamefont
  {Toneev}}]{KKG+Schulz-coal1983}%
  \BibitemOpen
  \bibfield  {author} {\bibinfo {author} {\bibnamefont {Schulz}, \bibfnamefont
  {H.}}, \bibinfo {author} {\bibfnamefont {G.}~\bibnamefont {{R\"{o}pke }and
  K.~Gudima}}, \ and\ \bibinfo {author} {\bibfnamefont {V.}~\bibnamefont
  {Toneev}}} (\bibinfo {year} {1983}),\ \href@noop {} {\bibfield  {journal}
  {\bibinfo  {journal} {Physics Letters}\ }\textbf {\bibinfo {volume} {134B}},\
  \bibinfo {pages} {458}}\BibitemShut {NoStop}%
\bibitem [{\citenamefont {Schumacher}\ \emph {et~al.}(1982)\citenamefont
  {Schumacher}, \citenamefont {Adams}, \citenamefont {Ingham}, \citenamefont
  {Matthews}, \citenamefont {Sapp}, \citenamefont {Turley}, \citenamefont
  {Owens},\ and\ \citenamefont {Roberts}}]{Schumacher}%
  \BibitemOpen
  \bibfield  {author} {\bibinfo {author} {\bibnamefont {Schumacher},
  \bibfnamefont {R.}}, \bibinfo {author} {\bibfnamefont {G.}~\bibnamefont
  {Adams}}, \bibinfo {author} {\bibfnamefont {D.}~\bibnamefont {Ingham}},
  \bibinfo {author} {\bibfnamefont {J.}~\bibnamefont {Matthews}}, \bibinfo
  {author} {\bibfnamefont {W.}~\bibnamefont {Sapp}}, \bibinfo {author}
  {\bibfnamefont {R.}~\bibnamefont {Turley}}, \bibinfo {author} {\bibfnamefont
  {R.}~\bibnamefont {Owens}}, \ and\ \bibinfo {author} {\bibfnamefont
  {B.}~\bibnamefont {Roberts}}} (\bibinfo {year} {1982}),\ \href@noop {}
  {\bibfield  {journal} {\bibinfo  {journal} {Physical Review C}\ }\textbf
  {\bibinfo {volume} {25}},\ \bibinfo {pages} {2269}}\BibitemShut {NoStop}%
\bibitem [{\citenamefont {Shen}\ \emph {et~al.}(1989)\citenamefont {Shen},
  \citenamefont {Wang}, \citenamefont {Feng}, \citenamefont {Zhan},
  \citenamefont {Zhu},\ and\ \citenamefont {Feng}}]{Shen}%
  \BibitemOpen
  \bibfield  {author} {\bibinfo {author} {\bibnamefont {Shen}, \bibfnamefont
  {W.}}, \bibinfo {author} {\bibfnamefont {B.}~\bibnamefont {Wang}}, \bibinfo
  {author} {\bibfnamefont {J.}~\bibnamefont {Feng}}, \bibinfo {author}
  {\bibfnamefont {W.}~\bibnamefont {Zhan}}, \bibinfo {author} {\bibfnamefont
  {Y.}~\bibnamefont {Zhu}}, \ and\ \bibinfo {author} {\bibfnamefont
  {E.}~\bibnamefont {Feng}}} (\bibinfo {year} {1989}),\ \href@noop {}
  {\bibfield  {journal} {\bibinfo  {journal} {Nuclear Physics A}\ }\textbf
  {\bibinfo {volume} {491}},\ \bibinfo {pages} {130}}\BibitemShut {NoStop}%
\bibitem [{\citenamefont {Sihver}\ \emph
  {et~al.}(2014{\natexlab{a}})\citenamefont {Sihver}, \citenamefont {Kohama},
  \citenamefont {Iida}, \citenamefont {Hashimoto}, \citenamefont {Iwase},\ and\
  \citenamefont {Niita}}]{SihverPHITS}%
  \BibitemOpen
  \bibfield  {author} {\bibinfo {author} {\bibnamefont {Sihver}, \bibfnamefont
  {L.}}, \bibinfo {author} {\bibfnamefont {A.}~\bibnamefont {Kohama}}, \bibinfo
  {author} {\bibfnamefont {K.}~\bibnamefont {Iida}}, \bibinfo {author}
  {\bibfnamefont {S.}~\bibnamefont {Hashimoto}}, \bibinfo {author}
  {\bibfnamefont {H.}~\bibnamefont {Iwase}}, \ and\ \bibinfo {author}
  {\bibfnamefont {K.}~\bibnamefont {Niita}}} (\bibinfo {year}
  {2014}{\natexlab{a}}),\ \href@noop {} {\bibfield  {journal} {\bibinfo
  {journal} {Nuclear Instruments and Methods in Physics Research B}\ }\textbf
  {\bibinfo {volume} {334}},\ \bibinfo {pages} {34}}\BibitemShut {NoStop}%
\bibitem [{\citenamefont {Sihver}\ \emph
  {et~al.}(2014{\natexlab{b}})\citenamefont {Sihver}, \citenamefont {Lantz},\
  and\ \citenamefont {Kohama}}]{Sihver2014}%
  \BibitemOpen
  \bibfield  {author} {\bibinfo {author} {\bibnamefont {Sihver}, \bibfnamefont
  {L.}}, \bibinfo {author} {\bibfnamefont {M.}~\bibnamefont {Lantz}}, \ and\
  \bibinfo {author} {\bibfnamefont {A.}~\bibnamefont {Kohama}}} (\bibinfo
  {year} {2014}{\natexlab{b}}),\ \href@noop {} {\bibfield  {journal} {\bibinfo
  {journal} {Physical Review C}\ }\textbf {\bibinfo {volume} {89}},\ \bibinfo
  {pages} {067602}}\BibitemShut {NoStop}%
\bibitem [{\citenamefont {Sihver}\ \emph
  {et~al.}(2012{\natexlab{a}})\citenamefont {Sihver}, \citenamefont {Lantz},
  \citenamefont {{T.~B\"{o}hlen}}, \citenamefont {Mairani}, \citenamefont
  {Cerutti},\ and\ \citenamefont {Ferrari}}]{SihverGEANT}%
  \BibitemOpen
  \bibfield  {author} {\bibinfo {author} {\bibnamefont {Sihver}, \bibfnamefont
  {L.}}, \bibinfo {author} {\bibfnamefont {M.}~\bibnamefont {Lantz}}, \bibinfo
  {author} {\bibnamefont {{T.~B\"{o}hlen}}}, \bibinfo {author} {\bibfnamefont
  {A.}~\bibnamefont {Mairani}}, \bibinfo {author} {\bibfnamefont
  {A.}~\bibnamefont {Cerutti}}, \ and\ \bibinfo {author} {\bibfnamefont
  {A.}~\bibnamefont {Ferrari}}} (\bibinfo {year} {2012}{\natexlab{a}}),\
  \href@noop {} {\enquote {\bibinfo {title} {A comparison of total reaction
  cross section models used in {FLUKA{,} GEANT4{,} and PHITS}},}\ }\bibinfo
  {howpublished} {IEEE Aerospace Conference},\ \bibinfo {note} {{DOI:
  10.1109/AERO.2012.6187014}}\BibitemShut {NoStop}%
\bibitem [{\citenamefont {Sihver}\ \emph
  {et~al.}(2012{\natexlab{b}})\citenamefont {Sihver}, \citenamefont {Lantz},
  \citenamefont {Takechi}, \citenamefont {Kohama}, \citenamefont {Ferrari},
  \citenamefont {Cerutti},\ and\ \citenamefont {Sato}}]{SihverFLUKA}%
  \BibitemOpen
  \bibfield  {author} {\bibinfo {author} {\bibnamefont {Sihver}, \bibfnamefont
  {L.}}, \bibinfo {author} {\bibfnamefont {M.}~\bibnamefont {Lantz}}, \bibinfo
  {author} {\bibfnamefont {M.}~\bibnamefont {Takechi}}, \bibinfo {author}
  {\bibfnamefont {A.}~\bibnamefont {Kohama}}, \bibinfo {author} {\bibfnamefont
  {A.}~\bibnamefont {Ferrari}}, \bibinfo {author} {\bibfnamefont
  {F.}~\bibnamefont {Cerutti}}, \ and\ \bibinfo {author} {\bibfnamefont
  {T.}~\bibnamefont {Sato}}} (\bibinfo {year} {2012}{\natexlab{b}}),\
  \href@noop {} {\bibfield  {journal} {\bibinfo  {journal} {Advances in Space
  Research}\ }\textbf {\bibinfo {volume} {49}},\ \bibinfo {pages}
  {812}}\BibitemShut {NoStop}%
\bibitem [{\citenamefont {Sihver}\ \emph {et~al.}(1993)\citenamefont {Sihver},
  \citenamefont {Tsao}, \citenamefont {Silberberg}, \citenamefont {Kanai},\
  and\ \citenamefont {Barghouty}}]{Sihver}%
  \BibitemOpen
  \bibfield  {author} {\bibinfo {author} {\bibnamefont {Sihver}, \bibfnamefont
  {L.}}, \bibinfo {author} {\bibfnamefont {C.}~\bibnamefont {Tsao}}, \bibinfo
  {author} {\bibfnamefont {R.}~\bibnamefont {Silberberg}}, \bibinfo {author}
  {\bibfnamefont {T.}~\bibnamefont {Kanai}}, \ and\ \bibinfo {author}
  {\bibfnamefont {A.}~\bibnamefont {Barghouty}}} (\bibinfo {year} {1993}),\
  \href@noop {} {\bibfield  {journal} {\bibinfo  {journal} {Physical Review C}\
  }\textbf {\bibinfo {volume} {47}},\ \bibinfo {pages} {1225}}\BibitemShut
  {NoStop}%
\bibitem [{\citenamefont {Singleterry}(2012)}]{Singleterry}%
  \BibitemOpen
  \bibfield  {author} {\bibinfo {author} {\bibnamefont {Singleterry},
  \bibfnamefont {R.}}} (\bibinfo {year} {2012}),\ \href@noop {} {\enquote
  {\bibinfo {title} {Space travel and the long tent pole},}\ }\bibinfo
  {howpublished} {presentation at LANL, San Ildefonso Auditory},\ \bibinfo
  {note} {and private communication from Dr. Singleterry, 2012}\BibitemShut
  {NoStop}%
\bibitem [{\citenamefont {Souza}\ \emph {et~al.}(2013)\citenamefont {Souza},
  \citenamefont {Carlson}, \citenamefont {Donangelo}, \citenamefont {Lynch},\
  and\ \citenamefont {Tsang}}]{Souza2013}%
  \BibitemOpen
  \bibfield  {author} {\bibinfo {author} {\bibnamefont {Souza}, \bibfnamefont
  {S.}}, \bibinfo {author} {\bibfnamefont {B.}~\bibnamefont {Carlson}},
  \bibinfo {author} {\bibfnamefont {R.}~\bibnamefont {Donangelo}}, \bibinfo
  {author} {\bibfnamefont {W.}~\bibnamefont {Lynch}}, \ and\ \bibinfo {author}
  {\bibfnamefont {M.}~\bibnamefont {Tsang}}} (\bibinfo {year} {2013}),\
  \href@noop {} {\bibfield  {journal} {\bibinfo  {journal} {Physical Review C}\
  }\textbf {\bibinfo {volume} {88}},\ \bibinfo {pages} {014607}}\BibitemShut
  {NoStop}%
\bibitem [{\citenamefont {Strizhak}(1957)}]{Strizhak}%
  \BibitemOpen
  \bibfield  {author} {\bibinfo {author} {\bibnamefont {Strizhak},
  \bibfnamefont {V.}}} (\bibinfo {year} {1957}),\ \href@noop {} {\bibfield
  {journal} {\bibinfo  {journal} {Soviet Physics - JETP}\ }\textbf {\bibinfo
  {volume} {4}},\ \bibinfo {pages} {769}}\BibitemShut {NoStop}%
\bibitem [{\citenamefont {Taieb}\ \emph {et~al.}(2003)\citenamefont {Taieb},
  \citenamefont {Schmidt}, \citenamefont {Tissan-Got}, \citenamefont
  {Armstrong}, \citenamefont {Benlliure}, \citenamefont {Bernas}, \citenamefont
  {Boudard}, \citenamefont {Casarejos}, \citenamefont {Czajkowski},
  \citenamefont {Enqvist}, \citenamefont {Legrain}, \citenamefont {Leray},
  \citenamefont {Mustapha}, \citenamefont {Pravikoff}, \citenamefont {Rejmund},
  \citenamefont {St\'ephan}, \citenamefont {Volant},\ and\ \citenamefont
  {Wlazlo}}]{p1000U}%
  \BibitemOpen
  \bibfield  {author} {\bibinfo {author} {\bibnamefont {Taieb}, \bibfnamefont
  {J.}}, \bibinfo {author} {\bibfnamefont {K.-H.}\ \bibnamefont {Schmidt}},
  \bibinfo {author} {\bibfnamefont {L.}~\bibnamefont {Tissan-Got}}, \bibinfo
  {author} {\bibfnamefont {P.}~\bibnamefont {Armstrong}}, \bibinfo {author}
  {\bibfnamefont {J.}~\bibnamefont {Benlliure}}, \bibinfo {author}
  {\bibfnamefont {M.}~\bibnamefont {Bernas}}, \bibinfo {author} {\bibfnamefont
  {A.}~\bibnamefont {Boudard}}, \bibinfo {author} {\bibfnamefont
  {E.}~\bibnamefont {Casarejos}}, \bibinfo {author} {\bibfnamefont
  {S.}~\bibnamefont {Czajkowski}}, \bibinfo {author} {\bibfnamefont
  {T.}~\bibnamefont {Enqvist}}, \bibinfo {author} {\bibfnamefont
  {R.}~\bibnamefont {Legrain}}, \bibinfo {author} {\bibfnamefont
  {S.}~\bibnamefont {Leray}}, \bibinfo {author} {\bibfnamefont
  {B.}~\bibnamefont {Mustapha}}, \bibinfo {author} {\bibfnamefont
  {M.}~\bibnamefont {Pravikoff}}, \bibinfo {author} {\bibfnamefont
  {F.}~\bibnamefont {Rejmund}}, \bibinfo {author} {\bibfnamefont
  {C.}~\bibnamefont {St\'ephan}}, \bibinfo {author} {\bibfnamefont
  {C.}~\bibnamefont {Volant}}, \ and\ \bibinfo {author} {\bibfnamefont
  {W.}~\bibnamefont {Wlazlo}}} (\bibinfo {year} {2003}),\ \href@noop {}
  {\bibfield  {journal} {\bibinfo  {journal} {Nuclear Physics A}\ }\textbf
  {\bibinfo {volume} {724}},\ \bibinfo {pages} {413}}\BibitemShut {NoStop}%
\bibitem [{\citenamefont {Takechi}\ \emph {et~al.}(2009)\citenamefont
  {Takechi}, \citenamefont {Fukuda}, \citenamefont {Mihara}, \citenamefont
  {Tanaka}, \citenamefont {Chinda}, \citenamefont {Matsumasa}, \citenamefont
  {Nishimoto}, \citenamefont {Matsumiya}, \citenamefont {Nakashima},
  \citenamefont {Matsubara}, \citenamefont {Matsuta}, \citenamefont
  {Minamisono}, \citenamefont {Ohtsubo}, \citenamefont {Izumikawa},
  \citenamefont {Momota}, \citenamefont {Suzuki}, \citenamefont {Yamaguchi},
  \citenamefont {Koyama}, \citenamefont {Shinozaki}, \citenamefont {Takahashi},
  \citenamefont {Takizawa}, \citenamefont {Matsuyama}, \citenamefont
  {Nakajima}, \citenamefont {Kobayashi}, \citenamefont {Hosoi}, \citenamefont
  {Suda}, \citenamefont {Sasaki}, \citenamefont {Sato}, \citenamefont
  {Kanazawa},\ and\ \citenamefont {Kitagawa}}]{Takechi}%
  \BibitemOpen
  \bibfield  {author} {\bibinfo {author} {\bibnamefont {Takechi}, \bibfnamefont
  {M.}}, \bibinfo {author} {\bibfnamefont {M.}~\bibnamefont {Fukuda}}, \bibinfo
  {author} {\bibfnamefont {M.}~\bibnamefont {Mihara}}, \bibinfo {author}
  {\bibfnamefont {K.}~\bibnamefont {Tanaka}}, \bibinfo {author} {\bibfnamefont
  {T.}~\bibnamefont {Chinda}}, \bibinfo {author} {\bibfnamefont
  {T.}~\bibnamefont {Matsumasa}}, \bibinfo {author} {\bibfnamefont
  {M.}~\bibnamefont {Nishimoto}}, \bibinfo {author} {\bibfnamefont
  {R.}~\bibnamefont {Matsumiya}}, \bibinfo {author} {\bibfnamefont
  {Y.}~\bibnamefont {Nakashima}}, \bibinfo {author} {\bibfnamefont
  {H.}~\bibnamefont {Matsubara}}, \bibinfo {author} {\bibfnamefont
  {K.}~\bibnamefont {Matsuta}}, \bibinfo {author} {\bibfnamefont
  {T.}~\bibnamefont {Minamisono}}, \bibinfo {author} {\bibfnamefont
  {T.}~\bibnamefont {Ohtsubo}}, \bibinfo {author} {\bibfnamefont
  {T.}~\bibnamefont {Izumikawa}}, \bibinfo {author} {\bibfnamefont
  {S.}~\bibnamefont {Momota}}, \bibinfo {author} {\bibfnamefont
  {T.}~\bibnamefont {Suzuki}}, \bibinfo {author} {\bibfnamefont
  {T.}~\bibnamefont {Yamaguchi}}, \bibinfo {author} {\bibfnamefont
  {R.}~\bibnamefont {Koyama}}, \bibinfo {author} {\bibfnamefont
  {W.}~\bibnamefont {Shinozaki}}, \bibinfo {author} {\bibfnamefont
  {M.}~\bibnamefont {Takahashi}}, \bibinfo {author} {\bibfnamefont
  {A.}~\bibnamefont {Takizawa}}, \bibinfo {author} {\bibfnamefont
  {T.}~\bibnamefont {Matsuyama}}, \bibinfo {author} {\bibfnamefont
  {S.}~\bibnamefont {Nakajima}}, \bibinfo {author} {\bibfnamefont
  {K.}~\bibnamefont {Kobayashi}}, \bibinfo {author} {\bibfnamefont
  {M.}~\bibnamefont {Hosoi}}, \bibinfo {author} {\bibfnamefont
  {T.}~\bibnamefont {Suda}}, \bibinfo {author} {\bibfnamefont {M.}~\bibnamefont
  {Sasaki}}, \bibinfo {author} {\bibfnamefont {S.}~\bibnamefont {Sato}},
  \bibinfo {author} {\bibfnamefont {M.}~\bibnamefont {Kanazawa}}, \ and\
  \bibinfo {author} {\bibfnamefont {A.}~\bibnamefont {Kitagawa}}} (\bibinfo
  {year} {2009}),\ \href@noop {} {\bibfield  {journal} {\bibinfo  {journal}
  {Physical Review C}\ }\textbf {\bibinfo {volume} {79}},\ \bibinfo {pages}
  {061601}}\BibitemShut {NoStop}%
\bibitem [{\citenamefont {{Tarr\'{i}o}}\ \emph {et~al.}(2011)\citenamefont
  {{Tarr\'{i}o}}, \citenamefont {Tassan-Got}, \citenamefont {Audouin},
  \citenamefont {Berthier}, \citenamefont {Duran}, \citenamefont {Ferrant},
  \citenamefont {Isaev}, \citenamefont {Naour}, \citenamefont {Paradela},
  \citenamefont {Stephan}, \citenamefont {Trubert}, \citenamefont {Abbondanno},
  \citenamefont {Aerts}, \citenamefont {F.{ \'{A}lvarez}-Velarde},
  \citenamefont {Andriamonje}, \citenamefont {Andrzejewski}, \citenamefont
  {Assimakopoulos}, \citenamefont {Badurek}, \citenamefont {Baumann},
  \citenamefont {F.{ Be\v{c}v\'{a}\v{r} }}, \citenamefont {Belloni},
  \citenamefont {Berthoumieux}, \citenamefont {F.{ Calvi\~{n}o}}, \citenamefont
  {Calviani}, \citenamefont {Cano-Ott}, \citenamefont {Capote}, \citenamefont
  {{Carrapi\c{c}o }and A. Carrillo~de Albornoz}, \citenamefont {Cennini},
  \citenamefont {Chepel}, \citenamefont {Chiaveri}, \citenamefont {Colonna},
  \citenamefont {Cortes}, \citenamefont {Couture}, \citenamefont {Cox},
  \citenamefont {Dahlfors}, \citenamefont {David}, \citenamefont {Dillmann},
  \citenamefont {Dolfini}, \citenamefont {Domingo-Pardo}, \citenamefont
  {Dridi}, \citenamefont {Eleftheriadis}, \citenamefont {Embid-Segura},
  \citenamefont {Ferrari}, \citenamefont {Ferreira-Marques}, \citenamefont
  {Fitzpatrick}, \citenamefont {Frais-Koelbl}, \citenamefont {Fujii},
  \citenamefont {Furman}, \citenamefont {Goncalves}, \citenamefont
  {{Gonz\'{a}lez}-Romero}, \citenamefont {Goverdovski}, \citenamefont
  {Gramegna}, \citenamefont {Griesmayer}, \citenamefont {Guerrero},
  \citenamefont {Gunsing}, \citenamefont {Haas}, \citenamefont {Haight},
  \citenamefont {Heil}, \citenamefont {Herrera-Martinez}, \citenamefont
  {Igashira}, \citenamefont {Jericha}, \citenamefont {Kadi}, \citenamefont
  {{K\"{a}ppeler}}, \citenamefont {Karadimos}, \citenamefont {Karamanis},
  \citenamefont {Kerveno}, \citenamefont {Ketlerov}, \citenamefont {Koehler},
  \citenamefont {Konovalov}, \citenamefont {Kossionides}, \citenamefont
  {{Krti\v{c}ka}}, \citenamefont {Lampoudis}, \citenamefont {Leeb},
  \citenamefont {Lederer}, \citenamefont {Lindote}, \citenamefont {Lopes},
  \citenamefont {Losito}, \citenamefont {Lozano}, \citenamefont {Lukic},
  \citenamefont {Marganiec}, \citenamefont {Marques}, \citenamefont {Marrone},
  \citenamefont {{Mart\'{i}nez}}, \citenamefont {Massimi}, \citenamefont
  {Mastinu}, \citenamefont {Mendoza}, \citenamefont {Mengoni}, \citenamefont
  {Milazzo}, \citenamefont {Moreau}, \citenamefont {Mosconi}, \citenamefont
  {Neves}, \citenamefont {Oberhummer}, \citenamefont {O'Brien}, \citenamefont
  {Oshima}, \citenamefont {Pancin}, \citenamefont {Papachristodoulou},
  \citenamefont {Papadopoulos}, \citenamefont {Patronis}, \citenamefont
  {Pavlik}, \citenamefont {Pavlopoulos}, \citenamefont {Perrot}, \citenamefont
  {Pigni}, \citenamefont {Plag}, \citenamefont {Plompen}, \citenamefont
  {Plukis}, \citenamefont {Poch}, \citenamefont {Praena}, \citenamefont
  {Pretel}, \citenamefont {Quesada}, \citenamefont {Rauscher}, \citenamefont
  {Reifarth}, \citenamefont {Rosetti}, \citenamefont {Rubbia}, \citenamefont
  {Rudolf}, \citenamefont {Rullhusen}, \citenamefont {Salgado}, \citenamefont
  {Santos}, \citenamefont {Sarchiapone}, \citenamefont {Sarmento},
  \citenamefont {Savvidis}, \citenamefont {Tagliente}, \citenamefont {Tain},
  \citenamefont {Tavora}, \citenamefont {Terlizzi}, \citenamefont {Vannini},
  \citenamefont {Vaz}, \citenamefont {Ventura}, \citenamefont {Villamarin},
  \citenamefont {Vlachoudis}, \citenamefont {Vlastou}, \citenamefont {Voss},
  \citenamefont {Walter}, \citenamefont {Wendler}, \citenamefont {Wiescher},\
  and\ \citenamefont {Wisshak}}]{n+BiTOF}%
  \BibitemOpen
  \bibfield  {author} {\bibinfo {author} {\bibnamefont {{Tarr\'{i}o}},
  \bibfnamefont {D.}}, \bibinfo {author} {\bibfnamefont {L.}~\bibnamefont
  {Tassan-Got}}, \bibinfo {author} {\bibfnamefont {L.}~\bibnamefont {Audouin}},
  \bibinfo {author} {\bibfnamefont {B.}~\bibnamefont {Berthier}}, \bibinfo
  {author} {\bibfnamefont {I.}~\bibnamefont {Duran}}, \bibinfo {author}
  {\bibfnamefont {L.}~\bibnamefont {Ferrant}}, \bibinfo {author} {\bibfnamefont
  {S.}~\bibnamefont {Isaev}}, \bibinfo {author} {\bibfnamefont {C.~L.}\
  \bibnamefont {Naour}}, \bibinfo {author} {\bibfnamefont {C.}~\bibnamefont
  {Paradela}}, \bibinfo {author} {\bibfnamefont {C.}~\bibnamefont {Stephan}},
  \bibinfo {author} {\bibfnamefont {D.}~\bibnamefont {Trubert}}, \bibinfo
  {author} {\bibfnamefont {U.}~\bibnamefont {Abbondanno}}, \bibinfo {author}
  {\bibfnamefont {G.}~\bibnamefont {Aerts}}, \bibinfo {author} {\bibnamefont
  {F.{ \'{A}lvarez}-Velarde}}, \bibinfo {author} {\bibfnamefont
  {S.}~\bibnamefont {Andriamonje}}, \bibinfo {author} {\bibfnamefont
  {J.}~\bibnamefont {Andrzejewski}}, \bibinfo {author} {\bibfnamefont
  {P.}~\bibnamefont {Assimakopoulos}}, \bibinfo {author} {\bibfnamefont
  {G.}~\bibnamefont {Badurek}}, \bibinfo {author} {\bibfnamefont
  {P.}~\bibnamefont {Baumann}}, \bibinfo {author} {\bibnamefont {F.{
  Be\v{c}v\'{a}\v{r} }}}, \bibinfo {author} {\bibfnamefont {F.}~\bibnamefont
  {Belloni}}, \bibinfo {author} {\bibfnamefont {E.}~\bibnamefont
  {Berthoumieux}}, \bibinfo {author} {\bibnamefont {F.{ Calvi\~{n}o}}},
  \bibinfo {author} {\bibfnamefont {M.}~\bibnamefont {Calviani}}, \bibinfo
  {author} {\bibfnamefont {D.}~\bibnamefont {Cano-Ott}}, \bibinfo {author}
  {\bibfnamefont {R.}~\bibnamefont {Capote}}, \bibinfo {author} {\bibfnamefont
  {C.}~\bibnamefont {{Carrapi\c{c}o }and A. Carrillo~de Albornoz}}, \bibinfo
  {author} {\bibfnamefont {P.}~\bibnamefont {Cennini}}, \bibinfo {author}
  {\bibfnamefont {V.}~\bibnamefont {Chepel}}, \bibinfo {author} {\bibfnamefont
  {E.}~\bibnamefont {Chiaveri}}, \bibinfo {author} {\bibfnamefont
  {N.}~\bibnamefont {Colonna}}, \bibinfo {author} {\bibfnamefont
  {G.}~\bibnamefont {Cortes}}, \bibinfo {author} {\bibfnamefont
  {A.}~\bibnamefont {Couture}}, \bibinfo {author} {\bibfnamefont
  {J.}~\bibnamefont {Cox}}, \bibinfo {author} {\bibfnamefont {M.}~\bibnamefont
  {Dahlfors}}, \bibinfo {author} {\bibfnamefont {S.}~\bibnamefont {David}},
  \bibinfo {author} {\bibfnamefont {I.}~\bibnamefont {Dillmann}}, \bibinfo
  {author} {\bibfnamefont {R.}~\bibnamefont {Dolfini}}, \bibinfo {author}
  {\bibfnamefont {C.}~\bibnamefont {Domingo-Pardo}}, \bibinfo {author}
  {\bibfnamefont {W.}~\bibnamefont {Dridi}}, \bibinfo {author} {\bibfnamefont
  {C.}~\bibnamefont {Eleftheriadis}}, \bibinfo {author} {\bibfnamefont
  {M.}~\bibnamefont {Embid-Segura}}, \bibinfo {author} {\bibfnamefont
  {A.}~\bibnamefont {Ferrari}}, \bibinfo {author} {\bibfnamefont
  {R.}~\bibnamefont {Ferreira-Marques}}, \bibinfo {author} {\bibfnamefont
  {L.}~\bibnamefont {Fitzpatrick}}, \bibinfo {author} {\bibfnamefont
  {H.}~\bibnamefont {Frais-Koelbl}}, \bibinfo {author} {\bibfnamefont
  {K.}~\bibnamefont {Fujii}}, \bibinfo {author} {\bibfnamefont
  {W.}~\bibnamefont {Furman}}, \bibinfo {author} {\bibfnamefont
  {I.}~\bibnamefont {Goncalves}}, \bibinfo {author} {\bibfnamefont
  {E.}~\bibnamefont {{Gonz\'{a}lez}-Romero}}, \bibinfo {author} {\bibfnamefont
  {A.}~\bibnamefont {Goverdovski}}, \bibinfo {author} {\bibfnamefont
  {F.}~\bibnamefont {Gramegna}}, \bibinfo {author} {\bibfnamefont
  {E.}~\bibnamefont {Griesmayer}}, \bibinfo {author} {\bibfnamefont
  {C.}~\bibnamefont {Guerrero}}, \bibinfo {author} {\bibfnamefont
  {F.}~\bibnamefont {Gunsing}}, \bibinfo {author} {\bibfnamefont
  {B.}~\bibnamefont {Haas}}, \bibinfo {author} {\bibfnamefont {R.}~\bibnamefont
  {Haight}}, \bibinfo {author} {\bibfnamefont {M.}~\bibnamefont {Heil}},
  \bibinfo {author} {\bibfnamefont {A.}~\bibnamefont {Herrera-Martinez}},
  \bibinfo {author} {\bibfnamefont {M.}~\bibnamefont {Igashira}}, \bibinfo
  {author} {\bibfnamefont {E.}~\bibnamefont {Jericha}}, \bibinfo {author}
  {\bibfnamefont {Y.}~\bibnamefont {Kadi}}, \bibinfo {author} {\bibfnamefont
  {F.}~\bibnamefont {{K\"{a}ppeler}}}, \bibinfo {author} {\bibfnamefont
  {D.}~\bibnamefont {Karadimos}}, \bibinfo {author} {\bibfnamefont
  {D.}~\bibnamefont {Karamanis}}, \bibinfo {author} {\bibfnamefont
  {M.}~\bibnamefont {Kerveno}}, \bibinfo {author} {\bibfnamefont
  {V.}~\bibnamefont {Ketlerov}}, \bibinfo {author} {\bibfnamefont
  {P.}~\bibnamefont {Koehler}}, \bibinfo {author} {\bibfnamefont
  {V.}~\bibnamefont {Konovalov}}, \bibinfo {author} {\bibfnamefont
  {E.}~\bibnamefont {Kossionides}}, \bibinfo {author} {\bibfnamefont
  {M.}~\bibnamefont {{Krti\v{c}ka}}}, \bibinfo {author} {\bibfnamefont
  {C.}~\bibnamefont {Lampoudis}}, \bibinfo {author} {\bibfnamefont
  {H.}~\bibnamefont {Leeb}}, \bibinfo {author} {\bibfnamefont {C.}~\bibnamefont
  {Lederer}}, \bibinfo {author} {\bibfnamefont {A.}~\bibnamefont {Lindote}},
  \bibinfo {author} {\bibfnamefont {I.}~\bibnamefont {Lopes}}, \bibinfo
  {author} {\bibfnamefont {R.}~\bibnamefont {Losito}}, \bibinfo {author}
  {\bibfnamefont {M.}~\bibnamefont {Lozano}}, \bibinfo {author} {\bibfnamefont
  {S.}~\bibnamefont {Lukic}}, \bibinfo {author} {\bibfnamefont
  {J.}~\bibnamefont {Marganiec}}, \bibinfo {author} {\bibfnamefont
  {L.}~\bibnamefont {Marques}}, \bibinfo {author} {\bibfnamefont
  {S.}~\bibnamefont {Marrone}}, \bibinfo {author} {\bibfnamefont
  {T.}~\bibnamefont {{Mart\'{i}nez}}}, \bibinfo {author} {\bibfnamefont
  {C.}~\bibnamefont {Massimi}}, \bibinfo {author} {\bibfnamefont
  {P.}~\bibnamefont {Mastinu}}, \bibinfo {author} {\bibfnamefont
  {E.}~\bibnamefont {Mendoza}}, \bibinfo {author} {\bibfnamefont
  {A.}~\bibnamefont {Mengoni}}, \bibinfo {author} {\bibfnamefont
  {P.}~\bibnamefont {Milazzo}}, \bibinfo {author} {\bibfnamefont
  {C.}~\bibnamefont {Moreau}}, \bibinfo {author} {\bibfnamefont
  {M.}~\bibnamefont {Mosconi}}, \bibinfo {author} {\bibfnamefont
  {F.}~\bibnamefont {Neves}}, \bibinfo {author} {\bibfnamefont
  {H.}~\bibnamefont {Oberhummer}}, \bibinfo {author} {\bibfnamefont
  {S.}~\bibnamefont {O'Brien}}, \bibinfo {author} {\bibfnamefont
  {M.}~\bibnamefont {Oshima}}, \bibinfo {author} {\bibfnamefont
  {J.}~\bibnamefont {Pancin}}, \bibinfo {author} {\bibfnamefont
  {C.}~\bibnamefont {Papachristodoulou}}, \bibinfo {author} {\bibfnamefont
  {C.}~\bibnamefont {Papadopoulos}}, \bibinfo {author} {\bibfnamefont
  {N.}~\bibnamefont {Patronis}}, \bibinfo {author} {\bibfnamefont
  {A.}~\bibnamefont {Pavlik}}, \bibinfo {author} {\bibfnamefont
  {P.}~\bibnamefont {Pavlopoulos}}, \bibinfo {author} {\bibfnamefont
  {L.}~\bibnamefont {Perrot}}, \bibinfo {author} {\bibfnamefont
  {M.}~\bibnamefont {Pigni}}, \bibinfo {author} {\bibfnamefont
  {R.}~\bibnamefont {Plag}}, \bibinfo {author} {\bibfnamefont {A.}~\bibnamefont
  {Plompen}}, \bibinfo {author} {\bibfnamefont {A.}~\bibnamefont {Plukis}},
  \bibinfo {author} {\bibfnamefont {A.}~\bibnamefont {Poch}}, \bibinfo {author}
  {\bibfnamefont {J.}~\bibnamefont {Praena}}, \bibinfo {author} {\bibfnamefont
  {C.}~\bibnamefont {Pretel}}, \bibinfo {author} {\bibfnamefont
  {J.}~\bibnamefont {Quesada}}, \bibinfo {author} {\bibfnamefont
  {T.}~\bibnamefont {Rauscher}}, \bibinfo {author} {\bibfnamefont
  {R.}~\bibnamefont {Reifarth}}, \bibinfo {author} {\bibfnamefont
  {M.}~\bibnamefont {Rosetti}}, \bibinfo {author} {\bibfnamefont
  {C.}~\bibnamefont {Rubbia}}, \bibinfo {author} {\bibfnamefont
  {G.}~\bibnamefont {Rudolf}}, \bibinfo {author} {\bibfnamefont
  {P.}~\bibnamefont {Rullhusen}}, \bibinfo {author} {\bibfnamefont
  {J.}~\bibnamefont {Salgado}}, \bibinfo {author} {\bibfnamefont
  {C.}~\bibnamefont {Santos}}, \bibinfo {author} {\bibfnamefont
  {L.}~\bibnamefont {Sarchiapone}}, \bibinfo {author} {\bibfnamefont
  {R.}~\bibnamefont {Sarmento}}, \bibinfo {author} {\bibfnamefont
  {I.}~\bibnamefont {Savvidis}}, \bibinfo {author} {\bibfnamefont
  {G.}~\bibnamefont {Tagliente}}, \bibinfo {author} {\bibfnamefont
  {J.}~\bibnamefont {Tain}}, \bibinfo {author} {\bibfnamefont {L.}~\bibnamefont
  {Tavora}}, \bibinfo {author} {\bibfnamefont {R.}~\bibnamefont {Terlizzi}},
  \bibinfo {author} {\bibfnamefont {G.}~\bibnamefont {Vannini}}, \bibinfo
  {author} {\bibfnamefont {P.}~\bibnamefont {Vaz}}, \bibinfo {author}
  {\bibfnamefont {A.}~\bibnamefont {Ventura}}, \bibinfo {author} {\bibfnamefont
  {D.}~\bibnamefont {Villamarin}}, \bibinfo {author} {\bibfnamefont
  {V.}~\bibnamefont {Vlachoudis}}, \bibinfo {author} {\bibfnamefont
  {R.}~\bibnamefont {Vlastou}}, \bibinfo {author} {\bibfnamefont
  {F.}~\bibnamefont {Voss}}, \bibinfo {author} {\bibfnamefont {S.}~\bibnamefont
  {Walter}}, \bibinfo {author} {\bibfnamefont {H.}~\bibnamefont {Wendler}},
  \bibinfo {author} {\bibfnamefont {M.}~\bibnamefont {Wiescher}}, \ and\
  \bibinfo {author} {\bibfnamefont {K.}~\bibnamefont {Wisshak}}} (\bibinfo
  {year} {2011}),\ \href@noop {} {\bibfield  {journal} {\bibinfo  {journal}
  {Physical Review C}\ }\textbf {\bibinfo {volume} {83}},\ \bibinfo {pages}
  {044620}},\ \bibinfo {note} {{n\_TOF} Collaboration}\BibitemShut {NoStop}%
\bibitem [{\citenamefont {Taylor}\ \emph {et~al.}(1955)\citenamefont {Taylor},
  \citenamefont {{O.~L\"{o}nsj\"{o} }},\ and\ \citenamefont {Bonner}}]{Taylor}%
  \BibitemOpen
  \bibfield  {author} {\bibinfo {author} {\bibnamefont {Taylor}, \bibfnamefont
  {H.}}, \bibinfo {author} {\bibnamefont {{O.~L\"{o}nsj\"{o} }}}, \ and\
  \bibinfo {author} {\bibfnamefont {T.}~\bibnamefont {Bonner}}} (\bibinfo
  {year} {1955}),\ \href@noop {} {\bibfield  {journal} {\bibinfo  {journal}
  {Physical Review}\ }\textbf {\bibinfo {volume} {100}},\ \bibinfo {pages}
  {174}}\BibitemShut {NoStop}%
\bibitem [{\citenamefont {Toneev}\ and\ \citenamefont
  {Gudima}(1983)}]{Toneev1983}%
  \BibitemOpen
  \bibfield  {author} {\bibinfo {author} {\bibnamefont {Toneev}, \bibfnamefont
  {V.}}, \ and\ \bibinfo {author} {\bibfnamefont {K.}~\bibnamefont {Gudima}}}
  (\bibinfo {year} {1983}),\ \href@noop {} {\bibfield  {journal} {\bibinfo
  {journal} {Nuclear Physics}\ }\textbf {\bibinfo {volume} {A400}},\ \bibinfo
  {pages} {173c}}\BibitemShut {NoStop}%
\bibitem [{\citenamefont {Townsend}\ and\ \citenamefont
  {Wilson}(1988)}]{Townsend}%
  \BibitemOpen
  \bibfield  {author} {\bibinfo {author} {\bibnamefont {Townsend},
  \bibfnamefont {L.}}, \ and\ \bibinfo {author} {\bibfnamefont
  {J.}~\bibnamefont {Wilson}}} (\bibinfo {year} {1988}),\ \href@noop {}
  {\bibfield  {journal} {\bibinfo  {journal} {Physical Review C}\ }\textbf
  {\bibinfo {volume} {37}},\ \bibinfo {pages} {892}}\BibitemShut {NoStop}%
\bibitem [{\citenamefont {Tripathi}\ \emph {et~al.}(1996)\citenamefont
  {Tripathi}, \citenamefont {Cucinotta},\ and\ \citenamefont {Wilson}}]{NASAp}%
  \BibitemOpen
  \bibfield  {author} {\bibinfo {author} {\bibnamefont {Tripathi},
  \bibfnamefont {R.}}, \bibinfo {author} {\bibfnamefont {F.}~\bibnamefont
  {Cucinotta}}, \ and\ \bibinfo {author} {\bibfnamefont {J.}~\bibnamefont
  {Wilson}}} (\bibinfo {year} {1996}),\ \href@noop {} {\bibfield  {journal}
  {\bibinfo  {journal} {Nuclear Instruments and Methods in Physics Research B}\
  }\textbf {\bibinfo {volume} {117}},\ \bibinfo {pages} {347}}\BibitemShut
  {NoStop}%
\bibitem [{\citenamefont {Tripathi}\ \emph {et~al.}(1997)\citenamefont
  {Tripathi}, \citenamefont {Cucinotta},\ and\ \citenamefont {Wilson}}]{NASAn}%
  \BibitemOpen
  \bibfield  {author} {\bibinfo {author} {\bibnamefont {Tripathi},
  \bibfnamefont {R.}}, \bibinfo {author} {\bibfnamefont {F.}~\bibnamefont
  {Cucinotta}}, \ and\ \bibinfo {author} {\bibfnamefont {J.}~\bibnamefont
  {Wilson}}} (\bibinfo {year} {1997}),\ \href@noop {} {\bibfield  {journal}
  {\bibinfo  {journal} {Nuclear Instruments and Methods in Physics Research B}\
  }\textbf {\bibinfo {volume} {129}},\ \bibinfo {pages} {11}}\BibitemShut
  {NoStop}%
\bibitem [{\citenamefont {Tripathi}\ \emph {et~al.}(1999)\citenamefont
  {Tripathi}, \citenamefont {Cucinotta},\ and\ \citenamefont {Wilson}}]{NASAl}%
  \BibitemOpen
  \bibfield  {author} {\bibinfo {author} {\bibnamefont {Tripathi},
  \bibfnamefont {R.}}, \bibinfo {author} {\bibfnamefont {F.}~\bibnamefont
  {Cucinotta}}, \ and\ \bibinfo {author} {\bibfnamefont {J.}~\bibnamefont
  {Wilson}}} (\bibinfo {year} {1999}),\ \href@noop {} {\bibfield  {journal}
  {\bibinfo  {journal} {Nuclear Instruments and Methods in Physics Research B}\
  }\textbf {\bibinfo {volume} {155}},\ \bibinfo {pages} {349}}\BibitemShut
  {NoStop}%
\bibitem [{\citenamefont {Tsang}\ \emph {et~al.}(1990)\citenamefont {Tsang},
  \citenamefont {Srinivasan},\ and\ \citenamefont {Azziz}}]{Tsang90}%
  \BibitemOpen
  \bibfield  {author} {\bibinfo {author} {\bibnamefont {Tsang}, \bibfnamefont
  {H.}}, \bibinfo {author} {\bibfnamefont {G.}~\bibnamefont {Srinivasan}}, \
  and\ \bibinfo {author} {\bibfnamefont {N.}~\bibnamefont {Azziz}}} (\bibinfo
  {year} {1990}),\ \href@noop {} {\bibfield  {journal} {\bibinfo  {journal}
  {Physical Review C}\ }\textbf {\bibinfo {volume} {42}},\ \bibinfo {pages}
  {1598}}\BibitemShut {NoStop}%
\bibitem [{\citenamefont {Ugryumov}\ \emph {et~al.}(2004)\citenamefont
  {Ugryumov}, \citenamefont {Kuznetsov}, \citenamefont {Basybekov},
  \citenamefont {Bialkowski}, \citenamefont {Budzanowski}, \citenamefont
  {Duysebaev}, \citenamefont {Duysebaev}, \citenamefont {Zholdybaev},
  \citenamefont {Ismailov}, \citenamefont {Kadyrzhanov}, \citenamefont
  {Kalpakchieva}, \citenamefont {Kugler}, \citenamefont {Kukhtina},
  \citenamefont {Kushniruk}, \citenamefont {Kuterbekov}, \citenamefont
  {Mukhambetzhan}, \citenamefont {Penionzhkevich}, \citenamefont {Sadykov},
  \citenamefont {Skwirczynska},\ and\ \citenamefont {Sobolev}}]{Ugryumov2004}%
  \BibitemOpen
  \bibfield  {author} {\bibinfo {author} {\bibnamefont {Ugryumov},
  \bibfnamefont {V.}}, \bibinfo {author} {\bibfnamefont {I.}~\bibnamefont
  {Kuznetsov}}, \bibinfo {author} {\bibfnamefont {K.}~\bibnamefont
  {Basybekov}}, \bibinfo {author} {\bibfnamefont {E.}~\bibnamefont
  {Bialkowski}}, \bibinfo {author} {\bibfnamefont {A.}~\bibnamefont
  {Budzanowski}}, \bibinfo {author} {\bibfnamefont {A.}~\bibnamefont
  {Duysebaev}}, \bibinfo {author} {\bibfnamefont {B.}~\bibnamefont
  {Duysebaev}}, \bibinfo {author} {\bibfnamefont {T.}~\bibnamefont
  {Zholdybaev}}, \bibinfo {author} {\bibfnamefont {K.}~\bibnamefont
  {Ismailov}}, \bibinfo {author} {\bibfnamefont {K.}~\bibnamefont
  {Kadyrzhanov}}, \bibinfo {author} {\bibfnamefont {R.}~\bibnamefont
  {Kalpakchieva}}, \bibinfo {author} {\bibfnamefont {A.}~\bibnamefont
  {Kugler}}, \bibinfo {author} {\bibfnamefont {I.}~\bibnamefont {Kukhtina}},
  \bibinfo {author} {\bibfnamefont {V.}~\bibnamefont {Kushniruk}}, \bibinfo
  {author} {\bibfnamefont {K.}~\bibnamefont {Kuterbekov}}, \bibinfo {author}
  {\bibfnamefont {A.}~\bibnamefont {Mukhambetzhan}}, \bibinfo {author}
  {\bibfnamefont {Y.}~\bibnamefont {Penionzhkevich}}, \bibinfo {author}
  {\bibfnamefont {B.}~\bibnamefont {Sadykov}}, \bibinfo {author} {\bibfnamefont
  {I.}~\bibnamefont {Skwirczynska}}, \ and\ \bibinfo {author} {\bibfnamefont
  {Y.}~\bibnamefont {Sobolev}}} (\bibinfo {year} {2004}),\ \href@noop {}
  {\bibfield  {journal} {\bibinfo  {journal} {Nuclear Physics A}\ }\textbf
  {\bibinfo {volume} {734}},\ \bibinfo {pages} {E53}}\BibitemShut {NoStop}%
\bibitem [{\citenamefont {Ugryumov}\ \emph {et~al.}(2005)\citenamefont
  {Ugryumov}, \citenamefont {Kuznetsov}, \citenamefont {Bialkowski},
  \citenamefont {Kugler}, \citenamefont {Kuterbekov}, \citenamefont {Kuhtina},
  \citenamefont {Kushniruk}, \citenamefont {Lyapin}, \citenamefont {Maslov},
  \citenamefont {Penionzhkevich}, \citenamefont {Sobolev}, \citenamefont
  {Trzaska}, \citenamefont {Tjurin}, \citenamefont {Khlebnikov},\ and\
  \citenamefont {Yamaletdinov}}]{Ugryumov2005}%
  \BibitemOpen
  \bibfield  {author} {\bibinfo {author} {\bibnamefont {Ugryumov},
  \bibfnamefont {V.}}, \bibinfo {author} {\bibfnamefont {I.}~\bibnamefont
  {Kuznetsov}}, \bibinfo {author} {\bibfnamefont {E.}~\bibnamefont
  {Bialkowski}}, \bibinfo {author} {\bibfnamefont {A.}~\bibnamefont {Kugler}},
  \bibinfo {author} {\bibfnamefont {K.}~\bibnamefont {Kuterbekov}}, \bibinfo
  {author} {\bibfnamefont {I.}~\bibnamefont {Kuhtina}}, \bibinfo {author}
  {\bibfnamefont {V.}~\bibnamefont {Kushniruk}}, \bibinfo {author}
  {\bibfnamefont {V.}~\bibnamefont {Lyapin}}, \bibinfo {author} {\bibfnamefont
  {V.}~\bibnamefont {Maslov}}, \bibinfo {author} {\bibfnamefont
  {Y.}~\bibnamefont {Penionzhkevich}}, \bibinfo {author} {\bibfnamefont
  {Y.}~\bibnamefont {Sobolev}}, \bibinfo {author} {\bibfnamefont
  {W.}~\bibnamefont {Trzaska}}, \bibinfo {author} {\bibfnamefont
  {G.}~\bibnamefont {Tjurin}}, \bibinfo {author} {\bibfnamefont
  {S.}~\bibnamefont {Khlebnikov}}, \ and\ \bibinfo {author} {\bibfnamefont
  {S.}~\bibnamefont {Yamaletdinov}}} (\bibinfo {year} {2005}),\ \href@noop {}
  {\bibfield  {journal} {\bibinfo  {journal} {Physics of Atomic Nuclei}\
  }\textbf {\bibinfo {volume} {68}},\ \bibinfo {pages} {16}}\BibitemShut
  {NoStop}%
\bibitem [{\citenamefont {Uozumi}\ \emph {et~al.}(2007)\citenamefont {Uozumi},
  \citenamefont {Evtoukhovitch}, \citenamefont {Fukuda}, \citenamefont
  {Imamura}, \citenamefont {Iwamoto}, \citenamefont {Kalinikov}, \citenamefont
  {Kallies}, \citenamefont {Khumutov}, \citenamefont {Kin}, \citenamefont
  {Koba}, \citenamefont {Koba}, \citenamefont {Kuchinski}, \citenamefont
  {Moisenko}, \citenamefont {Mzavia}, \citenamefont {Nakano}, \citenamefont
  {Samoilov}, \citenamefont {Tsamalaidze}, \citenamefont {Wakabayashia},\ and\
  \citenamefont {Yamashita}}]{Uozumi}%
  \BibitemOpen
  \bibfield  {author} {\bibinfo {author} {\bibnamefont {Uozumi}, \bibfnamefont
  {Y.}}, \bibinfo {author} {\bibfnamefont {P.}~\bibnamefont {Evtoukhovitch}},
  \bibinfo {author} {\bibfnamefont {H.}~\bibnamefont {Fukuda}}, \bibinfo
  {author} {\bibfnamefont {M.}~\bibnamefont {Imamura}}, \bibinfo {author}
  {\bibfnamefont {H.}~\bibnamefont {Iwamoto}}, \bibinfo {author} {\bibfnamefont
  {V.}~\bibnamefont {Kalinikov}}, \bibinfo {author} {\bibfnamefont
  {W.}~\bibnamefont {Kallies}}, \bibinfo {author} {\bibfnamefont
  {N.}~\bibnamefont {Khumutov}}, \bibinfo {author} {\bibfnamefont
  {T.}~\bibnamefont {Kin}}, \bibinfo {author} {\bibfnamefont {N.}~\bibnamefont
  {Koba}}, \bibinfo {author} {\bibfnamefont {Y.}~\bibnamefont {Koba}}, \bibinfo
  {author} {\bibfnamefont {N.}~\bibnamefont {Kuchinski}}, \bibinfo {author}
  {\bibfnamefont {A.}~\bibnamefont {Moisenko}}, \bibinfo {author}
  {\bibfnamefont {D.}~\bibnamefont {Mzavia}}, \bibinfo {author} {\bibfnamefont
  {M.}~\bibnamefont {Nakano}}, \bibinfo {author} {\bibfnamefont
  {V.}~\bibnamefont {Samoilov}}, \bibinfo {author} {\bibfnamefont
  {Z.}~\bibnamefont {Tsamalaidze}}, \bibinfo {author} {\bibfnamefont
  {G.}~\bibnamefont {Wakabayashia}}, \ and\ \bibinfo {author} {\bibfnamefont
  {Y.}~\bibnamefont {Yamashita}}} (\bibinfo {year} {2007}),\ \href@noop {}
  {\bibfield  {journal} {\bibinfo  {journal} {Nuclear Instruments and Metods in
  Physics Research A}\ }\textbf {\bibinfo {volume} {571}},\ \bibinfo {pages}
  {743}}\BibitemShut {NoStop}%
\bibitem [{\citenamefont {Venables}\ \emph {et~al.}(2015)\citenamefont
  {Venables}, \citenamefont {Smith},\ and\ \citenamefont {the {R Core
  Team}}}]{R}%
  \BibitemOpen
  \bibfield  {author} {\bibinfo {author} {\bibnamefont {Venables},
  \bibfnamefont {W.}}, \bibinfo {author} {\bibfnamefont {D.}~\bibnamefont
  {Smith}}, \ and\ \bibinfo {author} {\bibnamefont {the {R Core Team}}}}
  (\bibinfo {year} {2015}),\ \href@noop {} {\enquote {\bibinfo {title} {An
  introduction to {R. Notes on R: A} programming environment for data analysis
  and graphics version 3.2.0},}\ }\bibinfo {howpublished}
  {cran.r-project.org/doc/manuals/r-release/R-intro.pdf}\BibitemShut {NoStop}%
\bibitem [{\citenamefont {Walt}\ and\ \citenamefont {Beyster}(1955)}]{Walt}%
  \BibitemOpen
  \bibfield  {author} {\bibinfo {author} {\bibnamefont {Walt}, \bibfnamefont
  {M.}}, \ and\ \bibinfo {author} {\bibfnamefont {J.}~\bibnamefont {Beyster}}}
  (\bibinfo {year} {1955}),\ \href@noop {} {\bibfield  {journal} {\bibinfo
  {journal} {Physical Review}\ }\textbf {\bibinfo {volume} {98}},\ \bibinfo
  {pages} {677}}\BibitemShut {NoStop}%
\bibitem [{\citenamefont {Warner}\ \emph {et~al.}(1996)\citenamefont {Warner},
  \citenamefont {Patty}, \citenamefont {Voyles}, \citenamefont {Nadasen},
  \citenamefont {Becchetti}, \citenamefont {Brown}, \citenamefont {Esbensen},
  \citenamefont {Galonsky}, \citenamefont {Kolata}, \citenamefont {Kruse},
  \citenamefont {Lee}, \citenamefont {Ronningen}, \citenamefont {Schwandt},
  \citenamefont {von Schwarzenberg}, \citenamefont {Sherrill}, \citenamefont
  {Subotic}, \citenamefont {Wang},\ and\ \citenamefont {Zecher}}]{Warner}%
  \BibitemOpen
  \bibfield  {author} {\bibinfo {author} {\bibnamefont {Warner}, \bibfnamefont
  {R.}}, \bibinfo {author} {\bibfnamefont {R.}~\bibnamefont {Patty}}, \bibinfo
  {author} {\bibfnamefont {P.}~\bibnamefont {Voyles}}, \bibinfo {author}
  {\bibfnamefont {A.}~\bibnamefont {Nadasen}}, \bibinfo {author} {\bibfnamefont
  {F.}~\bibnamefont {Becchetti}}, \bibinfo {author} {\bibfnamefont
  {J.}~\bibnamefont {Brown}}, \bibinfo {author} {\bibfnamefont
  {H.}~\bibnamefont {Esbensen}}, \bibinfo {author} {\bibfnamefont
  {A.}~\bibnamefont {Galonsky}}, \bibinfo {author} {\bibfnamefont
  {J.}~\bibnamefont {Kolata}}, \bibinfo {author} {\bibfnamefont
  {J.}~\bibnamefont {Kruse}}, \bibinfo {author} {\bibfnamefont
  {M.}~\bibnamefont {Lee}}, \bibinfo {author} {\bibfnamefont {R.}~\bibnamefont
  {Ronningen}}, \bibinfo {author} {\bibfnamefont {P.}~\bibnamefont {Schwandt}},
  \bibinfo {author} {\bibfnamefont {J.}~\bibnamefont {von Schwarzenberg}},
  \bibinfo {author} {\bibfnamefont {B.}~\bibnamefont {Sherrill}}, \bibinfo
  {author} {\bibfnamefont {K.}~\bibnamefont {Subotic}}, \bibinfo {author}
  {\bibfnamefont {J.}~\bibnamefont {Wang}}, \ and\ \bibinfo {author}
  {\bibfnamefont {P.}~\bibnamefont {Zecher}}} (\bibinfo {year} {1996}),\
  \href@noop {} {\bibfield  {journal} {\bibinfo  {journal} {Physical Review C}\
  }\textbf {\bibinfo {volume} {54}},\ \bibinfo {pages} {1700}}\BibitemShut
  {NoStop}%
\bibitem [{\citenamefont {Wellisch}\ and\ \citenamefont {Axen}(1996)}]{Axen}%
  \BibitemOpen
  \bibfield  {author} {\bibinfo {author} {\bibnamefont {Wellisch},
  \bibfnamefont {H.}}, \ and\ \bibinfo {author} {\bibfnamefont
  {D.}~\bibnamefont {Axen}}} (\bibinfo {year} {1996}),\ \href@noop {}
  {\bibfield  {journal} {\bibinfo  {journal} {Physical Review C}\ }\textbf
  {\bibinfo {volume} {54}},\ \bibinfo {pages} {1329}}\BibitemShut {NoStop}%
\bibitem [{\citenamefont {Wu}\ and\ \citenamefont {Chang}(1978)}]{wuchang}%
  \BibitemOpen
  \bibfield  {author} {\bibinfo {author} {\bibnamefont {Wu}, \bibfnamefont
  {J.}}, \ and\ \bibinfo {author} {\bibfnamefont {C.}~\bibnamefont {Chang}}}
  (\bibinfo {year} {1978}),\ \href@noop {} {\bibfield  {journal} {\bibinfo
  {journal} {Physical Review C}\ }\textbf {\bibinfo {volume} {17}},\ \bibinfo
  {pages} {1540}}\BibitemShut {NoStop}%
\bibitem [{\citenamefont {Yariv}(2008)}]{Isabel3}%
  \BibitemOpen
  \bibfield  {author} {\bibinfo {author} {\bibnamefont {Yariv}, \bibfnamefont
  {Y.}}} (\bibinfo {year} {2008}),\ \href@noop {} {\enquote {\bibinfo {title}
  {{ISABEL --- INC} model for high-energy hadron-nucleus reactions},}\
  }\bibinfo {howpublished} {Proc. Joint ICTP-IAEA Advanced Workshop on Model
  Codes for Spallation Reactions, ICTP Trieste, Italy, 4-8 February 2008},\
  \bibinfo {note} {{INDC(NDS)-0530 Distr. SC, IAEA, Vienna, August 2008, pp.
  15--28}}\BibitemShut {NoStop}%
\bibitem [{\citenamefont {Yariv}\ and\ \citenamefont {Frankel}(1979)}]{Isabel}%
  \BibitemOpen
  \bibfield  {author} {\bibinfo {author} {\bibnamefont {Yariv}, \bibfnamefont
  {Y.}}, \ and\ \bibinfo {author} {\bibfnamefont {Z.}~\bibnamefont {Frankel}}}
  (\bibinfo {year} {1979}),\ \href@noop {} {\bibfield  {journal} {\bibinfo
  {journal} {Physical Review C}\ }\textbf {\bibinfo {volume} {20}},\ \bibinfo
  {pages} {2227}}\BibitemShut {NoStop}%
\bibitem [{\citenamefont {Yariv}\ and\ \citenamefont
  {Frankel}(1981)}]{Isabel2}%
  \BibitemOpen
  \bibfield  {author} {\bibinfo {author} {\bibnamefont {Yariv}, \bibfnamefont
  {Y.}}, \ and\ \bibinfo {author} {\bibfnamefont {Z.}~\bibnamefont {Frankel}}}
  (\bibinfo {year} {1981}),\ \href@noop {} {\bibfield  {journal} {\bibinfo
  {journal} {Physical Review C}\ }\textbf {\bibinfo {volume} {24}},\ \bibinfo
  {pages} {488}}\BibitemShut {NoStop}%
\bibitem [{\citenamefont {Zeitlin}\ \emph {et~al.}(2007)\citenamefont
  {Zeitlin}, \citenamefont {Guetersloh}, \citenamefont {Heilbronn},
  \citenamefont {Miller}, \citenamefont {Fukumura}, \citenamefont {Iwata},\
  and\ \citenamefont {Murakami}}]{Zeitlin}%
  \BibitemOpen
  \bibfield  {author} {\bibinfo {author} {\bibnamefont {Zeitlin}, \bibfnamefont
  {C.}}, \bibinfo {author} {\bibfnamefont {S.}~\bibnamefont {Guetersloh}},
  \bibinfo {author} {\bibfnamefont {L.}~\bibnamefont {Heilbronn}}, \bibinfo
  {author} {\bibfnamefont {J.}~\bibnamefont {Miller}}, \bibinfo {author}
  {\bibfnamefont {A.}~\bibnamefont {Fukumura}}, \bibinfo {author}
  {\bibfnamefont {Y.}~\bibnamefont {Iwata}}, \ and\ \bibinfo {author}
  {\bibfnamefont {T.}~\bibnamefont {Murakami}}} (\bibinfo {year} {2007}),\
  \href@noop {} {\bibfield  {journal} {\bibinfo  {journal} {Physical Review C}\
  }\textbf {\bibinfo {volume} {76}},\ \bibinfo {pages} {014911}}\BibitemShut
  {NoStop}%
\end{thebibliography}%


\begin{thebibliography}{999}


\bibitem{MCNP6}                                           
T. Goorley, M. James, T. Booth, F. Brown, J. Bull, L. J. Cox, J. Durkee, 
J. Elson, M. Fensin, R. A. Forster, J. Hendricks, H. G. Hughes, R. Johns, 
B. Kiedrowski, R. Martz, S. Mashnik, G. McKinney, D. Pelowitz, R. Prael, 
J. Sweezy, L. Waters, T. Wilcox, and T. Zukaitis,
Nucl.\ Technol.\ {\bf 180}, 298 (2012).

\bibitem{CEM03.03}                                           
S. G. Mashnik and A. J. Sierk,
CEM03.03 User Manual,
LANL Report LA-UR-12-01364, Los Alamos, 2012;
https://mcnp.lanl.gov/.

\bibitem{CEM}                                                           
K. K. Gudima, S. G. Mashnik, and V. D. Toneev,
Nucl.\ Phys.\ {\bf A401}, 329 (1983).

\bibitem{Trieste08}                                              
S. G. Mashnik, K. K. Gudima, R. E. Prael, A. J. Sierk, M. I. Baznat, 
and N. V. Mokhov,
CEM03.03 and LAQGSM03.03 Event Generators for the MCNP6, MCNPX, and MARS15
Transport Codes,
LANL Report LA-UR-08-2931, Los Alamos, 2008;
arXiv:0805.0751.

\bibitem{LAQGSM}                                                         
K. K. Gudima, S. G. Mashnik, and A. J. Sierk,
User Manual for the Code LAQGSM,
LANL Report LA-UR-01-6804, Los Alamos, 2001;
https://mcnp.lanl.gov/.

\bibitem{LAQGSM03.03}                                                    
S. G. Mashnik,
K. K. Gudima, N. V. Mokhov, and R. E. Prael,
LAQGSM03.03 Upgrade and Its Validation,
LANL Report LA-UR-07-6198, Los Alamos, 2007; arXiv:0709.173.

\bibitem{BertiniINC}                                        
H. W. Bertini,
Phys.\ Rev.\ {\bf 131}, 1801 (1963);
{\it ibid.}
{\bf 188},  1711 (1969).

\bibitem{ISABEL}                                           
Y. Yariv and Z. Frankel,
Phys.\ Rev.\ C {\bf 20}, 2227 (1979). 

\bibitem{INCL4.2}                                                 
A. Boudard, J. Cugnon, S. Leray, and C. Volant,
Phys.\ Rev.\ C {\bf 66}, 044615 (2002).

\bibitem{Cugnon}                                                 
A. Boudard, J. Cugnon, S. Leray, and C Volant, 
Nucl.\ Phys.\ {\bf A740}, 195 (2004);
D. Mancusi, A. Boudard, J. Cugnon,
J.-C. David, P. Kaitaniemi, and S. Leray,
Phys.\ Rev.\ C {\bf 90}, 054602 (2014).

\bibitem{Konobeyev}                                            
A. V. Konobeyev, and Y. A. Korovin, 
Kerntechnik {\bf 60}, 14 (1995).
 
\bibitem{Uozumi}                                         
Y. Uozumi, P. Evtoukhovitch, H. Fukuda, M. Imamura, H. Iwamoto,
V. Kalinikov, W. Kallies, N. Khumutov, T. Kin, N. Koba, Y. Koba,
N. Kuchinski, A. Moisenko, D. Mzavia, M. Nakano, V. Samoilov,
Z. Tsamalaidze, G. Wakabayashia, and Y. Yamaxhita,
Nucl.\ Instrum.\ Meth.\ A {\bf 571}, 743 (2007).

\bibitem{SantaFe2002}                                     
S. G. Mashnik, A. J. Sierk, and K. K. Gudima,
Complex Particle and Light Fragment Emission in the Cascade-Exciton Model
of Nuclear Reactions,
LANL Report LA-UR-02-5185, Los Alamos (2002); arXiv: nucl-th/0208048.

\bibitem{Fermi}                                            
E.~Fermi, 
Prog.\ Theor.\ Phys.\ {\bf 5}, 570 (1950).

\bibitem{SMM}                                               
J. Bondorf, A. Botvina, A. Iljinov, I. Mishustin, and K. Sneppen,
Phys.\ Rep.\ {\bf 257}, 133 (1995).

\bibitem{QMD}                                                  
J. Aichelin,
Phys.\ Rep.\ {\bf 202}, 233 (1991).

\bibitem{Spall-Handbook}                                       
D. Filges and F. Goldenbaum,
{\it Handbook of Spallation Research: Theory, Experiments and
Applications}, 
(WILEY-VCH Verlag GmbH \& Co., 2009).

\bibitem{David2015}                                      
J.-C. David, 
Eur.\ Phys.\ J.\ A {\bf 51}, 157 (2015);
arXiv:1505.0382.

\bibitem{Machner}                                         
H. Machner, D. Aschman, K. Baruth-Ram, J. Carter,
A. Cowley, F. Goldenbaum, B. Nangu, J. Pilcher,
E. Sideras-Haddad, J. Sellschop, F. Smit, B. Spoelstra, and
D. Steyn, Phys.\ Rev.\ C {\bf 73}, 044606 (2006).

\bibitem{Budzanowski}                                     
A. Budzanowski,
M. Fidelus, D. Filges, F. Goldenbaum,
H. Hodde, L. Jarczyk, B. Kamys, M. Kistryn, S. Kistryn,
S. Kliczewski, A. Kowalczyk, E. Kozik, P. Kulessa,
H. Machner, A. Magiera, B. Piskor-Ignatowicz, K. Pysz,
Z. Rudy, R. Siudak, and M. Wojciechowski,
Phys.\ Rev.\ C {\bf 78}, 024603 (2008).

\bibitem{BudzanowskiNi}                                  
A. Budzanowski, 
M. Fidelus, D. Filges, F. Goldenbaum,
H. Hodde, L. Jarczyk, B. Kamys, M. Kistryn, S. Kistryn,
S. Kliczewski, A. Kowalczyk, E. Kozik, P. Kulessa,
H. Machner, A. Magiera, B. Piskor-Ignatowicz, K. Pysz,
Z. Rudy, R. Siudak, and M. Wojciechowski,
Phys.\ Rev.\ C {\bf 82}, 034605 (2010).

\bibitem{EPJ-Plus2011}                                      
S. G. Mashnik,
Eur.\ Phys.\ J.\ Plus {\bf 126}, 49 (2011); arXiv:1011.4978.

\bibitem{Tokyo2014}                                            
S. G. Mashnik,                          
JPS Conf.\ Proc.\ {\bf 6}, 030143 (2015); arXiv:1407.2832.

\bibitem{Barashenkov1972}                                    
V. S. Barashenkov and V. D. Toneev, 
{\it Interaction of High Energy Particle
and Nuclei with Atomic Nuclei} (Atomizdat, Moscow, 1972).

\bibitem{Barashenkov1973}                                     
V.~S.~Barashenkov,
A.~S.~Iljinov,  N.~M.~Sobolevskii, and V.~D.~Toneev,
Usp.\ Fiz.\ Nauk {\bf 109}, 91 (1973)
[Sov.\ Phys\ ~Usp.\ {\bf 16}, 31 (1973)].

\bibitem{MEM}                                                  
K.~K.~Gudima, G.~A.~Ososkov, and V.~D.~Toneev,
Yad.\ Fiz.\ {\bf 21}, 260  (1975)
[Sov.\ J.\ Nucl.\ Phys.\ {\bf 21}, 138 (1975)].

\bibitem{MODEX}                                                 
S. G. Mashnik and V. D. Toneev,
MODEX---the Program for Calculation of the
Energy Spectra of Particles Emitted in the Reactions
of Pre-Equilibrium and Equilibrium Statistical Decays,
JINR Communication P4-8417, Dubna, USSR, 1974;
https://mcnp.lanl.gov/.

\bibitem{Amelin}                                              
N.~S.~Amelin, K.~K.~Gudima, and V.~D.~Toneev,
Yad.\ Fiz.\ {\bf 51}, 512 (1990)
[Sov.\ J.\ Nucl.\ Phys.\ {\bf 51}, 327 (1990)];
{\it ibid.}
{\bf 52}, 272 (1990)
[{\bf 52}, 172 (1990)].

\bibitem{GEM2}                                              
S. Furihata,
Nucl.\ Instr.\ Meth.\ B {\bf 171}, 252 (2000);
{\it Development of a Generalized Evaporation Model and Study of Residual
nuclei Production}, 
Ph.D. thesis, Tohoku University, Sendai, Japan, 2003.

\bibitem{GEANT4}                                               
I. Pshenichnov, A. Botvina, I. Mishustin, and W. Greiner,
Nucl.\ Instrum.\ Meth.\ B {\bf 268}, 604 (2010).

\bibitem{SHIELD}                                                
D. Hansen, A. L\"{u}hr, N. Sobolevsky, and N. Bassler,
Phys.\ in Medicine \& Biology {\bf 57}, 2393 (2012);
M. Hultqvist, M. Lazzeroni, A. Botvina, I. Gudowska,
N. Sobolevsky, and A. Brahme, 
{\it ibid.} {\bf 57}, 4369 (2012);
A. L\"{u}hr, D. Hansen, R. Teiwes, N. Sobolevsky, O. Jakel, and
N. Bassler, 
{\it ibid.} {\bf 57}, 5169  (2012).

\bibitem{PHITS}                                                
T. Ogawa, T. Sato, S. Hashimoto, and K. Niita,
Nucl.\ Instrum.\ Meth.\ A
{\bf 723}, 36  (2013);
T. Sato, K. Niita, N. Matsuda, S. Hashimoto, Y. Iwamoto, S. Noda, T. Ogawa,
H. Iwase, H. Nakashima, T. Fukahori, K. Okumura, T. Kai, S. Chiba, T. Furuta,
L. Sihver, 
J.\ Nucl.\ Sci.\ Technol.\ {\bf 50}, 913 (2013).

\bibitem{NIMA2014}                                             
S. G. Mashnik and L. M. Kerby, 
Nucl.\ Instrum.\ Meth.\ A {\bf 764}, 59 (2014); arXiv:1404.7820.

\bibitem{NIMB2015}                                             
L. M. Kerby and S. G. Mashnik,
Nucl.\ Instrum.\ Meth.\ B {\bf 356--357}, 135 (2015); arXiv:1505.00842.

\bibitem{T16}                                                   
S. G. Mashnik, A. J. Sierk, K. A. Van Riper, and W. B. Wilson,
Production and Validation of Isotope Production Cross Section Libraries
for Neutrons and Protons to 1.7 GeV,
LANL Report LA-UR-98-6000, Los Alamos, 1998; 
in: Proc.\ {\it  Fourth Int. Workshop on Simulating
Accelerator Radiation Environments (SARE-4),
Knoxville, TN, USA, September 13--16, 1998},
edited by T. A. Gabriel
(ORNL, Oak Ridge, USA, 1999, pp. 151--162); 
arXiv:nucl-th/9812071;
our T-16 Library ``T-16 Lib'' is updated permanently when new experimental 
data became available to us.

\bibitem{Green}                                               
R. E. L. Green, R. G. Korteling, J. M. D'Auria, K. P. Jackson, and
R. L. Helmer,
Phys.\ Rev.\ C {\bf 35}, 1341 (1987).

\bibitem{Benecke}                                              
J. Benecke, T. Chou, C. Yang, and E. Yen, 
Phys.\ Rev.\ {\bf 188}, 2159 (1969).

\bibitem{Fidelus}                                              
M. Fidelus, 
{\it Model description of proton induced
fragmentation of atomic nuclei}, 
Ph.D. thesis, Cracow University, Poland, 2010.

\bibitem{Leslie-thesis}                                        
L. M. Kerby,
{\it Precompound Emission of Energetic Light Fragments in Spallation Reactions}, 
Ph. D. thesis, University of Idaho, 2015.

\bibitem{RMPmanuscript}                                        
L. M. Kerby, S. G. Mashnik, K. K. Gudima, A. J. Sierk, J. S. Bull, 
and M. R. James,
Production of Energetic Heavy Clusters in CEM and MCNP6,
LANL Report LA-UR-15-29524, Los Alamos, 2015;
Trans. Am. Nucl. Soc.\ {\bf 114}, 649 (2016). 

\bibitem{KonFisch2014}                                          
A. Y. Konobeyev and U. Fischer, 
{\it Status of Evaluation of $^9$Be DPA and Gas Production Cross-Sections
at Neutron Incident Energies up to 200 MeV}, presentation
at the Fall 2014 Nuclear Data Week, 24--28 November
2014, NEA, Issy-le-Moulineaux, France; 
www.oecdnea.org/dbdata/meetings/nov2014/;
{\it Evaluation of Atomic Displacement and
Gas Production Cross-Section for $^9$Be
Irradiated with Neutrons at Energies
up to 200 MeV},
Institut f\"{u}r Neutronenphysik und Reaktortechnik,
Karlsruher Institut f\"{u}r Technologie (KIT) 
Technical Report, 2015.

\bibitem{ABLA}                                                 
A. R. Junghans,
M. de Jong, H.-G. Clerc, A. V. Ignatyuk, 
G. A. Kudyaev, and K.-H. Schmidt, 
Nucl.\ Phys.\ {\bf A629}, 635 (1998);
J.-J. Gaimard and K.-H. Schmidt, 
{\it ibid.}
{\bf A531}, 709 (1991).

\bibitem{TALYS}                                               
A. Koning, S. Hilaire, and M. Duijvestijn 
in {\it Proc. Int. Conf. on Nuclear Data for Sci. \& Techn. (ND2004), 
September 26--October 1, 2004, Santa Fe, NM, USA}, edited by
R. Haight, M. Chadwick, T. Kawano, and P. Talou, (AIP
Conf.\ Proc.\ {\bf 769}, 1154 (2005);
http://www.talys.eu/.

\bibitem{Meier1989}                                            
M. Meier, W. Amian, C. Goulding, G. Morgan, and C. Moss,
 Nucl.\ Sci.\ Eng.\ {\bf 110}, 289 (1992).

\bibitem{Meier1992}                                            
M. Meier,  D. Clark, C. Goulding, J. McClelland, G. Morgan,
C. Moss, and W. Mian,
Nucl.\ Sci.\ Eng.\ {\bf 102}, 310  (1989).

\bibitem{Ericson}                                                 
T. Ericson,
Adv.\ Phys.\ {\bf 9}, 425 (1960).

\bibitem{Williams}                                               
F. C. Williams Jr.,
Phys.\ Lett.\ B {\bf 31}, 184 (1970).

\bibitem{Williams2}                                              
F. C. Williams Jr.,
Nucl.\ Phys.\ {\bf A161}, 231 (1971).

\bibitem{Ribansky}                                              
I. Ribansky, P. Oblozinsky, and E. Betak,
Nucl.\ Phys.\ {\bf A205}, 545 (1973).

\bibitem{Mantzouranis}                                            
G. Mantzouranis, H. A. Weidenm\"uller, and D. Agassi,
Z.\ Phys.\ A {\bf 276}, 145 (1976).

\bibitem{Kalbach88}                                              
C. Kalbach,
Phys.\ Rev.\ C {\bf 37}, 2350 (1988).

\bibitem{MCNPX}                                              
D. B. Pelowitz, J. W. Durkee, J. S. Elson,
M. L. Fensin, J. S. Hendricks, M. R. James,
R. C. Johns, G. W. McKinney, S. G. Mashnik,
J. M. Verbeke, L. S. Waters, and T. A. Wilcox,
MCNPX 2.7.0 Extensions,
LANL Report LA-UR-11-02295, Los Alamos, 2011
and refrences therein;
https://mcnpx.lanl.gov/.

\bibitem{MARS15}                                              
N. V. Mokhov, K. K. Gudima, C. C. James, M. A. Kostin,
S. G. Mashnik, E. Ng, J.-F. Ostiguy, I. L. Rakhno, A. J. Sierk, 
and S. I. Striganov,
Rad.\ Prot.\ Dosim.\ {\bf 116}, 99 (2005);
http://www-ap.fnal.gov/MARS/.

\bibitem{Dostrovsky}                                              
I. Dostrovsky, Z. Frankel, and G. Friedlander,
Phys.\ Rev.\ {\bf 116}, 683 (1959);
I. Dostrovsky, P. Rabinowitz, and R. Bivins,
{\it ibid.}
{\bf 111}, 1659 (1958).

\bibitem{NASA}                                              
R. Tripathi, F. Cucinotta, and J. Wilson, 
Nucl.\ Instrum.\ Meth.\ B {\bf 117}, 347  (1996);
{\it ibid.}
{\bf 129}, 11 (1997);
{\it ibid.}
{\bf 155}, 349 (1999).

\bibitem{BP}                                              
V. S. Barashenkov and A. Polanski, 
Electronic Guide for Nuclear Cross-Sections, 
JINR Communication E2-94-417, JINR, Dubna, Russia, 1994.

\bibitem{Kalbach}                                              
C. Kalbach, 
J.\ Phys.\ G {\bf 24}, 847  (1998).

\bibitem{Tsang}                                              
H. Tsang, G. Srinivasan, and N. Azziz, 
Phys.\ Rev.\ C {\bf 42}, 1598 (1990).

\bibitem{CEM03.02}                                              
S. G. Mashnik, R. E. Prael, and K. K. Gudima,
Implementation of CEM03.01 into MCNP6 and
its Verification and Validation Running through MCNP6.
CEM03.02 Upgrade,
LANL Report LA-UR-06-8652, Los Alamos, 2007;
https://mcnp.lanl.gov/.

\bibitem{Prael}                                              
R. Prael, A. Ferrari, R. Tripathi, and A. Polanski,
Comparison of nucleon cross section parametrization
methods for medium and high energies,
in: Proc.\ {\it Fourth Int.\ Workshop on Simulating
Accelerator Radiation Environments (SARE-4),
Knoxville, TN, USA, September 13--16, 1998},
edited by T. A. Gabriel
(ORNL, Oak Ridge, USA, 1999, pp. 171--181). 

\bibitem{FY2014}                                              
L. M. Kerby and S. G. Mashnik, 
LANL Fiscal Year 2014 Report,
LANL Report LA-UR-14-27533, Los Alamos, 2014;
www.osti.gov/scitech/biblio/1162152.

\bibitem{Carlson}                                               
R. Carlson, 
At.\ Data Nucl.\ Data Tables {\bf 63}, 93 (1996).

\bibitem{CC}                                               
V. S. Barashenkov, 
{\it Cross sections of interaction
of particles and nuclei with nuclei}
(JINR, Dubna, Russia, 1993);
M. Mazarakis and W. Stephens, 
Phys.\ Rev.\ C {\bf 7}, 1280 (1973);
M. Takechi {\it et al.},
{\it ibid.} {\bf 79}, 061601 (2009).

\bibitem{TCC}                                               
A. Golovchenko, J. Skvarc, N. Yasuda, M. Giacomelli,
S. Tretyakova, R. Ilic, R. Bimbot, M. Toulemonde, and
T. Murakami, Phys.\ Rev.\ C {\bf 66}, 014609 (2002);
C. Zeitlin, S. Guetersloh, L. Heilbronn, J. Miller, A. Fukumura,
Y. Iwata, and T. Murakami 
{\it ibid.}
{\bf 76}, 014911 (2007).

\bibitem{WuChang}                                               
J. Wu and C. Chang,
Phys.\ Rev.\ C {\bf 17}, 1540 (1978);
P. E. Hodgson and E. Betak,   
Phys. Rep. {\bf 374}, 1 (2003).

\bibitem{Betak76}                                               
E. Betak, 
Acta Phys.\ Slovaca {\bf 26}, 21 (1976).

\bibitem{Blideanu}                                               
V. Blideanu, F. Lecolley, J. Lecolley, T. Lefort, N. Marie,
A. Atac and G. Ban, B. Bergenwall, J. Blomgren, S. Dangtip, 
K. Elmgren, P. Eudes, Y. Foucher, A. Guertin,
F. Haddad, A. Hildebrand, C. Johansson, O. Jonsson,
M. Kerveno, T. Kirchner, J. Klug, C. Brun, C. Lebrun,
M. Louvel, P. Nadel-Turonski, L. Nilsson, N. Olsson,
S. Pomp, A. Prokoev, P.-U. Renberg, G. Riviere,
I. Slypen, L. Stuttge, U. Tippawan, and M.  Osterlund,
Phys.\ Rev.\ C {\bf 70}, 014607 (2004).

\bibitem{NumericalRecipes}                                      
W. Press, W. Vetterling, S. Teukolsky, and B. Flannery
{\it Numerical Recipes in C: The Art of Scientic Computing} 
(Cambridge University Press, 1992).

\bibitem{DCM}                                                   
V.~D.~Toneev and K.~K.~Gudima,
Nucl.\ Phys.\ {\bf A400}, 173c (1983) .

\bibitem{Gudima:83a}                                             
K.~K.~Gudima,
 G.~R\"opke, H.~Schulz, and V.~D.~Toneev,
The Coalescence Model and Pauli Quenching in High-Energy Heavy-Ion Collisions,
JINR Preprint JINR-E2-83-101, Dubna, 1983;
H.~Schulz, G.~R\"opke, K.~K.~Gudima, and V.~D.~Toneev,
Phys.\ Lett.\ B {\bf 124}, 458 (1983).

\bibitem{Coal2015}                                             
L. M. Kerby and S. G. Mashnik,
An Expanded Coalescence Model within the Intranuclear 
Cascade of CEM,
LANL Report LA-UR-15-20322, Los Alamos, 2015.

\bibitem{Green480}                                               
R. E. L. Green, R. G. Korteling, and  K. P. Jackson, 
Phys.\ Rev.\ C {\bf 29}, 1806 (1984). 

\bibitem{Varenna06}                                              
K. K. Gudima and S. G. Mashnik,
{\it Proc.\ 11th Int.\ Conf.\ on Nuclear Reaction Mechanisms,
Varenna, Italy, June 12--16, 2006},
edited by E. Gadioli (2006, 
pp. 525--534);
E-print: nucl-th/0607007.

\bibitem{Amelin86}                                                
N.~S.~Amelin,
Simulation of Nuclear Collisions at High Energy in the Framework of 
the Quark-Gluon String Model,
JINR Communication JINR-86-802, Dubna, 1986;
A.~B.~Kaidalov,
Yad.\ Fiz.\ {\bf 45}, 1452  (1987)
[Sov.\ J.\ Nucl.\ Phys.\ {\bf 45}, 902 (1987)].

\bibitem{Mocko}                                         
M. Mocko, 
{\it Rare isotope production},
Ph. D. thesis, Michigan State University, 2006;
M. Mocko, B. Tsang, L. Andronenko, M. Andronenko, F. Delaunay,
M. Famiano, T. Ginter, V. Henzl, D. Henzlova,
H. Hua, S. Lukyanov, W. Lynch, A. Rogers, M. Steiner,
A. Stolz, O. Tarasov, M.-J. van Goethem, G. Verde,
W. Wallace, and A. Zalessov,
Phys.\ Rev.\ C {\bf 74}, 054612 (2006).

\bibitem{Heilbronn2007}                                
L. Heilbronn, C. Zeitlin, Y. Iwata, T. Murakami, H. Iwase,
T. Nakamura, T. Nunomiya, H. Sato, H. Yashima, R. M.
Ronningen, and K. Ieki,
Nucl.\ Sci.\ Eng.\ {\bf 157}, 142 (2007);
T. Nakamura and L. Heilbronn,
{\it Handbook on Secondary Particle Production and Transport by 
High-Energy Heavy Ions} 
(World Scientic, Singapore, 2006).

\bibitem{Frankel79}                                                     
S. Frankel, W. Frati, M. Gazzaly,
Y. D. Bayukov, V. I. Efremenko, G. A. Leksin, N. A. Nikiforov, 
V. I. Tchistilin, Y. M. Zaitsev, and C. F. Perdrisat,  
Phys.\ Rev.\ C {\bf 20}, 2257 (1979).

\bibitem{Jacak1987}                                                        
B. Jacak, G. Westfall, G. Crawley, D. Fox, C. Gelbke, and
L. Harwood, 
Phys.\ Rev.\ C {\bf 35}, 1751 (1987).

\bibitem{Lemaire78}                                                          
M.-C. Lemaire, S. Nagamiya, O. Chamberlain, G. Shapiro, S. Schnetzer, 
H. Steiner, and I. Tanihata, 
Tables of Light-Fragment Inclusive Cross Sections 
in Relativistic Heavy Ion Collisions. Part I. C + C, C + Pb, Ne + NaF, 
Ne + Cu, Ne + Pb $\to \pi^{\pm}$, p, d, t, $^3$He; E$_{BEAM} = 800$ MeV/A,
Lawrence Berkeley National Laboratory Report LBL-8463, 1978;
S. Nagamiya, M. -C. Lemaire, 
E. Moeller, S. Schnetzer, G. Shapiro, H. Steiner, and I. Tanihata,
Phys.\ Rev.\ C {\bf 24}, 971 (1981).

\bibitem{Toppi2016}                                                        
M. Toppi et al.,
Phys.\ Rev.\ C {\bf 93}, 064601 (2016).

\bibitem{Franz}                                                            
J. Franz, J., P. Koncz, E. Roessle, C. Sauerwein, H. Schmitt,
K. Schmoll, J. Eroe, Z. Fodor, J. Kecskemeti, Z. Kovacs,
and Z. Seres,
Nucl.\ Phys.\ {\bf A510}, 774 (1990).

\bibitem{Schumacher}                                                   
R. Schumacher, R., G. Adams, D. Ingham, J. Matthews,
W. Sapp, R. Turley, R. Owens, and B. Roberts,
Phys.\ Rev.\ C {\bf 25}, 2269 (1982).

\bibitem{Nakamoto}                                                     
T. Nakamoto, K. Ishibashi, N. Matsufuji, N. Shigyo, K. Maehata,
H. Arima, S. Meigo, H. Takada, S. Chiba, and M. Numajiri,
J.\ Nucl.\ Sci.\ Technol.\ {\bf 34}, 860 (1997).

\bibitem{Benlliure}                                                     
J. Benlliure,  P. Armbruster, M. Bernas, A. Boudard, J. Dufour,
T. Enqvist, R. Legrain, S. Leray, B. Mustapha, F. Rejmund,
K.-H. Schmidt, C. Stephan, L. Tasaan-Got, and C. Volant,
Nucl.\ Phys.\ {\bf A683}, 513 (2001);
F. Rejmund, B. Mustapha, P. Armbruster, J. Benlliure,
M. Bernas, A. Boudard, J. Dufour, T. Enqvist, R. Legrain,
S. Leray, K.-H. Schmidt, C. Stephan, J. Taieb, L. Tassan-Got, and C. Volant,
{\it ibid.} {\bf A683}, 540 (2001).

\bibitem{Fomichev}                                                      
A. Fomichev, V. Dushin, S. Soloviev, A. Fomichev, and
S. Mashnik,  
Neutron Induced Fission Cross Sections for $^{240}$Pu, $^{243}$Am, 
$^{209}$Bi, $^{nat}$W Measured Relative to $^{235}$U in the Energy
Range 1--350 MeV,
V. G. Khlopin Radium Institute Preprint RI-262, St. Petersburg, Russia, 2004;
LANL Report LA-UR-05-1533, Los Alamos, 2005;
https://mcnp.lanl.gov/.

\bibitem{Tarrio}                                                     
D. Tarr\'{i}o {\it et al.} (n\_TOF Collaboration),
Phys.\ Rev.\ C {\bf 83}, 044620 (2011).

\bibitem{GENXS}                                                  
R. Prael, 
Tally Edits for the MCNP6 GENXS Option,
LANL Report, LA-UR-11-02146, Los Alamos, 2011;
https://mcnp.lanl.gov/.

\bibitem{GENXS-extend}                                                  
L. M. Kerby, S. G. Mashnik, and J. S. Bull,
GENXS Expansion to Include Fragment Spectra of Heavy Ions,
LANL Report LA-UR-15-24006, Los Alamos (2015);
http://www.osti.gov/scitech/biblio/1183395/.

\end{thebibliography}
\end{document}